# Earth-strength magnetic-field axis modulates non-spatial probabilistic decision-making in humans

In-Taek Oh[1], Yong-Hwan Kim[2], Soo-Chan Kim[3], Yongkuk Kim[4,*], Kwon-Seok Chae[1,5,*]

[1] Brain Science and Engineering Institute, Kyungpook National University, Daegu 41566, Republic of Korea.

[2] Neuroscience Program, Boise State University, Boise, ID 83725, United States.

[3] School of Electronic and Electrical Engineering, Institute of IT Convergence, Hankyong National University, Anseong 17579, Republic of Korea.

[4] Department of Mathematics, Kyungpook National University, Daegu 41566, Republic of Korea.

[5] Department of Biology Education, Kyungpook National University, Daegu 41566, Republic of Korea.

[*] Correspondence: yongkuk@knu.ac.kr, kschae@knu.ac.kr

## Abstract

Magnetoreception has been primarily studied in orientation and navigation, but whether Earth-strength magnetic fields influence non-spatial decisions remains unclear. Evidence for human magnetic responses is limited, and the transduction mechanism is unknown. Using a Go stone selection paradigm in which near-zero geomagnetic field and Larmor-frequency conditions previously showed lower black stone rates, we tested whether non-spatial choice depends on magnetic-field geometry. Fasted men played in pairs under ten static two-seat configurations varied component-wise at near-constant north-seat intensity. The north-seat black stone rate was lower under the two configurations with horizontal fields transverse to the geomagnetic field than under the other eight, which did not differ detectably. The difference emerged promptly, without evidence of outcome-driven learning or configuration-dependent associations with the partner's preceding actions, and process analyses linked it to the joint parity–choice outcome without isolating either step. A Larmor-frequency difference observed over the ambient field was not detected over horizontal-only or vertical-only backgrounds, suggesting a potential magnetoresponsive axis, although the background difference was statistically unresolved. Frequency, light-spectrum, and posture analyses support a radical-pair mechanism as a leading working hypothesis. These findings associate the geomagnetic field axis with human non-spatial choice, providing a paradigm linking magnetic sensing to decision processes.

## Introduction

Magnetoreception—the use of the geomagnetic field (GMF) in orientation, navigation, and context-dependent behavior—is documented across many taxa, yet its underlying sensory mechanism remains unresolved.[1–4] Three main candidate mechanisms have been proposed: light-dependent radical-pair chemistry, magnetite-based transduction, and electromagnetic induction.[1,2,4] Among these, the radical-pair framework is the most extensively developed. Spin-correlated radical pairs generated in photoactive proteins, particularly cryptochromes, can undergo magnetic-field-dependent singlet–triplet interconversion at Earth-strength fields, potentially producing direction- and frequency-specific effects.[5–8] In its canonical form, this mechanism is axial: complete reversal of the magnetic-field vector is not expected to alter the response, whereas reversal of an individual component can change the full-vector axis and thus the predicted response. Cryptochrome 4 from a migratory songbird is magnetically sensitive in vitro, but its relevance at GMF strengths in vivo and its downstream transduction are unresolved.[9] In contrast, pigeon studies have implicated a light-independent vestibular route consistent with electromagnetic induction[10,11] and a superparamagnetic macrophage-associated route in the liver.[12] Thus, multiple magnetosensory mechanisms remain plausible, and recent reviews caution against premature commitment to a single model.[13–15] Given this uncertainty, precisely defining the local magnetic-field vector and testing its behavioral consequences may provide a mechanism-neutral approach to investigating magnetoreception.

Evidence for magnetic responses in humans remains limited and heterogeneous.[4,13,16] Human visual thresholds have been reported to change when a weak magnetic field is aligned with the line of sight.[17,18] An electroencephalography study reported alpha-band suppression in response to controlled rotations of Earth-strength fields and interpreted it as compatible with a magnetite-based mechanism.[19] Our previous behavioral studies in fasted men reported blue light-dependent geomagnetic food orientation[20] and, in a separate orientation task, sensitivity to a radiofrequency (RF) field at the electron Larmor frequency together with behavior consistent with a non-canonical inclination compass.[21] A subsequent non-spatial behavioral study found GMF-associated shifts in probabilistic binary choice in fasted men and women.[22] The light and RF dependencies observed in the orientation responses are consistent with a radical-pair mechanism,[2,5,6,20,21] although light-independent routes have also been proposed in humans.[16] These findings suggest that humans may respond to controlled magnetic-field manipulations under specific experimental conditions, but they do not establish a common receptor or a general human magnetic sense and require independent replication. In particular, whether the geometry of Earth-strength magnetic fields influences non-spatial decision-making remains largely unexplored.

Decision-making under uncertainty is commonly modeled as probabilistic evidence accumulation or Bayesian inference.[23–26] The task in the present study is a repeated two-player parity game derived from Go stone selection, for which the mixed-strategy equilibrium predicts equal choice probabilities. Deviations from this expectation can arise through processes described by behavioral game theory and reinforcement learning,[27,28] as well as representativeness, over-alternation, streak misperception, and probability matching.[29–33] Notably, abstract decision-making can recruit map-like representations that reuse neural circuitry associated with spatial navigation,[34–36] providing a possible computational context through which directional sensory input could influence a non-spatial choice. In our earlier studies, black stone rates—the proportion of trials in which a player obtains the black stone—were reduced under near-zero GMF and Larmor-frequency RF conditions.[22,37] A separate large-scale analysis of Go stone selection records found that black stone rate differed with professional rank and amateur Dan level.[38] However, deviation from the theoretical probability of 0.5 alone does not demonstrate a magnetic-field effect, and a more direct test requires comparing choice probabilities across controlled magnetic-field configurations. The two-step structure of the task also allows us to examine the processes contributing to the final outcome: participants reported that, as player 1 (P1), they could not intentionally control whether the concealed number of stones was odd or even, whereas, as player 2 (P2), they intentionally chose one or two stones.[22,37]

Here, we tested whether non-spatial probabilistic choice depends on the geometry of controlled Earth-strength magnetic-field configurations. Using the Go stone selection paradigm,[22,37,38] fasted men were tested under ten two-seat static-field configurations in which the north-seat total field intensity was held approximately constant while individual field components were systematically altered. We found that the north-seat black stone rate was lower specifically under the two configurations in which the horizontal field was oriented approximately transverse to the GMF than under the other eight configurations, among which no difference was detected. We further examined the temporal and decision-process characteristics of this difference and compared the geometric dependence with near-zero, Larmor-frequency RF, and light-spectrum manipulations to constrain potential magnetosensory mechanisms.

## Results

### Component-wise dissection of Earth-strength field geometry

To determine whether non-spatial probabilistic choice depends on magnetic-field geometry, we used the two-player Go stone selection paradigm from our earlier studies[22,37] within a shared three-axis Helmholtz coil system (Fig. 1a, b). In each trial, P1 concealed a handful of white stones and P2 selected one or two black stones before the odd–even parity was revealed. The match or mismatch between parity and P2's choice determined which player received the black stones. Given the reduced black stone rate previously observed under near-zero GMF,[22] we systematically altered individual components of an Earth-strength field while maintaining north-seat total intensity at approximately 45 μT. The resulting ten static configurations formed five symmetric pairs: ambient control (Con) and vertical-component inversion, horizontal-only north (mN) and south (mS), transverse east (mE) and west (mW), nominal ±30° rotations (mNE and mNW), and vertical-only down and up (Fig. 1c, d). Three-axis vectors were measured at the glabella position of each seat (Table 1). Because the shared coil altered the fields at both seats, south-seat intensity varied, and each condition therefore represents a complete two-seat magnetic-field configuration rather than an isolated north-seat exposure. The ten-configuration dataset comprised 454 north-seat 20-trial sets from 47 male participants, representing 66 unique participant pairs and 182 pair-sessions (Fig. 1e; Supplementary Table 1). The primary endpoint was the north-seat black stone rate for each 20-trial set, and the primary comparison evaluated the transverse pair (mE and mW) against the eight non-transverse configurations (Table 2). Near-zero, Larmor-frequency (RF1, 1.260 MHz), and non-resonant (RF2, 1.890 MHz) exposures lay outside this intensity-matched family and were analyzed separately as reference conditions.

**Lower black stone rate specific to the transverse pair**

Inverting or removing the vertical component left the rate close to the ambient level (Con, 50.7%; vertical-component inversion, 49.8%; mN, 51.3%), and reversing the horizontal-only field had little effect (mS, 49.4%; Fig. 2a). In contrast, the rate was 46.7% under both transverse rotations of the horizontal field (nominal 90°; mE and mW). Smaller ±30° rotations did not lower the rate (mNE, 52.1%; mNW, 50.7%), nor did removing the horizontal component (vertical-only down, 51.3%; up, 50.9%). Thus, eight of the ten configurations yielded rates between 49.4% and 52.1%, whereas only the two transverse configurations fell below this range. Because the field dissection provided no basis for selecting a single reference configuration, we compared the black stone rate of the transverse configurations (mE and mW) with that of the eight non-transverse configurations weighted equally. To account for repeated participation and shared-session dependence, we incorporated north-role participant, south-role participant, and pair-session random intercepts into a crossed-role set-level binomial Hamiltonian Monte Carlo (HMC) analysis, with cluster-robust, wild-cluster, and participant-aggregated procedures as sensitivity analyses. This primary analysis of all ten configurations yielded a transverse − non-transverse covariate-standardized difference of −4.05 percentage points (95% posterior interval, −6.64 to −1.44; odds ratio, 0.85; Fig. 2b; Table 3), with similar estimates for mE (−4.13) and mW (−3.98). The posterior probability that mE and mW occupied the two lowest ranks among the ten configurations was 0.61, compared with 0.022 expected if the configurations were exchangeable, consistent with the two transverse configurations together standing apart from the others. By contrast, no difference was detected among the eight non-transverse configurations (cluster-robust Wald test, $P = 0.95$; likelihood-ratio test, $P = 0.97$), and equivalence tests (±5 percentage points) and point-null Bayes factors supported similarity (Supplementary Table 2). Because this equivalence margin exceeds the observed transverse difference, however, these analyses alone cannot exclude differences of that size among the eight non-transverse configurations, nor does non-detection show that the field had no influence under them.

We next accounted for the fact that the transverse pair was identified in the same data[37] analyzed here. Among all 45 pairs of configurations, mE/mW produced the most negative contrast with the remaining eight (−4.07 percentage points; next, −2.39; Fig. 2c). Referring this contrast to the permutation distribution of the most extreme pair contrast gave a selection-adjusted two-sided $P$ = 0.015 among the five symmetric pairs defined by the dissection. Among all 45 pairs, a more conservative set, the corresponding $P$ was 0.096. The negative transverse contrast was retained across alternative likelihood, prior, dependence, dispersion, and nuisance specifications. Four prior specifications yielded estimates from −4.08 to −4.00 percentage points, and an independent Laplace-approximated mixed model yielded −4.06 percentage points (parametric bootstrap 95% interval, −6.52 to −1.49). All 53 reported transverse-contrast estimates from hierarchical-model sensitivity analyses were negative, with 51 intervals excluding zero (Supplementary Fig. 1; Supplementary Note 1), although these analyses share the underlying data and cannot exclude unmeasured bias. Forty-three of the 47 north-seat participants contributed both transverse and non-transverse sets, and the within-north-participant estimate was −4.10 percentage points (bias-reduced cluster-robust (CR2) 95% confidence interval, −6.72 to −1.48; $P$ = 0.003). In the primary model, omitting cohort 13 attenuated the contrast to −3.10 percentage points (−6.16 to −0.04, an upper limit near zero across seeds), and no heterogeneity was detected across cohorts, test periods, or experimenters (Supplementary Table 3). Within the primary model, the comparison with mN alone was more sensitive to cohort coverage: it was −4.54 percentage points (−8.73 to −0.42) in the full dataset but −1.08 percentage points (−5.73 to +3.52) after excluding cohort 13, in which the mN rate was unusually high (57.9%; Supplementary Table 3).

Finally, we asked whether the lower rate was specific to the transverse geometry rather than to removal or reversal of an individual field component. The eight non-transverse configurations comprised three geometric classes: horizontal-only fields at other azimuths (mN, mS, mNE, and mNW), vertical-only fields (down and up), and inclined fields containing both horizontal and vertical components (Con and vertical-component inversion). The transverse pair was lower than each class by a similar amount (−4.17, −4.38, and −3.54 percentage points, respectively; $P$ = 0.002, 0.007, and 0.018), whereas the three non-transverse classes did not differ from one another ($P$ = 0.79; Fig. 2d). Thus, the reduction was not simply associated with removal of either field component or polarity reversal. Across the ten intensity-matched configurations, a lower north-seat black stone rate was detected only for the transverse pair. The direction of this contrast was consistent across analytical specifications, although its magnitude and statistical resolution varied with model specification and cohort coverage.

**Early transverse difference without evidence of learning**

In our earlier study,[22] near-zero and Larmor-frequency conditions predominantly reduced black stone rates in the odd-parity cases. We therefore examined which components of the sequence accompany the configuration difference. Conditional on odd parity, the black stone rate under the transverse pair was lower than that of the non-transverse family in 8 of the first 12 north-seat trials, including the first two (Fig. 3a). In fixed-denominator analyses of the north-seat P1 trials among these, the ‘black-in-odd’ component of the black stone rate was lower under the transverse pair (−3.94 percentage points; bootstrap 95% confidence interval, −7.53 to −0.31; $P = 0.034$), with negative point estimates relative to each of the three geometric classes, whereas the black-in-even component did not differ (+1.45 percentage points; −2.29 to +5.24; Fig. 3b, c). Because the black-in-odd measure is a joint outcome of P1 parity generation and P2 one/two-stone choice, we next analyzed these processes separately (Supplementary Note 2). The transverse pair did not reliably alter P1 odd generation at either seat or P2 choice at the north seat. South-seat P2 one-stone probability was higher by +3.91 percentage points (set-clustered 95% confidence interval, +0.65 to +7.18), although multiplicity-adjusted support varied across covariance structures (Holm-adjusted $P$ = 0.035–0.077; Table 3). Because the magnetic-field vectors at both seats changed together, this result does not identify the sensitive seat or a single process responsible for the overall effect. The odd–even component difference was not statistically resolved (Supplementary Fig. 2a). We then tested whether outcome-driven learning or strategic adjustment within a set could account for the transverse difference.[28,38] Within-person cross-lag models showed no detectable configuration-dependent lag effects (Supplementary Fig. 2b). Obtaining the black stone on the preceding trial was not associated with subsequent P1 odd generation (odds ratio, 1.05; 95% posterior interval, 0.96 to 1.16) or P2 one-stone choice (1.01; 0.92 to 1.12). Neither these dependences nor those on the running score and preceding actions differed between transverse and non-transverse configurations (interaction odds ratios for the previous outcome, 0.91 and 1.12; Holm-adjusted $P$ = 0.48; Fig. 3d; Supplementary Table 3). Consistent with an early rather than progressively learned effect, the transverse difference was already present during the first five trials (−5.4 percentage points) and did not increase, although it declined numerically, across subsequent five-trial blocks (−4.3, −3.9, and −2.7 percentage points; transverse × trial-position odds ratio per standard deviation of trial position, 1.04; 95% posterior interval, 0.94 to 1.14; Fig. 3e). The difference arose almost entirely from changes in within-trial parity–choice coincidence rather than in their frequencies and was numerically larger with the north-seat participant as P2 (Supplementary Note 2). Together, the component and process analyses indicate that, in trials in which the north-seat participant was P1, the reduction arose mainly in odd-parity outcomes, without resolving the odd–even difference, and do not identify a single generating process, establish causal mediation, or demonstrate information transfer between participants.

**Field descriptors and mechanism-linked comparisons**

To relate the black stone rate pattern to magnetic-field geometry, we characterized the configurations using continuous physical descriptors. North-seat total field intensity separated the near-zero condition from the other conditions but did not distinguish those configurations clustered near 45 μT (Fig. 4a). For the six horizontal profiles, the horizontal-axis conflict descriptor, hOrthConflict = $1 - |\cos \theta|$, where $\theta$ is the angle between the condition and ambient horizontal vectors, was near 0 for fields aligned with the ambient horizontal axis (mN, 0.00; mS, 0.02) and increased to 0.79 for the approximately transverse fields (mE, mW) when calculated from the measured vectors (Supplementary Table 4). Black stone rates were lower only at the transverse value (Fig. 4b), although the three clusters of descriptor values cannot distinguish a graded from a threshold-like dependence. We also constructed a directional extractability index (DEI) after inspection of the observed pattern. The DEI combines a saturating field-intensity term, which approaches zero as the field is removed, with a transverse-axis term that decreases as a predominantly horizontal field turns transverse to the ambient horizontal vector (Methods). The index placed near-zero and the transverse pair at its low end and correlated with the 11 condition means ($r = 0.89$; Fig. 4c; Supplementary Table 4), although it did not outperform a two-level classification separating these three conditions from the others ($r = 0.93$). Thus, similarly reduced rates occurred when the field was removed (near-zero, recorded in the earlier series) and when its horizontal axis was turned approximately transverse. The mid-plane descriptor, $1 - |\cos \varphi|$, where $\varphi$ is the angle between the field vector and the geographic north–vertical plane, ranges from 0 within that plane to 1 for east–west fields. It primarily separated the transverse pair and thus restated the transverse contrast (Fig. 4d). Because each descriptor has only one value per condition, these associations are condition-level summaries and provide no independent participant-level inference. South-seat vectors under both vertical-only conditions were within 5° and 2 μT in intensity of those under the transverse pair (Table 1), yet the transverse pair was lower than the vertical-only class (−4.38 percentage points; $P = 0.007$) and than vertical-only down in a dyadic-overlap comparison (−5.32 percentage points; 95% confidence interval, −10.49 to −0.15; $P = 0.044$), which does not support an explanation based only on the south-seat field.

We next examined the RF conditions relevant to potential magnetosensory mechanisms. RF1 at 1.260 MHz, the electron Larmor frequency for a 45 μT static field,[2,22,37] was 4.1 percentage points lower than Con (two-sided Welch $P = 0.051$, versus one-sided $P = 0.026$ in the source analysis;[22,37] bootstrap 95% confidence interval, −8.2 to −0.1 percentage points), whereas RF2 at 1.890 MHz, off resonance for that background, showed no corresponding difference (Fig. 4e; Supplementary Table 5). In a direct comparison, RF1 was 5.4 percentage points lower than RF2 (two-sided Welch $P = 0.027$; bootstrap 95% confidence interval, −10.0 to −0.7). Unexpectedly, four additional RF1 conditions over horizontal-only mN and vertical-only down backgrounds produced little change from their respective RF-off references (−0.8 to +1.6 percentage points; all 95% confidence intervals included zero), including over mN, where the antenna axes were largely perpendicular to the static field (Fig. 4e, f). Averaged across the two modified backgrounds, the RF1 effect was +0.3 percentage points (95% confidence interval, −2.9 to +3.4). Although this interval excludes a reduction as large as the point estimate observed over the ambient field (Con), the difference between the RF1 effect over Con and that averaged over the modified backgrounds was not statistically resolved (−4.4 percentage points; 95% confidence interval, −9.5 to +0.8; $P = 0.10$; participant-level model, −4.5; 95% posterior interval, −9.8 to +0.8; Supplementary Table 5). These results are thus consistent with, but do not establish, dependence of the RF1 response on static-field geometry. Near-zero, the other mechanism-relevant condition, also resembled the transverse pair in its joint outcomes. It reduced the first-12 black-in-odd component relative to the non-transverse family (−4.94 percentage points; bootstrap 95% confidence interval, −9.53 to −0.41; $P = 0.033$) without changing the black-in-even component (Supplementary Fig. 3a).

Finally, because light dependence is expected for a radical-pair mechanism,[1,2,5] we examined whether the response varied with the spectrum of light reaching the eye under the mN configuration. Relative to non-filtered goggles, the north-seat black stone rate was higher when long-pass goggles worn by the south-seat participant removed wavelengths below 500 nm (+5.4 percentage points; two-sided Welch $P = 0.021$; bootstrap 95% confidence interval, +0.9 to +9.9 percentage points; Supplementary Table 6), but not when they removed wavelengths below 400 nm (Supplementary Fig. 4a). Because the two participants' outcomes are complementary, the higher north-seat rate corresponds to a lower black stone rate for the filtered participant. This pattern is compatible with a contribution of blue (short-wavelength) light, as expected for a cryptochrome-based radical-pair mechanism, although the filter also reduced photon flux and therefore does not isolate spectral composition from light intensity. The joint-outcome pattern also differed from that of the transverse configurations, possibly because only the south-seat participant wore the filter (Supplementary Fig. 4b; Supplementary Note 3).

## Discussion

In fasted men, the north-seat black stone rate was lower under the two transverse configurations, mE and mW, than under the other eight Earth-strength configurations. Unlike the initial interpretation of these observations as a direct interaction between the two players,[37] we consider them here within a local measured-vector and controlled-manipulation framework. This difference extends the study of human magnetic responses beyond the spatial outcomes emphasized in magnetoreception research[1–4,13,16] to a task requiring no directional report. Axis-dependent responses are well established in the avian magnetic compass, which responds to field axis rather than polarity,[2,39] and human visual thresholds have also been reported to depend on field direction.[17,18] In our data, a human non-spatial decision outcome differed with the axis of an Earth-strength field at approximately constant intensity. Field axis here denotes the line of the field vector irrespective of polarity, and the dependence was selective rather than general. Inverting the vertical component turned the north-seat axis by approximately 88° in three dimensions from the ambient axis while keeping it close to the geographic north–vertical plane, and did not lower the rate, whereas the transverse fields, approximately 82° from the ambient axis with their horizontal components nearly perpendicular to that plane, lowered it (Table 1; Fig. 4d). In this design, the lower rate was associated not with the size of the axis change but with alignment of the field along the east–west (nominally interaural) axis. Thus, this framework raises three testable questions: which task processes contribute to the effect, how exposure at each seat contributes, and which reference frame organizes the response.

The only nominally detectable primitive-process difference was an increased probability of a one-stone choice by the south-seat participant as P2, whose support depended on the covariance structure, whereas P1 odd generation did not change reliably at either seat. Because odd parity and one-stone choices each occurred in about half of the trials, such a shift moves outcomes between the odd and even components rather than lowering the overall rate. The odd-case step-1 reduction previously observed under near-zero GMF and RF1 likewise combined P1 parity generation and P2 choice,[22] so neither study identifies a single generating process. Because the transverse difference was present from the first trials without evidence of outcome-driven learning, and 43 of the 47 north-seat participants contributed both transverse and non-transverse sets, it is also difficult to attribute to strategic adjustment or to stable individual strategies, such as those that may underlie the differences in black stone rate with playing level among ranked players,[38] particularly because our participants had little or no familiarity with Go. Incentives for winning and for a higher monthly black stone rate were also identical across configurations (Methods), so differences in assigned incentives cannot explain the within-participant contrast (Table 3).

The second question concerns the contribution of the two seat-specific exposures. Approximately constant north-seat intensity constrains a simple intensity-based account. South-seat intensity varied substantially among the eight non-transverse configurations without a corresponding change in black stone rate, while the transverse-pair intensities fell within this range. In addition, the south-seat vectors under the transverse and both vertical-only conditions were closely matched, yet their black stone rates differed. Near-zero also had a closely matched south-seat vector and a similar rate, but its north-seat field also differed. These observations weaken, but do not exclude, an explanation based solely on the south-seat field (Table 1). At the north seat in nominal forward-facing posture, the east–west fields of mE/mW lie close to the interaural axis ($|B_y|/|B| \approx 1.00$ versus 0.12–0.15 under mN and Con; Table 1). One possible behavioral interpretation, termed 'spatial-state blockade', is that sustained transverse exposure reduces the usability of directional magnetic information as a reference for the task or body. The similar mean rate and joint-outcome pattern under near-zero, which removes rather than redirects the north-seat field, are consistent with this interpretation. This interpretation does not by itself account for the absence of a detectable RF1 effect over mN. Earlier orientation and choice assays[20–22] and the recruitment of map-like representations in abstract choice[34–36] suggest testing this hypothesis, although magnetic modulation of such representations has not been demonstrated in humans. An alternative behavioral explanation is that magnetic exposure alters arousal, sensory gating, or response bias without supplying directional information. Measuring magnetic-direction discrimination and stone selection in the same participants could distinguish the two interpretations: the first predicts that rate-lowering geometries also impair direction discrimination, whereas the second predicts rate changes without such impairment.

How magnetic input might influence choice is distinct from the physical process that detects the field. Light-dependent radical-pair chemistry is a specific candidate because it can generate orientation-dependent responses and because previous human orientation experiments showed light- and RF-dependent effects.[5–9,20,21] No polarity difference was detected under horizontal reversal (mS − mN, −1.9 percentage points), although an effect of that size cannot be excluded. This result, together with the lower rates under transverse exposure, is compatible with an axis-sensitive receptor such as a radical-pair system,[2,39] whereas polarity-sensitive alpha-band responses in humans[19] and polarity-encoding vestibular neurons in pigeons[40] have been reported. However, the present findings do not establish axial symmetry or exclude magnetite-based sensing, because canonical radical-pair axiality concerns complete reversal of the magnetic-field vector, whereas reversal of an individual component changes the full-vector axis.[2] In the earlier orientation task, vertical-component inversion reversed the orientation, whereas whole-field reversal abolished it, suggesting a non-canonical inclination compass.[21] Because the present north-seat reversals were approximate and did not reverse the complete two-seat configuration (Table 1), complete reversal of the ambient field would test whether this response extends to non-spatial choice. For frequency dependence, raising the static background shifted the disruptive frequency from RF1 to RF2 in the earlier orientation study,[21] indicating that RF2 is off resonance for the ambient background rather than universally inactive. The direct RF1–RF2 contrast in the stone selection data[22] raises the possibility that the effective frequency scales with static field strength, as reported in the avian compass when the static field was doubled.[39] In the present analyses, the RF1 point estimates suggest dependence on static-field geometry but do not establish it (Supplementary Table 5). Adequately powered experiments that manipulate static-field direction, RF direction, frequency, and amplitude independently within a single design will be required. Avian RF studies[6,39,41–43] provide useful mechanistic comparisons but should not be interpreted as defining a human response threshold (Supplementary Note 4).

Electromagnetic induction provides a contrasting candidate mechanism, in which head rotation through a static magnetic field induces an electric field in the semicircular canals that could stimulate electroreceptive hair cells.[11] In pigeons, a static field without movement did not activate the vestibular nuclei.[10] Because the human canals are nearly orthogonal,[44] induction depends mainly on the head's rotation axis relative to the field. In the observation-based posture model (Methods; Supplementary Note 5), under the observed cue-detection mixture, estimated induction before choices under the transverse fields fell within the range of the other configurations and exceeded that under vertical-only fields, which did not lower the rate. Reaching the detection threshold reported for elasmobranch electroreceptors[45] would also require canal radii at least approximately 1.6-fold larger than those of humans[46] for choice-related movements (Supplementary Note 5). Furthermore, induction provides a less straightforward account of the frequency and light-spectrum observations: at matched amplitude, RF2 would induce a larger electric field than RF1 (Methods, 'Observed task posture and induction estimates'), whereas the lower rate was associated with RF1, and vestibular induction is light-independent in pigeons.[10] Magnetite-based receptors, the best-studied form of the second candidate, are not expected to directly track weak megahertz fields at frequencies far below their gigahertz ferromagnetic resonances.[6] Taken together, the field-geometric and frequency patterns, the light-spectrum observation, and the modeled induction estimates support a light-dependent radical-pair mechanism as a leading working hypothesis, although the undetected RF1 effect over mN, where the RF field was largely perpendicular to the static field, runs counter to a simple perpendicular-component account, and induction cannot be strictly excluded without timing head movements against the task steps.

The third question is the reference frame in which the response is organized. The transverse-specific reduction, together with the pattern of the RF1 point estimates across static backgrounds, suggests a potential magnetoresponsive axis on the receiving side, distinct from the field axis itself. One hypothesis is that this axis is fixed relative to the participant's head rather than the laboratory. Anisotropic hyperfine interactions can make radical-pair yields depend on magnetic-field direction relative to the receptor molecules,[2,5–8] but predicting a head-fixed response requires specifying how those molecules are aligned and how their signals are integrated downstream. The present experiment cannot distinguish head-fixed from laboratory-fixed accounts, because head posture was not recorded, and geographic north, the participant/dyad axis, the visual alignment axis, and body orientation were nominally aligned and never manipulated independently (Fig. 1a). Task-related head turns were also directed toward the east and west, along the transverse field axis. In the observation-based posture model (Methods), the transverse fields were closer to the interaural axis than the other configurations over the trial, with mNW intermediate, implying that a head-fixed response would require a steep, threshold-like angular dependence between mNW and the transverse fields (Supplementary Note 5). Ordered retinal molecules are plausible candidate substrates,[5] but the lower rate under transverse left–right fields than under vertical-only up–down fields does not itself establish retinal anisotropy because head posture was unrecorded and downward gaze toward the board altered field orientation relative to the visual axis.

A head-fixed axis could also reconcile the present results with the earlier orientation studies, in which participants rotating freely on a chair[20,21] oriented correctly even in trials that began with the horizontal field component transverse to them.[20] Thus, any transverse field–head relationship was transient, whereas in the present task, seat and body orientation relative to the field were nominally fixed throughout a set. Given the blue-light dependence in the orientation paradigms,[20,21] a shared light-dependent magnetic input could support directional orientation during free rotation while influencing choice during sustained transverse exposure. A head-fixed response can be distinguished from a laboratory-fixed one by varying field direction and participant orientation independently: rotating the magnetic field and the participant together preserves their alignment, and crossing the field-head alignments of the two participants could further separate their contributions. Such experiments should measure local magnetic vectors at each head location and record head and eye movements. An automated opponent responsive to human choice, or participants linked over the internet with parity concealed until P2 commits,[37] could isolate a single exposed participant or allow tests under the natural GMF, without invoking magnetic signaling between participants.

Overall, the findings identify magnetic-field geometry as a candidate determinant of human non-spatial probabilistic choice, but their generality remains uncertain. Because the observations derive from a single laboratory and apparatus, independent preregistered replication across laboratories is essential.[47–49] Such replications, and behavioral studies of probabilistic choice more generally, would benefit from reporting the orientation of the testing room and seating and the local field vector at the participant's head, because, if the present dependence generalizes, an unreported field axis could add unexplained variance to choice probabilities. In addition, all directional-field records came from young adult men tested after an approximately 20 h fast, the condition under which our previous magnetic responses in men were observed.[20–22] Whether the difference generalizes to women, other ages, non-fasted participants, other decision tasks, or independently manipulated reference frames remains unknown. Moreover, the estimand was defined after the data had been examined, condition order was not randomized, and the family contrast was sensitive to the omission of one cohort. By allowing parity generation and deliberate choice to be examined separately while retaining a shared behavioral outcome between participants, this paradigm provides a testable framework for extending human magnetoreception research from spatial orientation to non-spatial decision-making.

## Methods

### Participants, eligibility, and ethics

The source participant pool, recruitment, and eligibility criteria (no reported physical disability or mental disorder, including color blindness or claustrophobia) were described previously.[22] Briefly, the source pool included 55 male Korean undergraduate volunteers at Kyungpook National University (21–27 years; mean, 23 years; body mass index, 20–34 kg $m^{-2}$; mean, 24 kg $m^{-2}$) with little or no familiarity with Go. Participants were recruited in successive volunteer groups (cohorts). Because some participants took part in more than one group, cohort denotes the group under which a set was recorded and is a set-level covariate. Cohorts are labeled by their program-wide recruitment number (for example, cohort 13). After duplicate participant records were reconciled across cohorts, the ten-configuration family involved 47 of these 55 men, each of whom contributed both north-role and south-role sets. The other eight did not take part in this family. The sex-specific scope is reported in accordance with the Sex and Gender Equity in Research (SAGER) guidelines.[50] The study was approved by the Institutional Review Board of Kyungpook National University (KNU-2021-0153). All procedures complied with the approved human-subject protocol, and written informed consent was obtained from all participants.

### Pre-experiment preparation and physiological state

The primary directional experiments followed the previously described approximately 20 h fasting and pre-test preparation protocol,[20–22] which included water-only fasting, at least 6 h of night sleep, approximately 15 min of seated rest before testing, and removal of colored lenses and detachable metallic or electromagnetic objects. The start of the fast was scheduled according to each session's start time (for example, from 22:00 on the previous evening for sessions at 18:00). Blood glucose was measured in both participants before the first game of each test visit. As in the earlier study,[22] a value deviating by more than 15% from the participant's fasting reference (Supplementary Methods 1) prompted review and, when appropriate, postponement of testing (approximately 2% of experiments). Glucose did not differ detectably across the ten configurations, and the transverse contrast remained negative after adjustment for both participants' glucose values or exclusion of the 12 sets in which a participant's value deviated by more than 15% from his mean (Supplementary Methods 1; Supplementary Table 3).

### Task familiarization and incentives

Familiarization, task instructions, and incentives followed published protocols,[22,38] including two preliminary games (not analyzed), a minimum of 10 white stones in step 1, and avoidance of predetermined one/two-stone patterns in step 2. In the post-experiment questionnaire, participants reported that they could not intentionally control step-1 parity, whereas step-2 choice was intentional.[22,37] Participation compensation and rewards for winning a game or achieving a higher monthly black stone rate were restated before the first game of each visit and were identical across configurations. Directional experiments used an opaque screen and auditory masking under diffuse LED illumination of approximately 280 lx at the glabella.[22] Head posture was neither constrained nor recorded, so head-angle distributions may have differed across conditions (see ‘Observed task posture and induction estimates’).

**Stone selection procedure and field configurations**

Two participants occupied the north and south seats of a shared three-axis Helmholtz coil system. Each game consisted of two 20-trial seat-specific sets. P1 concealed a handful of white stones, after which P2 selected one or two black stones before P1 revealed whether the concealed number was odd or even. Matching parity resulted in retention of the original stone colors, whereas mismatching parity resulted in exchange of colors. P1 and P2 roles alternated across trials, and participants exchanged seats before the second 20-trial set. In a few games (mixed-condition games), the two sets were recorded under different conditions. Trials began approximately 10–15 s after the field configuration was applied. The ten static configurations formed five symmetric pairs (Fig. 1c; Table 1). North-seat total intensity remained near 45 μT, whereas south-seat intensity ranged from 76% to 112% of the corresponding north-seat value. Near-zero, RF1, and RF2 served as reference conditions outside this intensity-matched family. Con and these three reference conditions came from the earlier study.[22] Con sets involved 44 of the 47 participants, all of whom also contributed geometry sets, and were recorded within the calendar span of the geometry experiments and in every cohort, but not in the same visits as the geometry sets.

**Experimental condition allocation and masking**

A written protocol standardized recruitment, fasting, seating, screening, and trial procedures before the geometry experiments. Seven of the nine manipulated field-geometry conditions (mN, mE, mW, mS, vertical-only up and down, and vertical-component inversion) were prospectively organized as a balanced geometry family, with dated schedules prepared before the corresponding sessions. Only the ±30° rotations (mNE and mNW) were added to the family after results of other configurations had been reviewed. Two eligible men were assigned to visits containing two coded conditions, subject to participant availability and an intended ≥3-day interval. The target was 50 sets per condition, as in the earlier stone selection study,[22] in which this number had detected condition differences. Some conditions fell short of this target because scheduled sessions could not be completed owing to participant availability or the end of the study period. Conditions and initial player order were assigned before session outcomes were known, and participants exchanged seats within each game.

The overall condition sequence, however, was not determined by unrestricted randomization: no formal random-number algorithm was used, and the introduction or ordering of some later conditions was informed by earlier observations. Accordingly, condition-label reallocation was treated as a scheduling sensitivity analysis rather than exact randomization inference (Supplementary Note 1). Only the principal investigator held the code linking settings to field identity. He did not conduct any analyzed session. He ran only some of the preliminary familiarization games, which were not analyzed. He validated the coil settings before and after each experimental day without contact with the participants. Experimenters implemented coded settings without knowing their decoded field identity, physical interpretation, or predicted effect. In the absence of automation, they cued each task step in real time with standardized hand signals specified in the written protocol. Participants were unaware of field or control status, and initial data processing was condition-blinded. The light-spectrum conditions were not part of this geometry family, and most of their sets were added after review of the geometry findings. Double-wrapped coils supported masking by using antiparallel currents to cancel the field while minimizing non-magnetic differences between active and sham settings,[51] although equivalence of heating, noise, and vibration across configurations was not directly demonstrated. Within the family, however, the coil-generated east–west field component at the north seat (condition − Con) was largest under mE and mW (+51.7 and −38.5 μT) but reached 69% of the mW value under mNW (−26.6 μT), which did not lower the rate, and the largest coil-generated fields (mS and vertical-only up, approximately 83 μT) did not lower it either (Table 1).

**Session scheduling and condition distribution**

Eleven confirmed intervals between successive visits of the same participant were shorter than the intended 3 days, and these visits were retained. The ten configurations were tested across 110 dates by five experimenters (identity unrecorded for the earliest 20 sets), and a planned personnel change divided testing into early and late test periods. A pair-session was defined by test date and the unordered participant pair and could therefore contain more than one two-set game. The dataset comprised 182 pair-sessions, including 29 containing both transverse and non-transverse sets. These within-session observations contributed directly to the primary contrast. The remaining comparisons relied on hierarchical adjustment across sessions. Recorded set start times ranged from 10:09 to 19:24 and were similar for the transverse pair and the other eight configurations (medians, 14:59 and 14:53), and set-level black stone rates were not associated with start time (Spearman $\rho = -0.025$, $P = 0.59$; 454 sets). Sensitivity analyses addressed repeated participation, session order, prior exposure, and inter-session interval, although unmeasured session-level influences cannot be excluded.

**Geomagnetic-field modulation and measurement**

Magnetic-field modulation used the previously described three-axis, double-wrapped Helmholtz coil system[22] with an ambient field of approximately 45 μT. The ambient condition (Con) was not an unenergized ambient baseline. Instead, its double-wrapped windings were driven with equal and antiparallel currents of the same magnitude as in the modulated settings, so that the coils were energized while the net coil field was canceled. This sham-energized implementation, used in the source experiments but not explicitly reported previously,[22] is specified here since it is relevant to interpretation of non-magnetic cues. The north-seat participant sat at the coil center, and the south-seat participant sat 1.30 m away on the north–south axis, outside the coil system. Condition-specific three-axis vectors were measured at both glabella locations with a three-axis magnetometer (MGM 3AXIS; AlphaLab, Salt Lake City, UT, USA) and are reported using +x = geographic north, +y = geographic east, and +z = downward (Table 1). Before and after each field application, the experimenter confirmed the setpoints on the controller display, and the output logs after each experiment matched the setpoints. For near-zero, all three coil pairs were driven to cancel the ambient field at the north-seat glabella, where each residual component was recorded as 0.0 μT (Table 1). Exposure was defined by these paired glabella vectors, which characterize the local field direction and intensity at each seat and together define the manipulated two-seat configuration.

**Radiofrequency and light-spectrum manipulations**

The original RF1 (1.260 MHz, the electron Larmor frequency at 45 μT) and RF2 (1.890 MHz) exposures used the published apparatus and calibration procedure,[21,22] delivering vertical oscillatory fields of approximately 100 nT at the north-seat glabella and <1 nT at the south seat (root-mean-square values). The four added RF1 conditions crossed horizontal-only mN and vertical-only down backgrounds with vertical and 37°-tilted antenna positions, matched at 100 nT in the forward-facing reference posture (Supplementary Note 4; Supplementary Methods 2; Supplementary Fig. 3b). RF1 effects over these backgrounds were estimated relative to their RF-off references and compared with the RF1 effect over Con (see 'Source-level contrasts and equivalence analyses'). These RF-off references were the mN and vertical-only down sets of the ten-configuration family rather than sets recorded in the same visits. The added RF1 sets were recorded within the calendar span of the geometry experiments, overlapping that of the RF1 sets over Con. In a separate light-spectrum experiment over the mN background, the south-seat participant wore non-filtered goggles or long-pass filters transmitting above 400 or 500 nm.[20,21,37] Glabella illuminance was 130, 108, and 72 lx, respectively, so spectral composition and light intensity varied together.

**Endpoint, estimand, and primary hierarchical model**

The primary observation unit was the 20-trial north-seat exposure set, and the primary endpoint was its north-seat black stone rate, modeled as a binomial count. The analysis table comprised 600 sets from 47 men in 209 pair-sessions: 454 sets from the ten static Earth-strength configurations and 146 reference sets (near-zero, $n$ = 50; RF1, $n$ = 50; RF2, $n$ = 46). The RF-background and light-spectrum series were analyzed separately. Source records, indexed before the present analysis, were linked to original trial sheets and stable pseudonymous participant IDs, and excluding the 31 recovered-record or 21 label-reconciled sets did not change the sign of the transverse contrast (Supplementary Methods 3; Supplementary Fig. 1; Supplementary Table 3). The estimand was the covariate-standardized difference in north-seat black stone probability between the transverse configurations (mE and mW) and the eight non-transverse configurations, with each configuration weighted equally over the observed seat-phase and cohort distribution. The estimand and analysis hierarchy were defined during the present reanalysis after condition labels had been decoded and were not preregistered. The transverse configurations mE and mW were combined because they share a horizontal field transverse to the ambient axis and differ only in its polarity. This grouping was first made in the preprint that disclosed these observations,[37] after the condition means had been examined, so its inference is accompanied by selection-adjusted tests (see 'Selection-aware permutation'). For set j, $Y_j \sim \text{Binomial}(20, p_j)$, with $\text{logit}(p_j)$ including indicators for the ten configurations (coding baseline, Con), seat phase, and cohort. Crossed random intercepts represented north-role participant, south-role participant, and pair-session. This model extends the paired Go stone selection framework used previously[38] by representing participant propensities separately across roles. Fixed effects used Normal(0,1.5) priors and random-effect standard deviations HalfNormal(0.5) priors. Posterior standardized predictions were averaged over the observed rows, and the transverse contrast used equal configuration weights. Posterior sampling used the no-U-turn sampler (NUTS) in NumPyro. Sampling settings, longer-chain and multiple-seed analyses, and an independent R/lme4 implementation are detailed in Supplementary Methods 4 and Code availability. Convergence diagnostics are reported in the legend of Supplementary Fig. 1.

**Robustness and participant-aggregated sensitivity analyses**

Robustness analyses examined alternative likelihoods and priors; within-participant, cluster-robust, dyadic-overlap, and wild-cluster inference; participant-, pair-, pair-session-, cohort-, and experimenter-omission analyses; calendar effects; within-pair-session comparisons; and restricted physical-profile models (Supplementary Methods 4). Additional Bradley–Terry, beta-binomial, and Pólya–Gamma formulations tested dependence on model structure. Heterogeneity across cohorts, test periods, and experimenters was evaluated using bootstrap, mixed-model, and partial-pooling approaches. Participant-level analyses compared transverse and non-transverse rates using available-case and strict complete-case averages and a within-north-participant model adjusted for cohort and seat phase. These analyses differed in weighting and treatment of repeated participation and were considered sensitivity analyses rather than additional confirmatory tests.

**Primitive-process and joint-outcome analyses**

Person-linked trial data were analyzed for four role-specific processes: north-P1 odd generation, south-P2 one-stone choice, south-P1 odd generation, and north-P2 one-stone choice. Logistic models included configuration, seat phase, cohort, and actor effects. Transverse − non-transverse contrasts were evaluated under alternative dependence structures with Holm adjustment across the four endpoints. Mixed-model sensitivity analyses included crossed actor, partner, and set random intercepts (Table 3). Following the stepwise analysis reported previously,[22] component summaries used the first 12 trials. Joint-outcome summaries were treated as descriptive decompositions rather than causal mediation. Model details are given in Supplementary Methods 4.

### Within-person cross-lag and outcome-history models

Cross-lag models of trials 2–20 tested whether configuration modified sequential associations between a participant's successive P1 and P2 actions. Outcome-history models additionally considered the actor's and partner's preceding actions, the preceding black stone outcome, running score, and trial position. These observational analyses test sequential associations and within-set learning but do not establish interpersonal directionality, causal mediation, or information transfer. Full model specifications are provided in Supplementary Methods 4.

### Source-level contrasts and equivalence analyses

Source-level condition, parity, geometric-class, single-reference (mE or mW versus mN), near-zero, and RF contrasts used two-sided Welch tests, percentile bootstrap intervals, or both,[52] extending the source-level procedures reported previously.[22] Resampling was performed within configuration. The added RF1 conditions were compared with their corresponding RF-off backgrounds, and the RF-by-background contrast was additionally refitted with the crossed-role model after participant linkage. Equivalence among non-transverse configurations was evaluated against a ±5-percentage-point bound using two one-sided tests,[53] and evidence for no difference was quantified with Jeffreys–Zellner–Siow Bayes factors ($BF_{01}$).[54] These analyses were supportive and secondary to the primary hierarchical estimand. Resampling counts and additional specifications are provided in Supplementary Methods 4.

### Selection-aware permutation

Homogeneity among the eight non-transverse configurations was assessed using cluster-robust Wald and crossed-role likelihood-ratio tests. Because mE/mW had been identified as the transverse pair in the earlier analysis,[37] selection-aware inference recomputed every pair-versus-remainder contrast under condition-label permutation. The five symmetric pairs defined by the field dissection formed the design-based comparison family, whereas all 45 possible pairs provided a broader, more conservative comparison. Each of the five pairs joins a configuration with the counterpart that reverses or mirrors its manipulated component. The pairing followed the planned geometry family except for mNE and mNW, which were added after results of other configurations had been reviewed (see 'Experimental condition allocation and masking'). Both comparison sets were specified during the present reanalysis, after the transverse pair had been identified. Permutations within cohort × seat phase and within north participant were sensitivity analyses. We also compared the transverse pair with horizontal-only, vertical-only, and inclined non-transverse classes and calculated the posterior probability that mE and mW occupied the two lowest ranks (Supplementary Methods 4).

## Physical descriptors and condition ordering

Measured north-seat vectors were used to quantify total field intensity, horizontal-axis conflict relative to Con, and departure from the geographic north–vertical plane. Horizontal-axis conflict was evaluated in the six horizontal configurations. The north-transverse descriptor, referenced to the measured mN horizontal direction, was compared with south-seat intensity in predictive analyses (Table 3). The directional extractability index (DEI), which combined field intensity and horizontal-axis alignment, was defined after inspection of the condition pattern and used descriptively. Equations and conventions are provided in Supplementary Methods 4, and comparisons in Supplementary Note 3.

## Directional-gradient model

An idealized Biot–Savart model based on coil geometry and measured center vectors quantified spatial variation in field direction over a head-scale volume using a directional gradient index (DGI; Supplementary Methods 5), to assess potential apparatus-related directional gradients. Although DGI was largest for mE and mW, across the ten non-zero configurations the maximum modeled angular deviation within the primary ±50 mm region was ≤0.014° (Supplementary Fig. 5; Supplementary Table 7). Although only north-seat vectors were used to set the coil fields, the model reproduced the measured south-seat vectors of near-zero and the nine static configurations other than Con within 1.1–4.0 μT (vector difference) and 0.6–4.5°.

## Observed task posture and induction estimates

Head and eye movements were observed but not recorded during the original experiments. Posture-model inputs are therefore observational estimates, compiled after the experiments from interviews with all experimenters who conducted the sessions (Supplementary Methods 6). A trial-cycle model combined observed movement ranges, seat geometry, and measured field vectors to estimate field orientation relative to head-fixed axes and magnetic induction across the six semicircular canals, using published canal orientations[44] (Supplementary Methods 6). Decision-window and latency assumptions were varied in sensitivity analyses (Supplementary Note 5).

Peak induced electric fields were estimated as $E = \pi B f r$, where $B$ is the field intensity, $f$ is the head-rotation rate in turns per second, and $r$ is the canal radius.[11] For an oscillating field, the same relation applies with $f$ as the RF frequency and $B$ as the field component normal to the loop, so that $E$ has the same amplitude convention as $B$. At matched field amplitude and orientation, the induced field is 1.5 times larger for RF2 than for RF1. Assuming minimum-jerk profiles over the observed durations (Supplementary Methods 6), peak angular velocities of choice-related head turns and tilts were approximately 15–230° $s^{-1}$. These calculations compare expected induction across configurations under simplified rigid-rotation assumptions and do not predict behavior.

## Use of large language models

Claude (Anthropic, San Francisco, CA, USA) was used under the authors' direction to edit draft manuscript text, suggest some of the statistical analyses, and write analysis code. Statistical models, analyses, and interpretations were specified or approved by the authors. Y.K. verified the data, implementation, and agreement of all statistical analyses. The authors reviewed all AI-assisted text, code, and calculations. The authors take full responsibility for the content.

## Data availability

Condition-level descriptive statistics for all analyzed conditions and the values underlying the figures are provided in Supplementary Data 1 and 2. Test statistics, confidence intervals, and model specifications are reported in the Tables, Supplementary Tables, and Supplementary Methods. Pseudonymized set- and trial-level data are available from the corresponding author (kschae@knu.ac.kr) upon reasonable request. Names, calendar dates, and source-record identifiers, which could identify participants, are not shared.

## Code availability

The analysis code (Python 3.11 with NumPyro 0.22.0, JAX 0.10.2, statsmodels 0.15.0, and polyagamma 2.0.2; R 4.3.3 with lme4 1.1-35.1) is available from the corresponding author upon reasonable request, together with the complete list of package versions. It comprises the hierarchical, cluster-robust, permutation, and outcome-history analyses, the independent R recomputation, the Pólya–Gamma Gibbs sampler, the Biot–Savart coil model used to compute the directional gradient index, and the posture and induction models. A single script reproduces all reported results from the set- and trial-level data.

**Acknowledgements**

We thank the undergraduate students for technical assistance in performing the experiments and tabulating the raw data. This research was supported by the Basic Science Research Program through the National Research Foundation of Korea (NRF), funded by the Korea government (MSIT) (2021R1A2C1094765 to K.-S.C.).

**Author contributions**

K.-S.C. initiated the study and conceptualized the experiments; S.-C.K. and K.-S.C. constructed the experimental apparatus; I.-T.O. performed the experiments; K.-S.C. validated the coil settings; I.-T.O., Y.K., and K.-S.C. analyzed the data; Y.K., Y.-H.K., and K.-S.C. verified the data; I.-T.O. and K.-S.C. visualized the data; K.-S.C. wrote the original draft; Y.-H.K. reviewed and edited the manuscript; and all authors discussed and commented on the manuscript.

**Competing interests**

K.-S.C. is the inventor on patent applications filed by Kyungpook National University Industry-Academic Cooperation Foundation that relate to aspects of this work. The other authors declare no competing interests.

## Figures and legends

Fig. 1

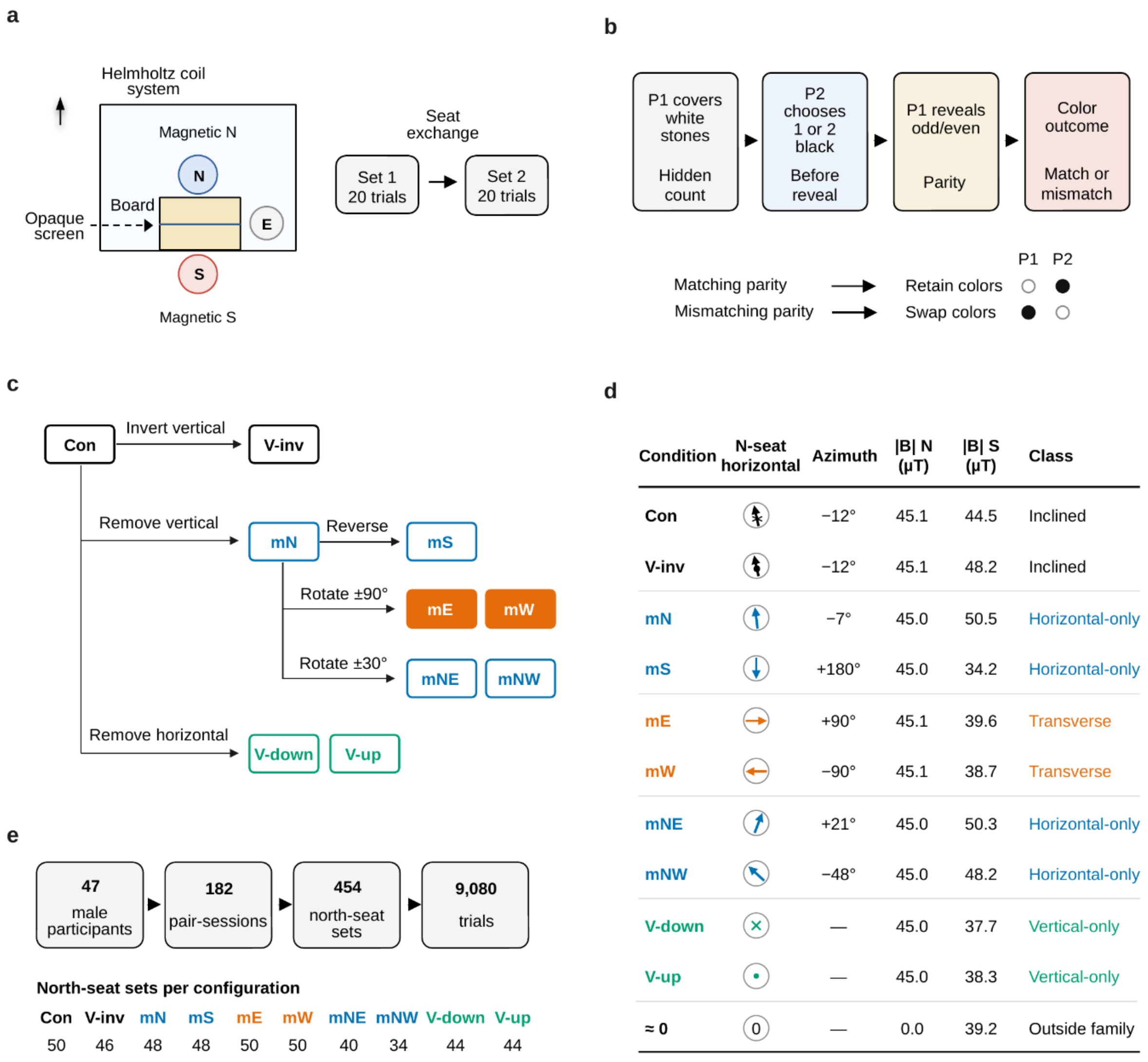

**Fig. 1. Experimental design and two-seat magnetic-field configurations.**

**a**, Shared-coil two-seat arrangement and repeated set structure. The north seat (N) was centered within the three-axis Helmholtz coil system, and the south seat (S) was positioned outside the coil system along the same dyad/visual axis. An opaque screen on the board separated the participants. The experimenter (E) sat to the east. The arrow indicates geographic north, and 'Magnetic N' and 'Magnetic S' mark the coil's magnetic north and south sides. Each game comprised two 20-trial sets, with participants exchanging seats before the second set. **b**, Two-step Go stone selection task. Player 1 (P1) covered a hidden handful of white stones, player 2 (P2) selected one or two black stones, and P1 then revealed whether the concealed count was odd or even. Matching parity retained the original stone colors, whereas mismatching parity exchanged them. **c**, Component-by-component dissection of the ambient magnetic field. Starting from the ambient inclined field (Con), the vertical component was inverted (V-inv) or removed (mN). The resulting horizontal-only field was reversed (mS) or rotated by nominal ±90° (mE, mW) or ±30° (mNE, mNW), while removal of the horizontal component generated vertical-only down and up fields (V-down and V-up). The five rows define symmetric configuration pairs. The transverse pair (mE, mW) was compared with the eight equally weighted non-transverse configurations. Near-zero, radiofrequency, and light-spectrum conditions were outside this intensity-matched family. **d**, Measured two-seat magnetic-field atlas, ordered as in **c**. Arrows indicate the north-seat horizontal azimuth (north up, east right; values referenced to geographic north, east positive). Crosses and dots mark downward and upward vertical fields, respectively. Numbers give the total field intensity (μT) at each seat. Full vectors are reported in Table 1. The near-zero row (≈ 0) is shown for reference. Colors denote geometric class throughout the figures (inclined, black; horizontal-only, blue; transverse, orange; vertical-only, green). **e**, Analysis population of the ten-configuration family and north-seat sets per configuration. Forty-three of the 47 north-seat participants contributed both transverse and non-transverse sets, and 38 met the strict complete-case criterion.

Fig. 2

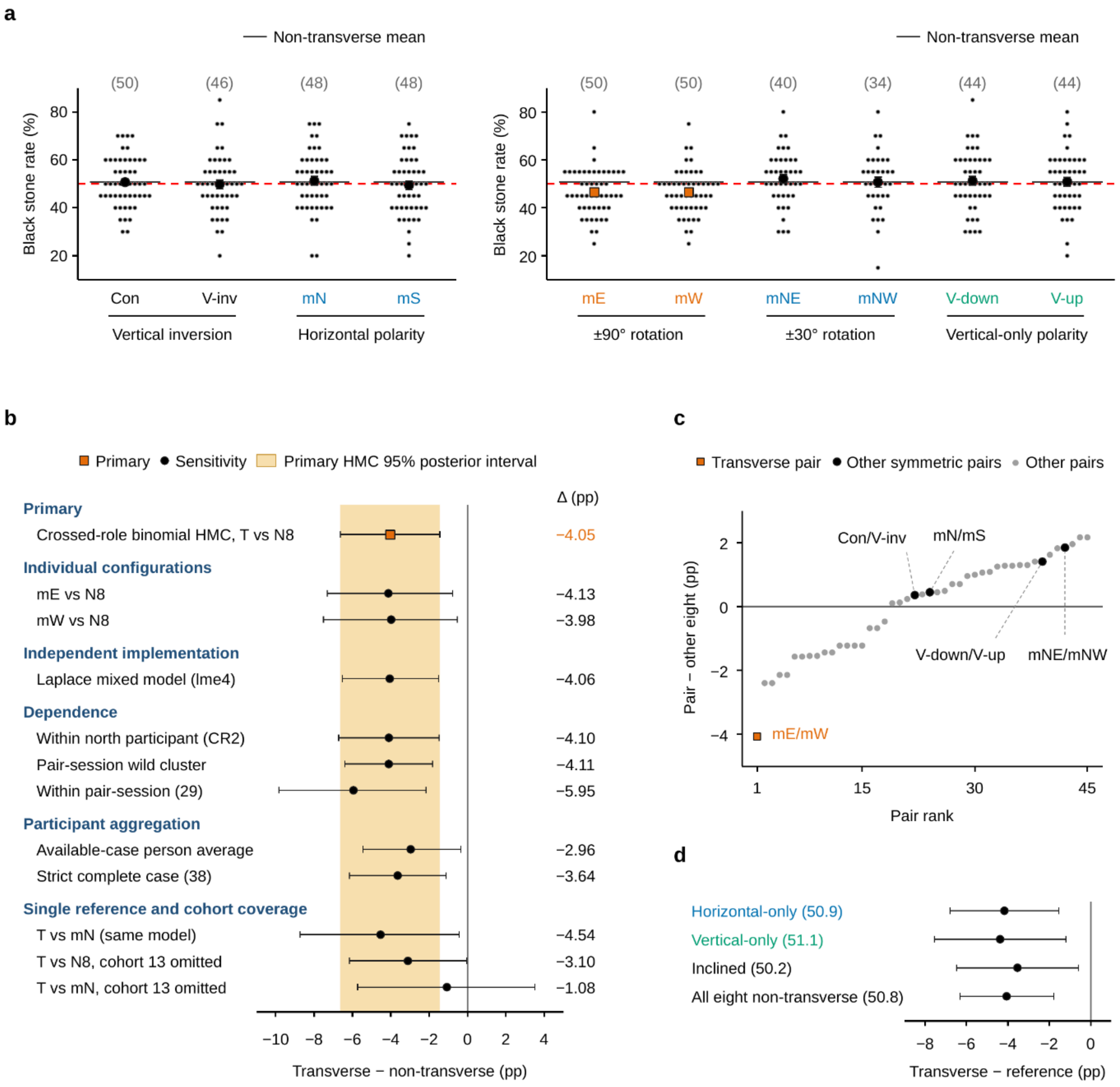

**Fig. 2. Lower black stone rate specific to the transverse pair in the ten-configuration Earth-strength field family.**

**a**, North-seat black stone rates for the ten configurations in dissection order. Small black dots represent individual 20-trial set rates, and larger symbols show condition means ± s.e.m. (black circles, non-transverse configurations; orange squares, transverse pair). Numbers in parentheses indicate north-seat set counts. The red dashed line denotes the theoretical probability (50%), and short solid lines show the equally weighted mean of the eight non-transverse configurations (50.8%). Lines below the condition labels identify the five symmetric pairs and their defining manipulations. **b**, Transverse pair (T) versus the eight non-transverse configurations (N8) or mN. The primary analysis is the crossed-role set-level binomial Hamiltonian Monte Carlo (HMC) model of all ten configurations (covariate-standardized difference, −4.05 pp; 95% posterior interval, −6.64 to −1.44). Additional rows show the individual transverse configurations and sensitivity analyses (CR2, bias-reduced cluster-robust), with further results in Table 3, Supplementary Table 3, and Supplementary Note 1. Horizontal lines indicate 95% posterior, confidence, or resampling intervals according to method, and the shaded band marks the primary HMC interval. **c**, Selection-aware comparison of all 45 possible configuration pairs. Each point shows a pair mean minus the mean of the remaining eight configurations, ranked from most negative (rank 1) to most positive (rank 45): orange square, transverse pair; black circles, the other four symmetric pairs; gray dots, the remaining pairs. The transverse contrast (−4.07 percentage points) was referred to the permutation distribution of the most extreme pair contrast (selection-adjusted two-sided $P = 0.015$ among the five symmetric pairs and 0.096 among all 45). **d**, Transverse pair versus each geometric class (horizontal-only, mN, mS, mNE, and mNW; vertical-only, V-down and V-up; inclined, Con and V-inv) and versus all eight non-transverse configurations, with reference means in parentheses. Points are equally weighted differences with bootstrap 95% confidence intervals, and the classes did not differ from one another ($P = 0.79$; condition-label permutation $P = 0.87$). Because the shared coil altered both seat vectors, all comparisons represent two-seat configuration contrasts. Δ, difference; pp, percentage points (likewise in subsequent figures).

**Fig. 3**

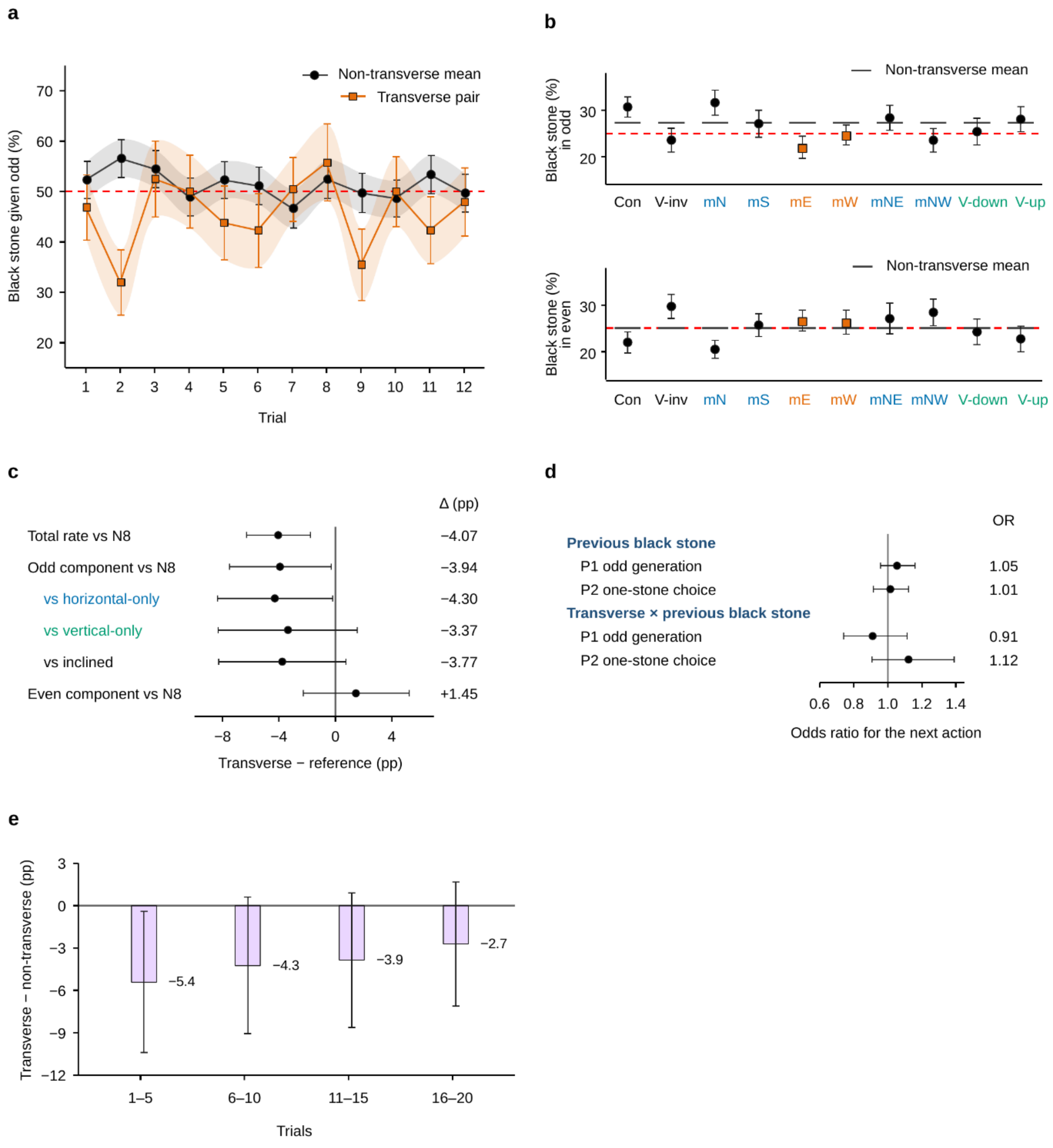

## Fig. 3. Components and time course of the transverse difference.

**a**, Black stone rate conditional on observed odd parity, P(black | odd), during the first 12 north-seat trials, regardless of the role held by the north-seat participant (denominator, sets with odd parity at that trial), for the transverse pair (orange squares) and the equally weighted mean of the eight non-transverse configurations (black circles). Error bars show the s.e. of each mean, computed from binomial variances within configurations, and shaded bands join these ranges across trials. The red dashed line denotes the theoretical probability (50%). **b**, Fixed-denominator black-in-odd (top) and

black-in-even (bottom) components over the first 12 trials, ordered as in Fig. 2a. Each set contributes six step-1 opportunities, and each component is the percentage of these opportunities producing that parity and a black outcome. Symbols show means ± s.e.m. (orange squares, transverse configurations; black circles, other configurations), and short horizontal lines indicate the equally weighted non-transverse means (27.3% and 25.0%). Red dashed lines mark 25%, expected if parity and one/two-stone choice are independent and equiprobable (0.5 × 0.5). Because these measures are joint outcomes, these panels do not isolate P1 parity generation. **c**, Source-level contrasts of the transverse pair with the eight non-transverse configurations (N8) for the total rate and the odd and even components, and odd-component comparisons with each non-transverse geometric class (indented rows). Horizontal lines show percentile bootstrap 95% confidence intervals. The components do not add directly to the total-rate contrast (Supplementary Note 2). **d**, Outcome-history analysis. Odds ratios (OR) show associations of obtaining the black stone on the preceding trial with subsequent P1 odd generation or P2 one-stone choice across the ten configurations, and their interactions with transverse exposure (Holm-adjusted $P = 0.48$). Horizontal lines show 95% posterior intervals. **e**, Transverse − non-transverse difference in north-seat black stone rate across successive five-trial blocks of the 20-trial set. Error bars show percentile bootstrap 95% confidence intervals (sets resampled within configuration; $10^5$ resamples).

Fig. 4

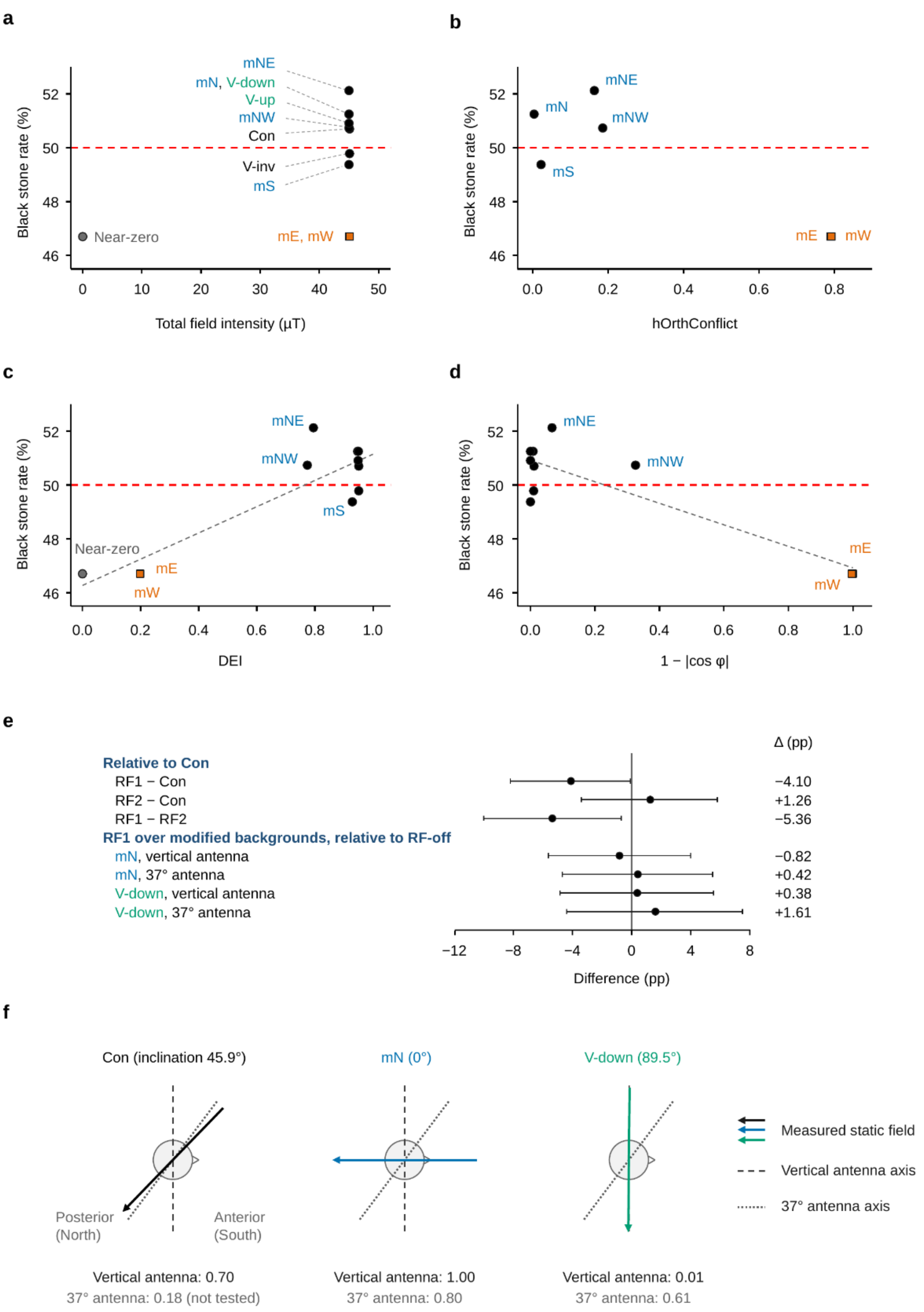

**Fig. 4. Magnetic-field descriptors and** radiofrequency comparisons**.**

**a**, Condition-mean north-seat black stone rate versus measured north-seat total field intensity. Near-zero (gray) is separated by intensity, whereas the other configurations cluster near 45 µT. The identical means for mE, mW, and near-zero (46.70%) arose from non-overlapping sets (467 of 1,000 trials each; Supplementary Methods 3). **b**, Mean rates for the six horizontal configurations versus horizontal-axis conflict, hOrthConflict = 1 − |cos θ|, where θ is the angle between the measured condition and ambient horizontal vectors (mE and mW, 0.79). Vertical-only and near-zero conditions were not assigned values (Supplementary Methods 4). **c**, Rates for 11 conditions versus the directional extractability index (DEI; Supplementary Note 3), with near-zero (gray) at its low-field limit of 0. The dashed line shows a descriptive fit ($r = 0.89$, $R^2 = 0.79$). No inferential $P$ value was assigned because DEI was defined post hoc. **d**, Rates for the ten configurations versus the mid-plane descriptor 1 − |cos φ|. The dashed line shows a descriptive fit (slope, −4.00 percentage points per unit; $R^2 = 0.77$) that largely restates the transverse contrast. In **a**–**d**, red dashed lines denote the theoretical probability (50%), and descriptor values are provided in Supplementary Table 4. **e**, Radiofrequency (RF) contrasts with percentile bootstrap 95% confidence intervals (CIs). Upper rows: RF1 (1.260 MHz) and RF2 (1.890 MHz) relative to Con, and RF1 versus RF2. Lower rows: RF1 over mN or vertical-only down backgrounds, from a vertical or 37°-tilted antenna, relative to the same-background RF-off reference. The RF1 effect over Con minus the effect averaged over the two modified backgrounds was −4.4 percentage points (95% CI, −9.5 to +0.8; $P = 0.10$; Supplementary Table 5). **f**, Sagittal-plane geometry of the measured static field at the north-seat glabella (north to the left, participant facing right) relative to the vertical and 37°-tilted antenna axes. Values indicate the fraction of RF amplitude perpendicular to the static field, assuming the RF vector follows the antenna axis. The tilted antenna axis lies approximately 11° from the measured Con vector, and this combination was not tested (Supplementary Note 4).

**Table 1. Measured magnetic-field vectors at the glabella by condition and seat.**

| Condition | Seat | $B_x$ (μT) | $B_y$ (μT) | $B_z$ (μT) | \|B\| (μT) | Relative total intensity (%) | Horizontal azimuth (°) |
|---|---|---|---|---|---|---|---|
| Con | N | 30.7 | −6.6 | 32.4 | 45.1 | 100 | −12 |
| | S | 29.3 | −4.1 | 33.2 | 44.5 | 99 | −8 |
| Vertical-component inversion | N | 30.6 | −6.5 | −32.5 | 45.1 | 100 | −12 |
| | S | 32.4 | −5.8 | 35.2 | 48.2 | 107 | −10 |
| Horizontal-only mN | N | 44.7 | −5.6 | 0.0 | 45.0 | 100 | −7 |
| | S | 37.6 | −6.7 | 33.1 | 50.5 | 112 | −10 |
| mS | N | −45.0 | 0.0 | 0.1 | 45.0 | 100 | +180 |
| | S | −2.0 | −6.1 | 33.6 | 34.2 | 76 | −108 |
| mE | N | 0.0 | 45.1 | 0.0 | 45.1 | 100 | +90 |
| | S | 17.7 | −8.2 | 34.5 | 39.6 | 88 | −25 |
| mW | N | −0.1 | −45.1 | 0.1 | 45.1 | 100 | −90 |
| | S | 17.0 | −3.4 | 34.6 | 38.7 | 86 | −11 |
| mNE | N | 42.0 | 16.2 | 0.2 | 45.0 | 100 | +21 |
| | S | 35.4 | −6.2 | 35.2 | 50.3 | 112 | −10 |
| mNW | N | 30.3 | −33.2 | 0.1 | 45.0 | 100 | −48 |
| | S | 32.1 | −4.0 | 35.7 | 48.2 | 107 | −7 |
| Vertical-only down | N | −0.4 | 0.0 | 45.0 | 45.0 | 100 | — |
| | S | 15.2 | −5.5 | 34.1 | 37.7 | 84 | −20 |
| Vertical-only up | N | 0.2 | 0.1 | −45.0 | 45.0 | 100 | — |
| | S | 15.7 | −5.6 | 34.5 | 38.3 | 85 | −20 |
| ≈ 0 (near-zero) | N | 0.0 | 0.0 | 0.0 | 0.0 | 0 | — |
| | S | 17.8 | −5.8 | 34.4 | 39.2 | 88 | −18 |

Values are means across three participants with the tallest, median, and shortest sitting heights among all participants. Components use +x geographic north, +y east, and +z downward, so positive $B_z$ is downward. |B|, the total field intensity, is the vector magnitude of the three components. Horizontal azimuth is the direction of the horizontal component ($B_x$, $B_y$), measured from geographic north with east positive. A dash (—) indicates a horizontal component of <1 μT. The south-seat azimuth under mS rests on a horizontal component of only 6.4 μT. Relative total intensity compares the south-seat |B| with the north-seat |B| of the same condition. For the ≈ 0 condition, the north- and south-seat values are instead compared with the corresponding seat in Con. Conditions are named with respect to the north seat. South-seat values provide physical context and were not independent primary-model covariates. Con, ambient control; N, north seat; S, south seat; mN, horizontal-only north-seat field near magnetic north; mS, mE, and mW, horizontal-only north-seat fields along the geographic south, east, and west axes; mNE/mNW, nominal ±30° rotations of mN. The nominal ±30° rotations were delivered at +33° and −35° about the ambient horizontal vector, equivalent to +28° and −40° relative to mN, because the coil-generated mN azimuth differs from the ambient horizontal azimuth by about 5°. mE, mW, and mS lie at +97°, −83°, and 187° relative to mN. Relative to the ambient horizontal vector, the reference of hOrthConflict, mE and mW lie at +102° and −78°, so hOrthConflict is 0.79 rather than 1. All descriptors reported in Supplementary Table 4 are computed from the measured vectors listed here rather than from the nominal settings.

**Table 2. Inference hierarchy of the analyses.**

| Level | Analysis | Role in inference | Reported in |
|---|---|---|---|
| **Primary analysis** | Crossed-role set-level binomial HMC; transverse (mE, mW) vs eight non-transverse configurations, equally weighted | Primary; retains the set-level binomial denominator and models role and pair-session dependence | Fig. 2b; Table 3 |
| **Family structure** | Homogeneity and equivalence of the eight non-transverse configurations; selection-aware permutation of the most extreme pair contrast; geometric classes | Separates the transverse pair from the family; the 45-pair set is broader and more conservative | Results; Fig. 2c, d; Supplementary Tables 2 and 3; Supplementary Note 1 |
| **Full-data structural robustness** | Laplace, four-prior HMC, independent Gibbs, Bradley–Terry reimplementation, beta-binomial cross-check, wild-cluster, nuisance, and leave-one-out analyses | Tests dependence, dispersion, nuisance adjustment, and influence | Fig. 2b; Supplementary Fig. 1; Table 3; Supplementary Table 3; Supplementary Note 1 |
| **Participant and cohort sensitivity analyses** | Available-case and strict person averages; within-north-participant configuration FE; cohort omission; single-reference comparison with mN | Consistently negative; magnitude and resolution depend on coverage and reference | Fig. 2b; Table 3; Supplementary Table 3; Supplementary Note 1 |
| **Secondary process** | Conditional and fixed-denominator components of the first 12 trials; role-specific primitive endpoints | Describes endpoint composition; the odd–even contrast is unresolved; does not localize a receptor or causal target | Fig. 3a–c; Supplementary Fig. 2a; Table 3; Supplementary Note 2 |
| **Sequence and outcome history** | Person-linked cross-lag models; previous-outcome and running-score models; trial blocks | Within-person role sequence; no evidence of an outcome-driven strategy shift; no interpersonal causal test | Fig. 3d, e; Supplementary Fig. 2b; Table 3; Supplementary Table 3; Supplementary Note 2 |
| **Mechanism-linked and descriptive** | RF, near-zero comparison, hOrthConflict, mid-plane, DGI, light spectrum, DEI, descriptor predictive comparison, and posture and induction models | Outside the primary inference; mechanism- and physics-linked context and targeted tests | Fig. 4; Table 3; Supplementary Figs 3–5; Supplementary Tables 4–7; Supplementary Notes 3–5 |

The pooled primary estimand and evidence hierarchy were defined during the present reanalysis, after condition labels had been decoded, and were not preregistered (Methods). Secondary and descriptive analyses were not adjusted for multiplicity across analyses. HMC, Hamiltonian Monte Carlo; FE, fixed effects; mN, horizontal-only north-seat field near magnetic north; mE and mW, horizontal-only north-seat fields along the geographic east and west axes (transverse pair); RF, radiofrequency; hOrthConflict, horizontal-axis conflict descriptor; mid-plane, mid-plane descriptor $1 - |\cos \varphi|$; DGI, directional gradient index; DEI, directional extractability index.

**Table 3. Key estimates for the transverse versus non-transverse contrast and process analyses.**

| Evidence role | Analysis / contrast | Endpoint | Estimate | 95% interval / inferential note |
|---|---|---|---|---|
| **Primary, full information** | Crossed-role HMC / transverse vs non-transverse, equally weighted | Total north-seat rate | −4.05 pp / OR 0.849 | −6.64 to −1.44 pp / OR 0.765–0.944; Pr(Δ < 0) = 0.999 |
| **Structural robustness** | Laplace GLMM in R (lme4), independent implementation | Total north-seat rate | −4.06 pp / OR 0.850 | Parametric bootstrap −6.52 to −1.49 pp; OR 0.768–0.940 |
| | Bradley–Terry reimplementation | Total north-seat rate | −4.10 pp / OR 0.847 | −6.72 to −1.55 pp; one propensity per participant |
| | Beta-binomial posterior-reweighting cross-check | Total north-seat rate | −4.05 pp | −6.64 to −1.34 pp; $BF_{\beta/bin}$ = 0.0043–0.044 across three priors |
| | Pair-session wild-cluster analysis | Total north-seat rate | −4.11 pp | −6.39 to −1.81 pp; two-sided *P* = 0.0011 |
| **Participant aggregation** | Available-case person average | Participant-average total rate | −2.96 pp | −5.44 to −0.35 pp; *n* = 43; two-sided reallocation *P* = 0.035 |
| | Strict complete case (mE, mW, and at least one non-transverse configuration) | Participant-average total rate | −3.64 pp | −6.15 to −1.12 pp; *n* = 38; two-sided sign-flip *P* = 0.0087 |
| **Participant and cohort sensitivity** | Within-north-participant configuration contrast | Total north-seat rate | −4.10 pp; cohort-13 omitted −3.29 pp | CR2 95% CI −6.72 to −1.48, *P* = 0.003; CR1 *P* = 0.005; dyadic-overlap *P* = 0.002; cohort-13 omitted CR2 *P* = 0.027 |
| **Secondary process** | Person-FE logistic / transverse vs non-transverse | North P1 odd probability | −1.55 pp / OR 0.939 | OR 0.826–1.068; *P* = 0.31–0.44 |
| | Person-FE logistic / transverse vs non-transverse | South P2 one probability | +3.91 pp / OR 1.175 | OR 1.027–1.345; *P* = 0.009–0.019; Holm-adjusted *P* = 0.035–0.077; crossed random-intercept GLMM *P* = 0.034 |
| | Person-FE logistic / transverse vs non-transverse | South P1 odd probability | +2.67 pp / OR 1.114 | OR 0.979–1.268; *P* = 0.10–0.24 |
| | Person-FE logistic / transverse vs non-transverse | North P2 one probability | +0.42 pp / OR 1.017 | OR 0.884–1.170; *P* = 0.81–0.85 |

**Table 3 (continued).**

| Evidence role | Analysis / contrast | Endpoint | Estimate | 95% interval / inferential note |
|---|---|---|---|---|
| **Sequence and outcome history** | Previous-outcome HMC / current P1 | Transverse × previous black stone | OR 0.91 | 0.74–1.11; Holm-adjusted $P$ = 0.48 |
| | Previous-outcome HMC / current P2 | Transverse × previous black stone | OR 1.12 | 0.91–1.39; Holm-adjusted $P$ = 0.48 |
| **Physical-profile comparison** | North-transverse vs south-intensity descriptors | Descriptor predictive error | Leave-one-profile-out RMSE 1.79 vs 1.98 pp | 11 profiles; bootstrap differences uncertain; does not separate north- from south-based accounts (Supplementary Note 3) |

Analyses estimate closely related total-rate contrasts but differ in weighting, aggregation, coverage, and assumptions. Intervals spanning zero do not resolve the contrast and do not by themselves indicate absence of an effect. Hierarchical-model intervals are posterior intervals, and other intervals are confidence or resampling intervals. In the four person-FE primitive-process rows, odds-ratio intervals use set-clustered covariance, $P$ values are unadjusted unless a Holm-adjusted value is given, and $P$ ranges span six covariance structures (Supplementary Methods 4). The GLMM $P$ value is from a separate crossed random-intercept model. HMC, Hamiltonian Monte Carlo; pp, percentage points; OR, odds ratio; Δ, equally weighted transverse − non-transverse difference; $\Pr(\Delta < 0)$, posterior probability that Δ is negative; mE and mW, horizontal-only north-seat fields along the geographic east and west axes (transverse pair); GLMM, generalized linear mixed model; $BF_{\beta/\mathrm{bin}}$, beta-binomial Bayes factor; $n$, number of participants; CR2, bias-reduced cluster-robust variance; CI, confidence interval; CR1, conventional cluster-robust variance; FE, fixed effects; P1 and P2, players 1 and 2; RMSE, root-mean-square error.

**Supplementary Information**

# Earth-strength magnetic-field axis modulates non-spatial probabilistic decision-making in humans

In-Taek Oh[1], Yong-Hwan Kim[2], Soo-Chan Kim[3], Yongkuk Kim[4,*], Kwon-Seok Chae[1,5,*]

**Contents**

Supplementary Methods 1. Glucose records and glucose-adjusted sensitivity analyses

Supplementary Methods 2. Radiofrequency resonance calculations and amplitude profile

Supplementary Methods 3. Record matching and label reconciliation

Supplementary Methods 4. Statistical analysis details

Supplementary Methods 5. Coil model parameters and validation

Supplementary Methods 6. Task posture observations and model inputs

Supplementary Data 1. Condition-level descriptive statistics

Supplementary Data 2. Values underlying the figures

Supplementary display items use the 454-set ten-configuration dataset or the explicitly identified 13-condition, earlier-study, RF-background, and light-spectrum datasets. Cohort is a set-level covariate, not part of participant identity. Statistical results are grouped by inferential role and type of computation. Participant aggregation, label reallocation, and trial-level process analyses are sensitivity analyses rather than preregistered confirmatory tests. Reference numbers in the Supplementary Information refer to the Supplementary References. Abbreviations follow the main text: Con, ambient control; mN and mS, horizontal-only north and south fields; mE and mW, transverse east and west fields; mNE and mNW, nominal ±30° rotations of mN; RF, radiofrequency; RF1 and RF2, 1.260 and 1.890 MHz RF fields; P1 and P2, player 1 and player 2; CR1 and CR2, conventional and bias-reduced cluster-robust variance; hOrthConflict, horizontal-axis conflict descriptor; DEI, directional extractability index; DGI, directional gradient index.

Supplementary Fig. 1

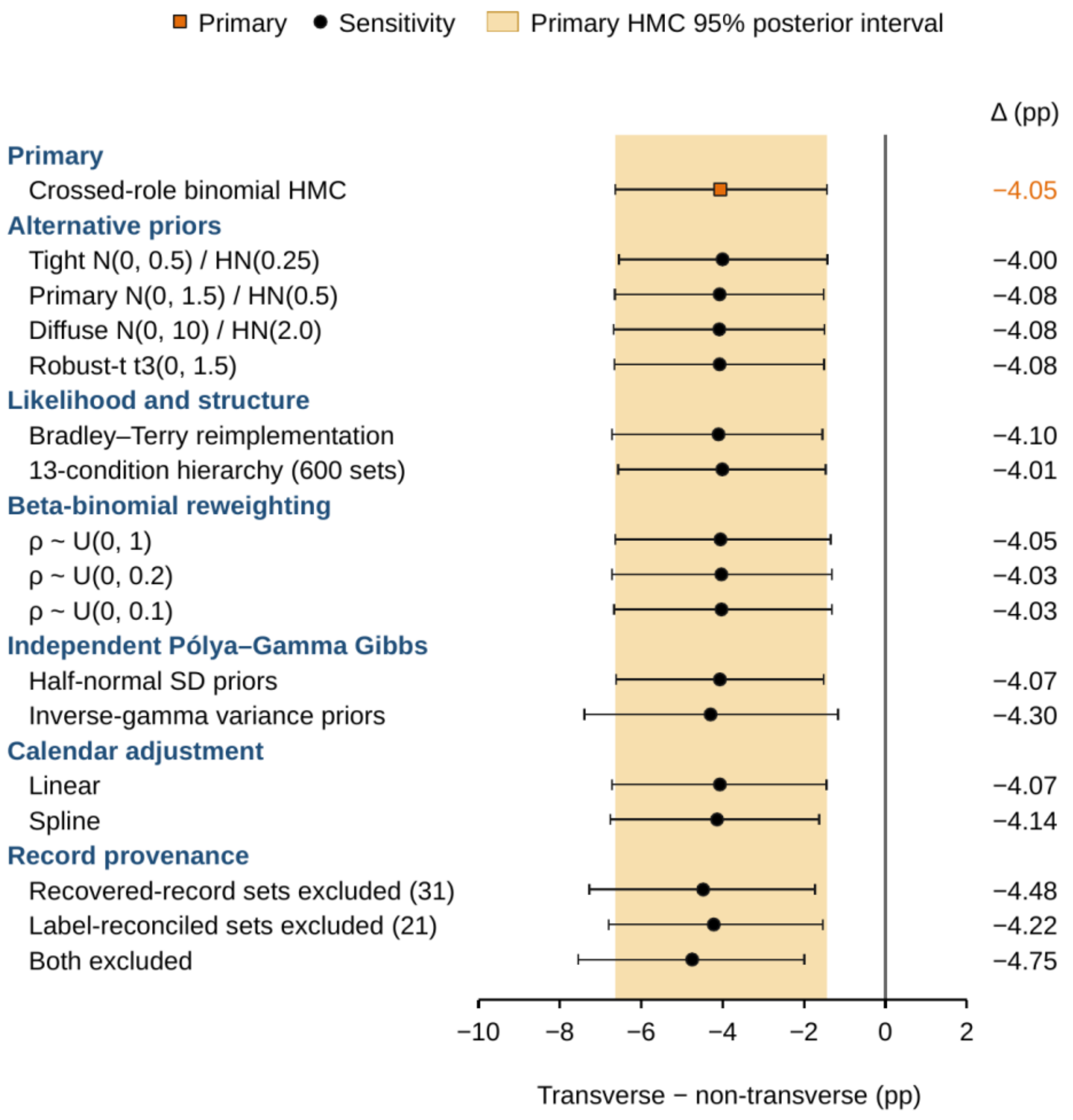

**Supplementary Fig. 1. Robustness and sensitivity of the primary transverse-field contrast.** Transverse versus non-transverse contrast under alternative priors, likelihoods and model structures, beta-binomial reweighting, independent Pólya–Gamma Gibbs sampling, calendar adjustment, and record-provenance exclusions, grouped as indicated. The orange square denotes the primary crossed-role Hamiltonian Monte Carlo (HMC) estimate, whereas filled circles denote sensitivity estimates. Horizontal lines show 95% posterior or resampling intervals, as appropriate, and the shaded band marks the primary HMC interval. All estimates were negative, and the displayed intervals excluded zero, although these analyses share the same underlying data and are not independent confirmatory tests. The primary HMC model showed satisfactory convergence, with no divergences, a maximum R-hat of 1.013, a minimum effective sample size of 370, and $\Pr(\Delta < 0) = 0.999$, where Δ denotes the transverse − non-transverse contrast. The prior-sensitivity rows were refitted with longer chains, so the primary prior appears there as −4.08 rather than −4.05 percentage points. This difference reflects Monte Carlo error. Longer-chain analyses and additional model diagnostics are detailed in Supplementary Note 1, and further sensitivity analyses are shown in Fig. 2b and Supplementary Table 3. HN, half-normal; N, normal; SD, standard deviation; $t_3$, Student's *t* with 3 degrees of freedom; U, uniform; ρ, intra-set correlation (beta-binomial); pp, percentage points (likewise in subsequent Supplementary Figures).

**Supplementary Fig. 2**

**a**

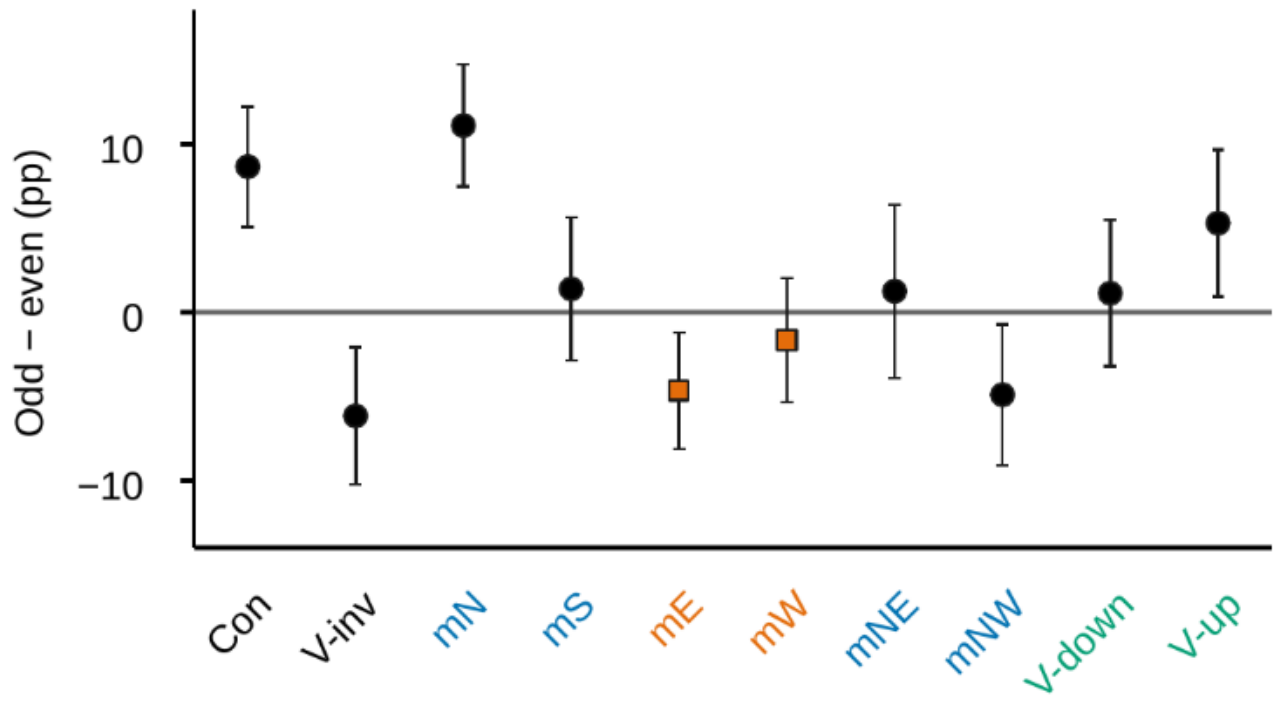

**b**

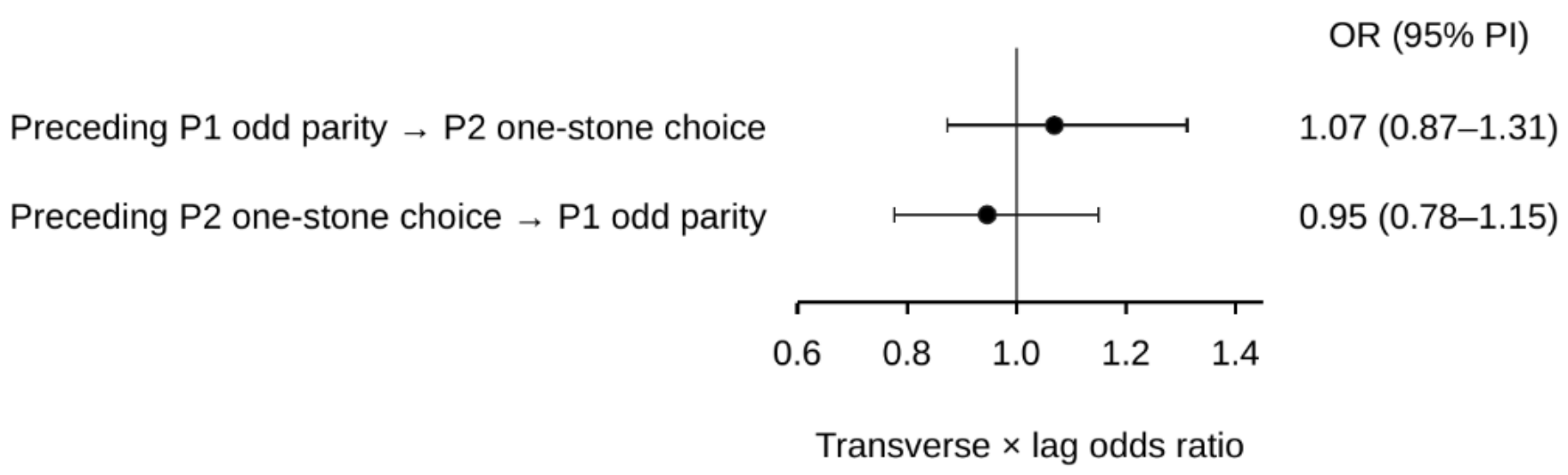

**Supplementary Fig. 2. Process decomposition and sequence sensitivity.**

**a**, Odd − even difference between the black-in-odd and black-in-even components during the first 12 trials for each configuration (mean ± s.e.m.; orange squares, transverse pair). The transverse − non-transverse difference in this quantity (the condition × parity interaction) was −5.39 percentage points (bootstrap 95% confidence interval, −11.13 to +0.28; $P = 0.062$), and the interaction values differed among the eight non-transverse configurations (homogeneity $P = 0.019$–$0.046$). Relative to mN alone, which had the largest odd − even difference, the transverse contrast was −14.28 percentage points (95% confidence interval, −22.94 to −5.81; $P = 0.0008$). However, the position of mN at the upper end of the odd component and the lower end of the even component lies within the range expected for one of eight configurations ($P = 0.49$ and 0.40). V-inv, vertical-component inversion; V-down and V-up, vertical-only down and up (likewise in Supplementary Fig. 5). **b**, Transverse × lag odds ratios (OR) with 95% posterior intervals (PI) from person-linked actor- and set-random-intercept HMC models (Pr(OR > 1) = 0.73 and Pr(OR < 1) = 0.70, upper and lower rows, respectively). The first trial of each set was excluded, leaving 8,626 trials, 47 actors, and 454 sets. Because roles alternate, each arrow links successive P1 and P2 actions of the same participant and thus represents a within-person sequence, not an influence of one participant on the other. Dependence on the partner's preceding action was examined separately using outcome-history models (Supplementary Note 2; Supplementary Table 3).

**Supplementary Fig. 3**

**a**

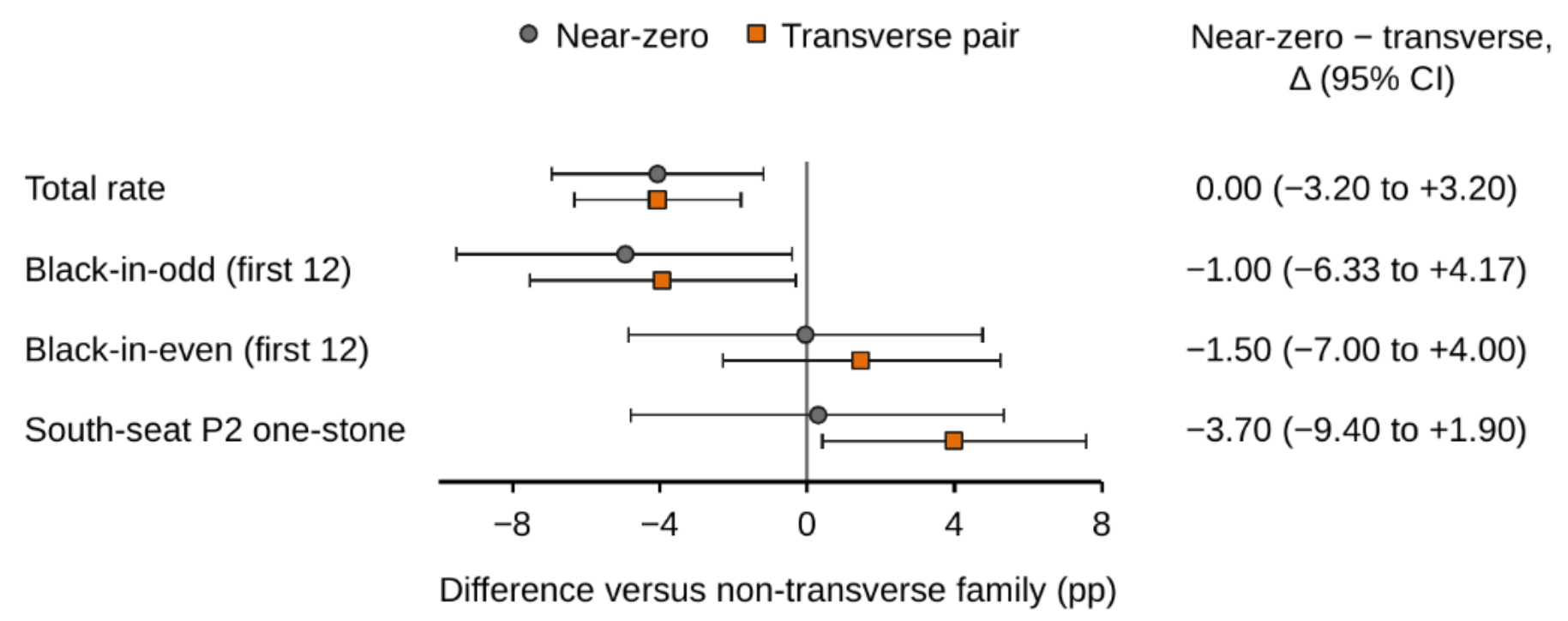

**b**

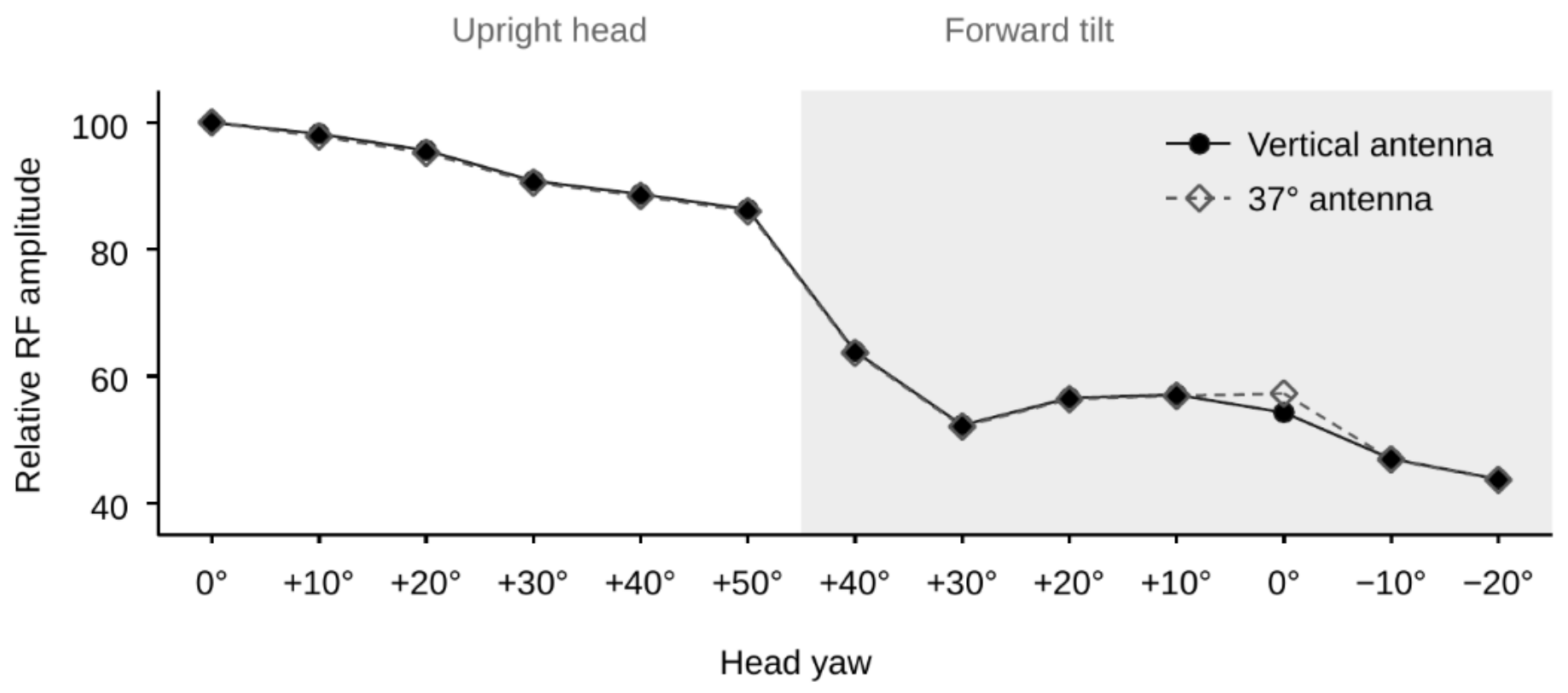

**Supplementary Fig. 3. Near-zero comparison and posture dependence of the radiofrequency amplitude.**

**a**, Near-zero (gray circles; 50 sets) and the transverse pair (orange squares) relative to the equally weighted non-transverse family for the total black stone rate, the first-12 black-in-odd and black-in-even components, and the set-level proportion of one-stone choices by the south-seat participant as P2 (the person-fixed-effects estimate is given in Table 3). Horizontal lines show percentile bootstrap 95% confidence intervals (CIs; $10^6$ set resamples within configuration), and the right column shows near-zero − transverse differences. In the earlier study,[1] near-zero reduced the odd-case step-1 rate (black stone outcomes in odd-parity trials with the north-seat participant as P1) relative to Con (−9.0 percentage points; bootstrap 95% CI, −14.7 to −3.0; $q_0 = 0.0019$; from the published trial-level files, see Supplementary Methods 3), although near-zero was predominantly tested after a Con game within the same visit. $q_0$ denotes the source-reported zero-percentile statistic under the original contrast coding. Values <0.025 or >0.975 indicate a two-sided difference, but $q_0$ is not a conventional $P$ value. **b**, Single-point radiofrequency (RF) amplitude at the glabella, normalized to the forward-facing posture (= 100), across 13 postures along a representative path reproducing task movements: head yaw toward the experimenter's hand cue to +50°, return through the forward-facing position, and yaw toward the stone container to −20°, with a progressively deepening forward head tilt to about 45° on the return limb (shaded). The two antenna positions produced closely similar profiles ($r$ = 0.9992). The profile is not a field map over the head or a dwell-time-weighted exposure dose (Supplementary Methods 2).

**Supplementary Fig. 4**

**a**

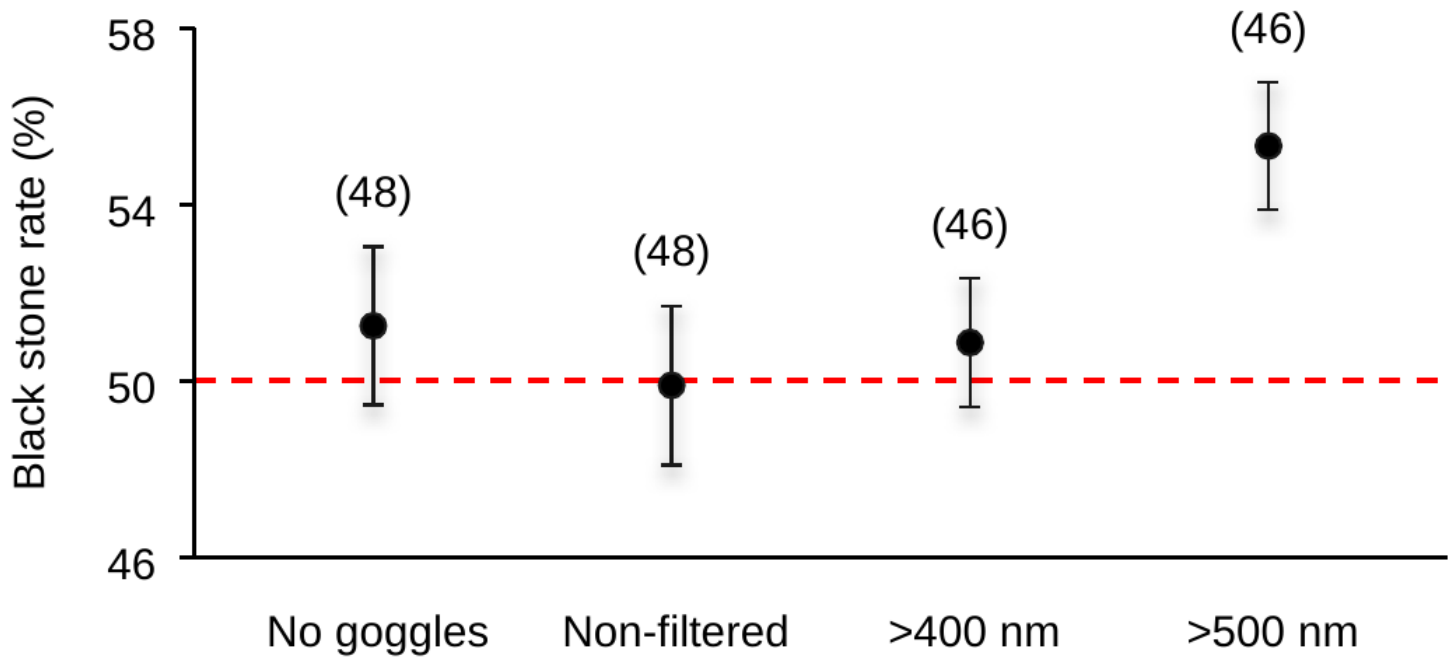

**b**

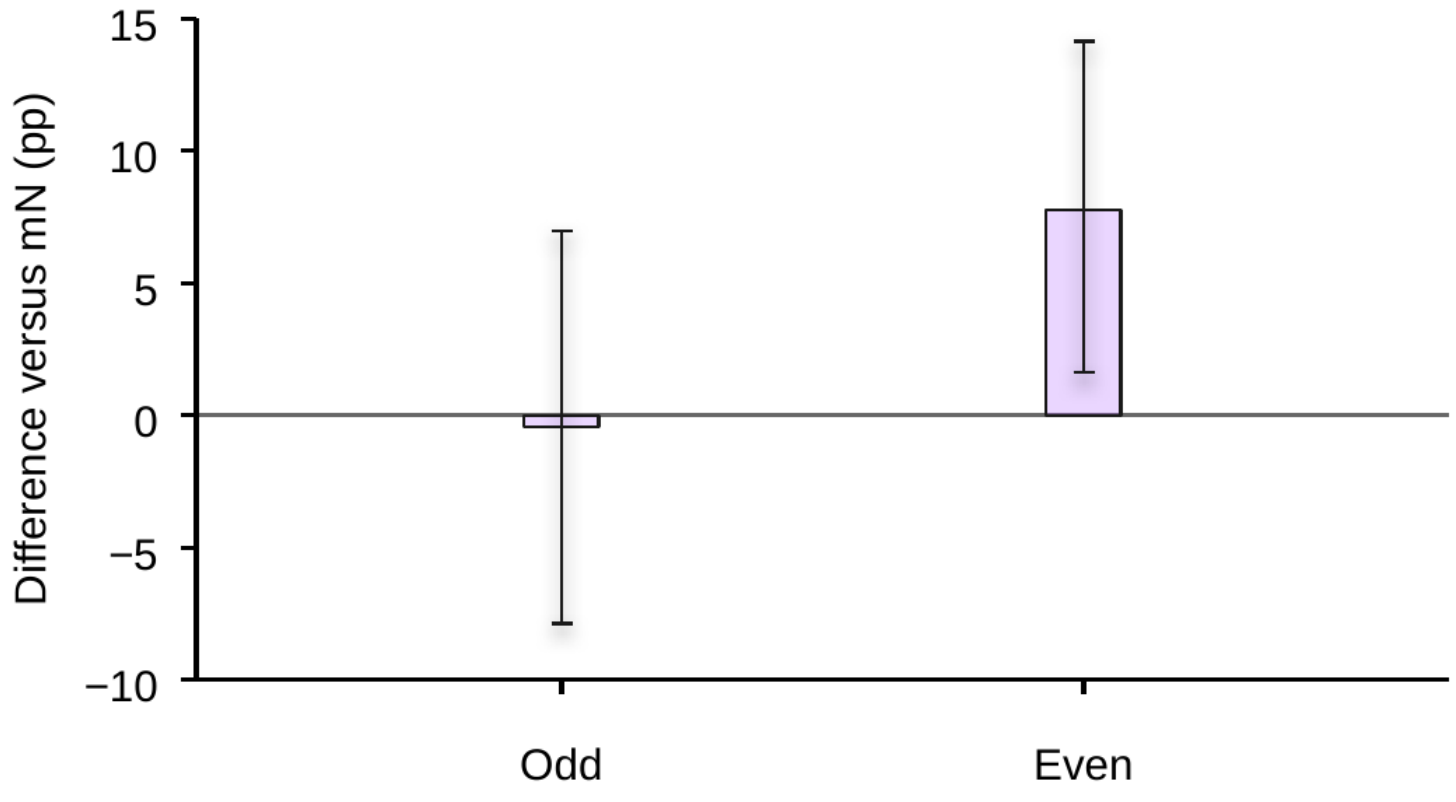

**Supplementary Fig. 4. Light-spectrum experiment.**

**a**, Source-level north-seat black stone rates under no goggles, non-filtered goggles, or long-pass goggles transmitting >400 nm or >500 nm, worn by the participant in the south seat (north-seat sets in parentheses). Error bars show the s.e.m. across the analyzed sets. The >500 nm goggles removed blue light, and the no-goggles condition comprised the same sets as mN. The red dashed line denotes the theoretical probability (50%), which lies within 0.1 percentage points of the non-filtered goggle condition (49.90%). **b**, Odd and even component contrasts of >500 nm versus mN, with percentile bootstrap 95% confidence intervals ($10^6$ set resamples within condition) based on the adopted goggle assignment, one of several consistent with the records (Supplementary Methods 3). Because mN lies at the low end of the even component among the ten configurations (Fig. 3b), this reference enlarges the even-component contrast. Rates, contrasts, and statistics are given in Supplementary Table 6. These conditions lie outside the primary inferential family, and illuminance was not matched across goggle conditions, so spectral and light-intensity effects cannot be separated (Supplementary Note 3).

**Supplementary Fig. 5**

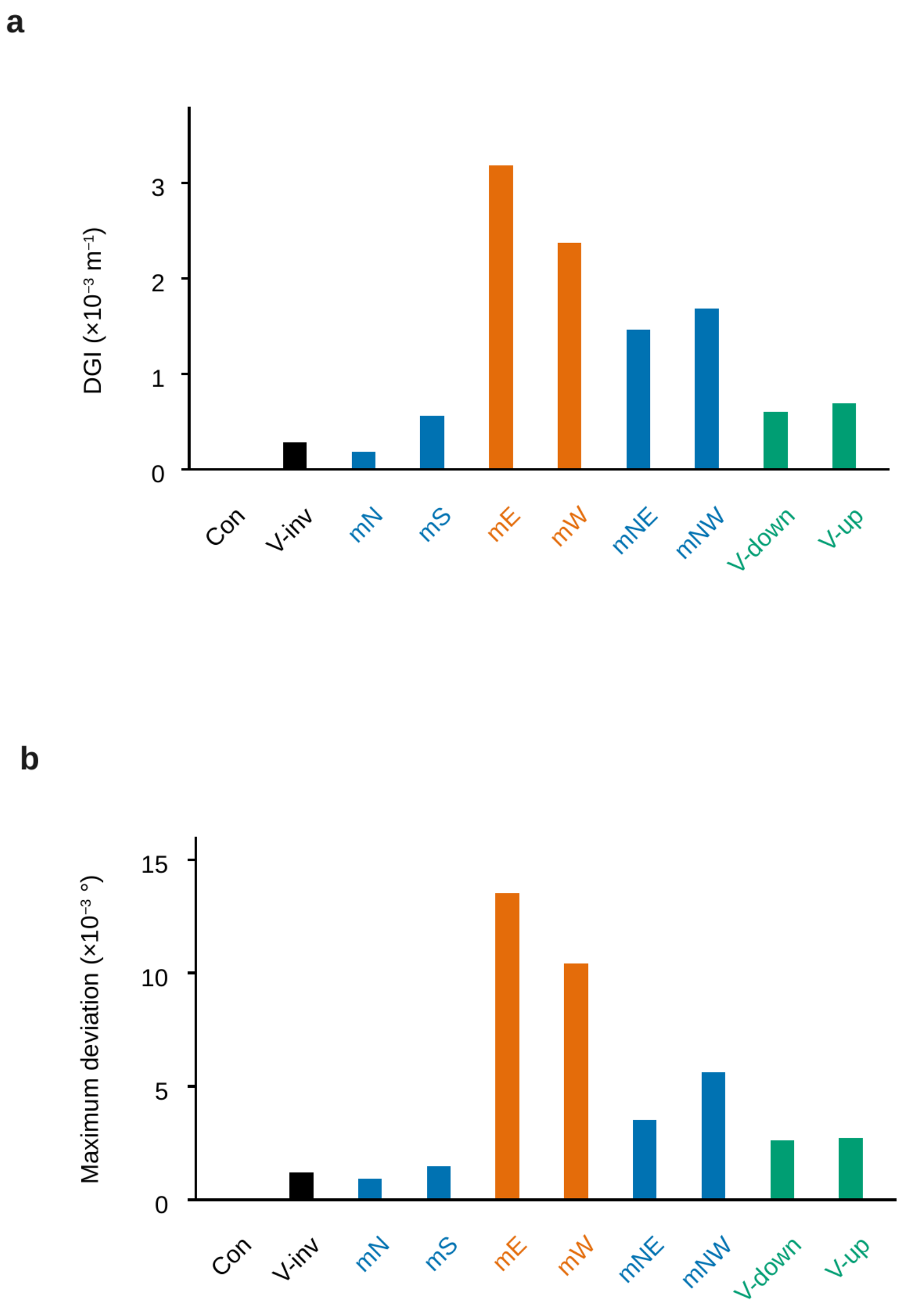

**Supplementary Fig. 5. Modeled spatial uniformity of the magnetic-field configurations.**

**a**, Directional gradient index (DGI) for each configuration from an idealized center-anchored Biot–Savart model of the coil geometry, averaged over a ±50 mm cube centered on the north-seat participant's glabella with 5 mm grid spacing. Bar colors follow the geometric classes defined in Fig. 1d. DGI is zero for Con under the model assumption of a spatially uniform ambient baseline. **b**, Maximum modeled angular deviation in field direction from the center vector within the same cube. Modeled values are not measured field maps. Region-of-interest and grid-spacing sensitivity analyses are listed in Supplementary Table 7 and pair-separation sensitivity in Supplementary Methods 5.

**Supplementary Table 1. Pair and pair-session structure of the ten-configuration family.**

| Item | Value | Interpretation |
|---|---|---|
| Pairs | 66 | Unique person-level pairs; 27 contributed both transverse and non-transverse sets |
| Pair-sessions | 182 | On 110 test dates |
| Pair-sessions with transverse and non-transverse sets | 29 (15.9%) | mE + mS 9; mNE + mW 9; mS + mW 5; vertical-component inversion + mE 3; mN + mW 2; vertical-only up + mE 1 |
| Seat phase | 233 phase 1; 221 phase 2 | First or second 20-trial set within a game |

Counts are person-level and refer to the 454 north-seat sets of the ten-configuration family (Fig. 1e). A pair-session groups a test date and dyad, and can include more than two sets.

**Supplementary Table 2. Equivalence and Bayesian evidence among the non-transverse configurations.**

| Contrast | Δ (pp) | TOST (±5 pp) | $BF_{01}$ |
|---|---|---|---|
| Con vs other seven | −0.04 | *P* = 0.0034 (equivalent) | 6.1 |
| Vertical-component inversion vs other seven | −1.09 | *P* = 0.020 (equivalent) | 5.0 |
| Horizontal-only mN vs other seven | +0.60 | *P* = 0.0090 (equivalent) | 5.7 |
| mS vs other seven | −1.57 | *P* = 0.032 (equivalent) | 4.3 |
| mNE vs other seven | +1.57 | *P* = 0.044 (equivalent) | 4.2 |
| mNW vs other seven | 0.00 | *P* = 0.010 (equivalent) | 5.2 |
| Vertical-only down vs other seven | +0.59 | *P* = 0.011 (equivalent) | 5.5 |
| Vertical-only up vs other seven | +0.20 | *P* = 0.0065 (equivalent) | 5.7 |
| mE vs all eight (for comparison) | −4.03 | *P* = 0.29 (equivalence not demonstrated) | 0.55 |
| mW vs all eight (for comparison) | −4.03 | *P* = 0.29 (equivalence not demonstrated) | 0.54 |

Each non-transverse configuration was compared with the pooled sets of the other seven. The transverse configurations, compared with all eight, are shown for reference. Their Δ is set-pooled, and the equally weighted transverse contrast is −4.07 percentage points. Equivalence was assessed by two one-sided tests against a ±5-percentage-point margin. A value in parentheses states whether the margin was met. Bayes factors compare the null against the alternative, so values above one favor the null. In the ten-configuration HMC model, the posterior probability that each non-transverse configuration lay within ±5 percentage points of the mean of the other seven was 0.96–1.00. Equivalence outcomes are supportive and do not prove the absence of an effect. Δ, difference between the configuration and the compared configurations; pp, percentage points; TOST, two one-sided tests; $BF_{01}$, Bayes factor for the null against the alternative; HMC, Hamiltonian Monte Carlo.

**Supplementary Table 3. Additional sensitivity, heterogeneity, and scheduling analyses.**

| Evidence role | Analysis / contrast | Estimate (pp) | Interval / test | Interpretation |
|---|---|---|---|---|
| **Alternative estimands** | HMC; mE vs non-transverse | −4.13 | −7.32 to −0.78 | Individual configuration estimate |
| | HMC; mW vs non-transverse | −3.98 | −7.51 to −0.55 | Individual configuration estimate |
| | HMC; transverse vs seven non-transverse configurations excluding mN | −3.99 | −6.59 to −1.37 | mN-independent family |
| | HMC; set-weighted family contrast | −4.02 | −6.63 to −1.42 | Weights by set count |
| | HMC; typical-participant standardization (random effects set to zero) | −4.07 | −6.66 to −1.45 | Same fit, alternative standardization |
| **Reference and cohort coverage** | Source-level; mE vs mN (preprint single-reference comparison) | −4.55 | Bootstrap 95% CI −8.93 to −0.11; *P* = 0.045 | All cohorts; preprint comparison reproduced with two-sided inference |
| | Source-level; mW vs mN (preprint single-reference comparison) | −4.55 | Bootstrap 95% CI −8.89 to −0.17; *P* = 0.042 | All cohorts; preprint comparison reproduced with two-sided inference |
| | Source-level; mE vs mN, cohort 13 omitted | −1.46 | Bootstrap 95% CI −6.70 to +3.83; *P* = 0.59 | Single reference sensitive to cohort 13 (mN 57.9% in that cohort) |
| | Source-level; mW vs mN, cohort 13 omitted | −0.50 | Bootstrap 95% CI −5.67 to +4.71; *P* = 0.85 | Single reference sensitive to cohort 13 (mN 57.9% in that cohort) |
| | Source-level; transverse vs eight non-transverse configurations, cohort 13 omitted | −2.99 | Bootstrap 95% CI −5.62 to −0.30; *P* = 0.030 | Family reference attenuated but below zero |
| | Source-level; transverse vs seven non-transverse configurations, Con omitted | −4.08 | Bootstrap 95% CI −6.39 to −1.73; *P* = 0.0007 | Contrast does not depend on Con, the only non-transverse configuration from the earlier study |

**Supplementary Table 3 (continued).**

| Evidence role | Analysis / contrast | Estimate (pp) | Interval / test | Interpretation |
|---|---|---|---|---|
| **Structural robustness** | Cluster-robust GLM; mE vs mean of other nine | −3.69 | 95% CI −6.10 to −1.29; raw *P* = 0.0026; Holm *P* = 0.026 | Ten-configuration screen |
| | Cluster-robust GLM; mW vs mean of other nine | −3.61 | 95% CI −6.74 to −0.48; raw *P* = 0.024; Holm *P* = 0.21 | Ten-configuration screen |
| | Raw source rows; set-pooled | −4.03 | 95% CI −6.31 to −1.75; Welch *P* = 0.0006 | Source-level reproduction; does not model repeated participation |
| | Participant leave-one-out | −4.49 to −3.52 | Range across 47 deletions | No sign reversal |
| | Pair leave-one-out | −4.48 to −3.62 | Range across 66 deletions | No sign reversal |
| | Pair-session leave-one-out | −4.57 to −3.74 | Range across 182 deletions | No sign reversal |
| **Sampler / prior robustness** | Independent Pólya–Gamma Gibbs; half-normal priors | −4.07; OR 0.848 | −6.62 to −1.52; Pr(Δ < 0) = 0.999 | Independent implementation |
| | Independent Pólya–Gamma Gibbs; inverse-gamma priors | −4.30; OR 0.838 | −7.40 to −1.17; Pr(Δ < 0) = 0.996 | Independent implementation |
| **Nuisance / design sensitivity** | Order, history, gap, and combined adjustments | −4.44 to −3.90 | Range across six specifications; all intervals below zero | All remained negative |
| | Calendar adjustment (linear; spline) | −4.07; −4.14 | −6.72 to −1.45; −6.76 to −1.63 | All remained negative |
| | Five-trial blocks, seat phase 1 only (first set of a game, before the within-game seat exchange) | −6.0, −5.6, −4.4, −3.3 | Source-level, equally weighted; seat phase 2, −4.8, −2.9, −3.7, −2.2 pp | Early difference present in the first set of a game, before the within-game seat exchange |

**Supplementary Table 3 (continued).**

| Evidence role | Analysis / contrast | Estimate (pp) | Interval / test | Interpretation |
|---|---|---|---|---|
| **Physiological state** | HMC; glucose complete case, unadjusted (394 sets) | −4.42 | −7.32 to −1.72; Pr(Δ < 0) = 0.999 | Sets with both participants' values; all after the approximately 20 h fast |
| | HMC; adjusted for north- and south-seat glucose (394 sets) | −4.68 | −7.55 to −1.80; Pr(Δ < 0) = 0.999 | Transverse × north-seat participant's glucose, GLM $P$ = 0.88 |
| | HMC; sets with a participant outside ±15% of his mean excluded (12) | −3.68 | −6.36 to −1.01; Pr(Δ < 0) = 0.996 | Retrospective criterion (mean of all recorded values); still negative |
| **Heterogeneity** | Transverse × cohort | τ = 0.07 (log-odds) | LRT $P$ = 0.62; wild-cluster $P$ = 0.55 | No detected heterogeneity |
| | Transverse × test period | — | Wald $P$ = 0.69; LRT $P$ = 0.74 | Early period alone, −3.14 pp (−6.91 to +0.72) vs the seven non-transverse configurations observed (mNW absent) |
| | Transverse × experimenter | τ = 0.08 (log-odds) | Wild-cluster $P$ = 0.23 and 0.35; LRT $P$ = 0.52 | No detected heterogeneity |
| | Each experimenter or the unrecorded category omitted in turn | −5.02 to −2.95 | Largest experimenter omitted, −6.18 to +0.13 | Reduced sample size; still negative |
| **Outcome history** | Running score, per standard deviation (P1; P2) | OR 0.99; OR 0.99 | 0.95–1.04; 0.94–1.04 | No detected outcome-driven change |
| | Running score × transverse (P1; P2) | OR 1.02; OR 1.01 | 0.92–1.14; 0.90–1.12 | No detected condition dependence |
| | Own preceding action × transverse (P1; P2) | OR 0.95; OR 1.07 | 0.77–1.15; 0.88–1.32 | No detected condition dependence |
| | Partner's preceding action × transverse (P1; P2) | OR 1.10; OR 1.02 | 0.90–1.35; 0.82–1.26 | No detected condition dependence |

**Supplementary Table 3 (continued).**

| Evidence role | Analysis / contrast | Estimate (pp) | Interval / test | Interpretation |
|---|---|---|---|---|
| **Scheduling sensitivity** | Label reallocation | Observed −4.07 | Two-sided *P* = 0.0021 unrestricted; 0.0019 within cohort × seat phase; 0.0036 within north participant | Sensitivity *P* values; not exact randomization inference |
| **Full-family robustness** | 13-condition crossed HMC; all 600 set-level rows | −4.01 | −6.57 to −1.47; Pr(Δ < 0) > 0.999 | Direct person-level refit; 47 participants and 209 pair-sessions; all 600 rows linked to trial sheets, 599 exactly (Supplementary Methods 3) |
| **Record provenance** | Recovered-record sets excluded (31) | −4.48 | −7.28 to −1.73 | Direct refit |
| | Label-reconciled sets excluded (21) | −4.22 | −6.80 to −1.54 | Direct refit |
| | Both excluded | −4.75 | −7.55 to −1.99 | Direct refit |

Analyses fully reported in the Results, Table 3, Fig. 3, or Supplementary Fig. 2 are not repeated. Where an analysis plotted in Fig. 2b or Supplementary Fig. 1 has a row here, the row gives its numerical interval. HMC intervals are posterior intervals, and other intervals are confidence or resampling intervals. *P* values are unadjusted unless a Holm-adjusted value is given. HMC, Hamiltonian Monte Carlo; pp, percentage points; GLM, generalized linear model; CI, confidence interval; OR, odds ratio; Δ, transverse − non-transverse difference; τ, between-group standard deviation of the transverse log-odds effect; LRT, likelihood-ratio test; P1 and P2, players 1 and 2.

**Supplementary Table 4. Continuous physical descriptors and DEI.**

| Condition | hOrthConflict | DEI | 1 − \|cos φ\| |
|---|---|---|---|
| Con | 0.00 | 0.951 | 0.011 |
| Horizontal-only mN | 0.00 | 0.947 | 0.008 |
| Vertical-component inversion | 0.00 | 0.951 | 0.010 |
| mS | 0.02 | 0.929 | 0.000 |
| mNE | 0.16 | 0.795 | 0.067 |
| mNW | 0.19 | 0.774 | 0.326 |
| mE | 0.79 | 0.200 | 1.000 |
| mW | 0.79 | 0.198 | 0.997 |
| Vertical-only down | Not assigned | 0.950 | 0.000 |
| Vertical-only up | Not assigned | 0.949 | 0.000 |
| Near-zero | Undefined | 0.000 | — |

hOrthConflict is $1 - |\cos\theta|$, where $\theta$ is the angle between the condition's north-seat horizontal unit vector and the ambient-control horizontal vector. It is also defined for Con and vertical-component inversion but is not assigned when the horizontal component is nominally removed. DEI was defined after inspection of the condition pattern (Supplementary Note 3; Supplementary Methods 4). $1 - |\cos\varphi|$ is the mid-plane descriptor, where $\varphi$ is the angle between the measured field vector and the geographic north–vertical plane, which contains the seat axis. It is defined for every non-zero field. All values are computed from the measured north-seat vectors in Table 1. DEI, directional extractability index.

**Supplementary Table 5. Detailed radiofrequency results.**

| Condition | *n* | Rate (%) | Δ vs reference (pp) | Statistical note |
|---|---|---|---|---|
| Con | 50 | 50.70 | Reference | — |
| RF1 (1.260 MHz) | 50 | 46.60 | −4.10 | Bootstrap 95% CI −8.2 to −0.1; Welch *P* = 0.051 |
| RF2 (1.890 MHz) | 46 | 51.96 | +1.26 | Bootstrap 95% CI −3.4 to +5.8; Welch *P* = 0.60 |
| RF1 vs RF2, direct contrast | 50 vs 46 | — | −5.36 | Bootstrap 95% CI −10.0 to −0.7; Welch *P* = 0.027 |
| RF1, horizontal-only mN background, vertical antenna | 46 | 50.43 | −0.82 | Bootstrap 95% CI −5.6 to +4.0; Welch *P* = 0.74 |
| RF1, horizontal-only mN background, 37°-tilted antenna | 30 | 51.67 | +0.42 | Bootstrap 95% CI −4.7 to +5.5; Welch *P* = 0.87 |
| RF1, vertical-only down background, vertical antenna | 46 | 51.63 | +0.38 | Bootstrap 95% CI −4.8 to +5.5; Welch *P* = 0.89 |
| RF1, vertical-only down background, 37°-tilted antenna | 28 | 52.86 | +1.61 | Bootstrap 95% CI −4.4 to +7.5; Welch *P* = 0.60 |
| RF1, modified backgrounds averaged (mN and vertical-only down weighted equally; antenna positions pooled) | 150 vs 92 | — | +0.26 | Bootstrap 95% CI −2.9 to +3.4; Welch *P* = 0.87 |
| RF1 over Con minus RF1 averaged over modified backgrounds | — | — | −4.36 | Bootstrap 95% CI −9.5 to +0.8; Welch *P* = 0.10 |

Rates for Con, RF1, and RF2 are those reported in the preprint.[2] The four modified-background rows are from the additional RF-background series (Methods). All contrasts use two-sided Welch tests and percentile bootstraps (Supplementary Methods 4). Near the 0.05 boundary, Welch *P* values and percentile bootstrap intervals can disagree, as for RF1 versus Con. The source workbook reported one-sided *P* values for Con versus RF1 and RF1 versus RF2. The four RF1 conditions over modified backgrounds use mN (51.25%; 48 sets) and vertical-only down (51.25%; 44 sets) as RF-off references, and their 95% CIs ($10^6$ resamples drawn within condition) all include zero. The last two rows average the two modified backgrounds and compare that average with the RF1 effect over Con (Welch-type *t* statistic with Satterthwaite degrees of freedom; $10^6$ resamples). For this difference, a quasi-binomial model of the set counts gave $P = 0.12$ for the log-odds interaction, a participant-level crossed random-intercept model gave −4.46 pp (95% posterior interval, −9.77 to +0.84), cluster-robust binomial models gave −4.39 pp ($P = 0.067$ with pair-session clusters and 0.15 with north-participant clusters), and the within-north-participant fixed-effect model gave −3.43 pp (CR2 $P = 0.25$; Supplementary Methods 4). *n*, number of 20-trial north-seat sets; Rate, black stone rate; Δ, difference from the reference; Con, ambient control; RF, radiofrequency; pp, percentage points; CI, confidence interval.

**Supplementary Table 6. Detailed light-spectrum results.**

| Condition / contrast | *n* | Rate (%) | Difference (pp) | Note |
|---|---|---|---|---|
| No goggles | 48 | 51.25 | — | Same sets as mN; component contrast control |
| Non-filtered goggles | 48 | 49.90 | Control | Total-rate contrast control |
| >400 nm | 46 | 50.87 | +1.0 | Not significant (Welch $P$ = 0.67) |
| >500 nm | 46 | 55.33 | +5.4 | Welch $P$ = 0.021; bootstrap 95% CI +0.9 to +9.9; source $q_0$ > 0.975 |
| >500 nm vs mN, odd | — | — | −0.4 | Bootstrap 95% CI −7.9 to +7.0; not significant |
| >500 nm vs mN, even | — | — | +7.8 | Bootstrap 95% CI +1.6 to +14.1; $q_0$ > 0.975 |

Rates are those reported in the preprint.[2] These conditions lie outside the primary inferential family (Methods). Total-rate contrasts use the non-filtered-goggles condition as the control (for example, +5.4 = 55.33 − 49.90). The no-goggles condition comprises the same sets as mN. The odd and even component contrasts use mN as the control and are recomputed from the analyzed sets (mN even component, 20.49% rather than the preprint value of 20.83%), with percentile bootstrap 95% confidence intervals (CIs; $10^6$ set resamples within condition) as in Supplementary Fig. 4b. *n*, number of 20-trial north-seat sets; Rate, black stone rate; —, not applicable or not supplied; mN, horizontal-only north-seat field near magnetic north; pp, percentage points; $q_0$, source-reported zero-percentile statistic under the original contrast coding ($q_0$ > 0.975 indicates a two-sided difference in that convention but is not a conventional two-sided $P$ value).

**Supplementary Table 7. Sensitivity of the DGI for the transverse pair.**

| Condition | ROI / grid | DGI ($\times 10^{-3}$ m$^{-1}$) | Maximum angle ($\times 10^{-3}$ °) | Note |
|---|---|---|---|---|
| mE | ±30 mm; 5 mm | 1.83 | 4.6 | ROI sensitivity |
| mW | ±30 mm; 5 mm | 1.36 | 3.5 | ROI sensitivity |
| mE | ±80 mm; 5 mm | 5.38 | 39.2 | ROI sensitivity |
| mW | ±80 mm; 5 mm | 4.07 | 31.1 | ROI sensitivity |
| mE | ±50 mm; 2.5 mm | 3.27 | 13.5 | Grid sensitivity; 2.8% above the unrounded 5 mm value |
| mW | ±50 mm; 2.5 mm | 2.44 | 10.4 | Grid sensitivity; 2.8% above the unrounded 5 mm value |

DGI and the maximum angular deviation of field direction from the center vector are the mean and maximum over a cubic ROI centered on the north-seat participant's glabella, from an idealized Biot–Savart model of the coil geometry. Nominal values (±50 mm, 5 mm grid) for all ten configurations are shown in Supplementary Fig. 5, and the pair-separation sensitivity is given in Supplementary Methods 5. Across the ten non-zero configurations, the largest maximum angle within ±80 mm was that of mE. ROI, region of interest; DGI, directional gradient index.

**Supplementary Note 1. Full enumeration of robustness, dependence, and allocation analyses.**

This note details the specifications summarized in the main text. Their estimates are shown in Fig. 2b, Table 3, Supplementary Fig. 1, and Supplementary Table 3.

**1. Full-data structural specifications.** The primary Hamiltonian Monte Carlo (HMC) model was refitted under three alternative priors, with longer chains and multiple seeds, and was reimplemented with the alternative likelihoods, samplers, and dependence structures listed in Methods and Supplementary Methods 4. The 53 hierarchical-model estimates summarized in the Results comprised alternative estimands and standardizations (6), prior specifications (4), additional seeds and the 600-set model (10), nuisance adjustments (6), period, calendar, exclusion, covariate, experimenter-omission, and record-provenance analyses (24), and Bradley–Terry and Gibbs implementations (3). The two intervals that included zero are described in section 4. The cohort-13-omitted estimate is counted from the longer chains, because its upper limit lay within Monte Carlo error of zero at the default length. These estimates are not independent: they derive from 44 distinct posterior samples, several estimands were calculated from the same model fits, and three entries duplicate estimates reported under other headings. The model with the lowest effective sample size included adjustments for order, history, inter-session gap, and calendar day (maximum R-hat, 1.035; minimum effective sample size, 129). Refitted with longer chains (four chains of 2,000 warm-up and 5,000 retained draws), it gave a transverse contrast of −4.44 percentage points (95% posterior interval, −7.06 to −1.79; maximum R-hat, 1.001).

**2. Participant-aggregated sensitivity.** Participant-aggregated analyses remained negative but were less precise (Table 3), reflecting differences in weighting, participant coverage, and retained set-level information rather than a change in the sign of the estimate.

**3. Dependence structure of the within-north-participant contrast.** The within-north-participant configuration contrast retains set-level information while using each participant as his own control. In addition to the bias-reduced (CR2) interval reported in the Results, the conventional estimator (CR1) applied within north participants yielded a 95% confidence interval of −6.88 to −1.33 ($P$ = 0.005), and dyadic-overlap CR1, in which reuse of either seat participant induces dependence, yielded −6.64 to −1.57 ($P$ = 0.002). These are alternative covariance estimates for the same fitted contrast. The point estimate is therefore unchanged by construction, and their similar intervals indicate that the precision of the family contrast is insensitive to the assumed dependence structure. Within the 29 pair-sessions that contained both transverse and non-transverse sets, the difference was −5.95 percentage points (bootstrap 95% confidence interval, −9.83 to −2.16; sign-flip $P$ = 0.005).

**4. Cohort, test-period, and experimenter sensitivity.** Source-level comparisons of mE and mW, each with mN as in the preprint, are provided with and without cohort 13 in Supplementary Table 3.

When cohort 13 was omitted from the primary model (Results), the upper interval limit lay within Monte Carlo error of zero at the default chain length (−0.20 to +0.13 across 20 seeds), and with longer chains it ranged from −0.06 to −0.01 across three seeds ($\Pr(\Delta < 0) = 0.975$–$0.977$). The two intervals that included zero, for the early test period alone (compared with the seven non-transverse configurations observed in that period, because mNW was absent) and with the largest experimenter omitted, occurred in analyses with fewer sets.

**5. Allocation and scheduling sensitivity analyses.** Condition labels were reallocated without restriction, within cohort × seat phase, and within north participant (Methods; Supplementary Table 3). Selection-adjusted versions, which refer the transverse contrast to the most extreme of all pair contrasts, yielded two-sided $P = 0.015$, 0.014, and 0.029 among the five symmetric pairs and 0.096, 0.096, and 0.17 among all 45 pairs for these three schemes, respectively.

**6. Singular Laplace fits.** The R mixed models (Methods; Supplementary Methods 4) were singular. In the total-rate model, the pair-session standard deviation was estimated at zero, and the north-role and south-role standard deviations were 0.08 and 0.06 on the logit scale, respectively. In each of the four role-specific primitive-process models, one or two of the three random-intercept standard deviations were estimated at or near zero. A singular fit indicates that these variance components could not be distinguished from zero with the present data, but it does not by itself invalidate the fixed-effect contrast. The total-rate contrast agreed with the primary HMC fit, whose half-normal priors regularize the standard deviations without fixing them at the boundary, and with the fixed-effect cluster-robust analyses (Table 3). Because intervals from singular fits may underestimate uncertainty in the dependence structure, the primitive-process mixed-model $P$ values were treated as sensitivity results. The estimated standard deviations are provided with the analysis code (Code availability).

**Supplementary Note 2. Process decomposition and sequence analyses.**

This note details the measurements underlying the process analyses summarized in the Results.

**1. What the joint outcomes do and do not measure.** The fixed-denominator 'black-in-odd' measure represents the joint odd–two outcome, in which P1 generated an odd count and P2 placed two stones, rather than P1 odd generation itself (Fig. 3b). Because a change in either primitive can redistribute several joint outcomes, changes in joint outcomes alone cannot identify the responsible primitive. These outcomes are based on six north-P1 opportunities within the first 12 trials and therefore do not add directly to the total rate across all 20 trials and both roles. Across all 20 trials, the transverse − non-transverse difference in the north-seat black stone rate was −3.08 percentage points when the north-seat participant was P1 and −5.05 percentage points when he was P2 (first 12 trials, −2.48 and

−7.29 percentage points). Because odd parity and one-stone choices each occurred in about half of the trials, changes in their frequencies contributed less than 0.3 percentage points to either 20-trial difference, which therefore arose almost entirely from a change in how often parity and choice coincided within trials. Handful sizes did not differ between transverse and non-transverse configurations (mean, 15.8 and 15.7 stones). The source of this change is not identified by these data and may include sequential patterns not captured by the one-step history models or cues not removed by the opaque screen and auditory masking.

**2. Sequence models.** Parity was revealed only after P2 chose (Fig. 1b), so it could inform only later trials. Because roles alternated, the cross-lag models related successive actions of the same participant and are therefore within-person sequence sensitivity analyses rather than tests of directional coupling between participants or of responses to previous stone outcomes (Supplementary Fig. 2b). In the outcome-history models (Supplementary Methods 4; Fig. 3d), which also included each participant's own and the partner's preceding actions, the running score showed no detectable effect on subsequent choices, and none of these dependences differed between transverse and non-transverse configurations (Supplementary Table 3). A small dependence of P2 one-stone choice on the partner's preceding choice (odds ratio, 0.87) did not differ between configurations (interaction odds ratio, 1.02) and was not robust to clustering by participant ($P = 0.045$ with set clustering and 0.23 with participant clustering).

**3. Near-zero comparison.** Under near-zero, the black-in-even component did not differ from that of the non-transverse family (−0.05 percentage points; bootstrap 95% confidence interval, −4.85 to +4.76; Supplementary Fig. 3a). The set-level proportion of south-seat P2 one-stone choices was not higher under near-zero (+0.30 percentage points; −4.78 to +5.33), although it did not differ reliably from that under the transverse pair (−3.70 percentage points; −9.40 to +1.90).

## Supplementary Note 3. Continuous descriptors, predictive comparison, and light-spectrum experiment.

These analyses examine whether a continuous descriptor or index summarizes the full condition family, whether north-seat geometry predicts condition means better than south-seat intensity, and whether the observed pattern varies with light-spectrum conditions. They lie outside the primary inferential family and do not provide independent confirmation of the primary contrast. Section 1 provides numerical details for the continuous descriptors summarized in the Results.

**1. Continuous descriptors.** Leaving hOrthConflict unassigned for the nominal vertical-only conditions (Supplementary Methods 4) is an analytical choice rather than a necessity, because their small residual horizontal components define a direction, albeit an unstable one. For near-zero, $\theta$ is undefined. Regression on the six horizontal condition means yielded a slope of −5.53 percentage points per unit ($R^2 = 0.76$). The mid-plane descriptor $1 - |\cos \varphi|$ is referenced to the geographic north–vertical plane rather than to the ambient horizontal vector, so the two descriptors are not identical even for horizontal fields. However, the mid-plane descriptor is defined for every non-zero vector. For hOrthConflict, a Gaussian ordinary least-squares model with an intercept and hOrthConflict slope was compared with a six-level categorical ordinary least-squares model containing an intercept and five condition indicators. The Akaike information criterion (AIC)[3] was calculated from the maximized Gaussian likelihood on the identical 270 set-level rows, and the resulting difference was AIC(categorical) − AIC(hOrthConflict) = +5.7. Because hOrthConflict takes a single value per condition, this is a descriptive condition-structured comparison rather than participant-level inference. The mid-plane descriptor was compared with categorical models on the same 270 set-level rows from six horizontal conditions and on 454 rows from ten non-zero conditions, yielding AIC(categorical) − AIC(descriptor) = +6.0 and +13.6, respectively.

**2. A single continuous ordering index.** Because the directional extractability index (DEI; Supplementary Methods 4) was defined after inspection of the pattern, its association with the 11 condition means (Results; Fig. 4c) carries no confirmatory *P* value. The index also inherits both confounds that limit the main-text descriptors. Near-zero occupies an extreme of the index and was collected in a distinct order, and the nominal geographic, participant, visual, and body axes were never independently rotated, so the index cannot identify the operative reference frame. DEI is therefore retained only as a descriptive heuristic and hypothesis generator rather than as an independent inferential test, mutual-information estimate, or first-principles receptor model (Supplementary Table 4). Here, 'extractability' refers to this descriptive ordering and does not denote measured biological decodability or information transfer between participants.

**3. Leave-one-profile-out predictive comparison.** This analysis tested whether a north-seat descriptor predicts condition means better than a south-seat descriptor. The comparison used 11 physical profiles: the ten non-zero conditions and near-zero. The north-transverse descriptor, defined as the squared fraction of the full field vector lying along the horizontal axis perpendicular to mN (Supplementary Methods 4), yielded a lower leave-one-profile-out root-mean-square error (RMSE, 1.79 percentage points) than a linear south-intensity descriptor (1.98). Quadratic, post hoc 40 µT threshold, and nested threshold forms of the south descriptor did not improve on the north-transverse descriptor (RMSE, 1.84–2.16), and pair-bootstrap differences between forms were highly uncertain (Table 3). Because the number of profiles is small and the functional forms were selected post hoc, these results show neither that the north descriptor is superior nor that south exposure is absent.

**4. Dependence on the visible spectrum.** A retinal light-dependent route predicts that removing the wavelengths that drive the putative receptor should alter the response. The >500 nm joint outcomes differed in pattern from those of the transverse configurations, increasing the north-seat even-parity component relative to mN rather than decreasing the odd-parity component (Supplementary Fig. 4b). Because the filter altered the light input of the south-seat participant only, whereas the transverse configurations altered the field at both seats, this difference does not argue against a light-dependent route. Illuminance also decreased with filtering (Methods). Separately varying wavelength and photon flux or irradiance would distinguish spectral dependence from a change in light intensity, whereas matching illuminance alone would not equate photon flux across spectra. Similarly, avian magnetic compass studies compare wavelengths at matched photon flux or irradiance rather than matched illuminance and treat light intensity as a separate variable.[4,5] The >400 nm filter reduced illuminance by 17% without a comparable change in rate, but the larger reduction under >500 nm filtering was not separately controlled. The light-spectrum records did not undergo the full source-sheet, order, and hierarchical checks applied to the ten-configuration family. The comparison therefore identifies targeted tests rather than providing confirmatory evidence (Supplementary Fig. 4; Supplementary Table 6).

**Supplementary Note 4. Radiofrequency exposure geometry and amplitude dependence in the avian magnetic compass.**

This note details the radiofrequency exposure geometry and the avian literature on radiofrequency amplitude dependence referred to in the Discussion.

**1. Exposure geometry.** For an ideal vertical radiofrequency field, the fraction of amplitude perpendicular to the static background is maximal over a horizontal-only background, approximately 0.70 over Con (measured inclination, 45.9°), and near zero over a vertical-only background. For the antenna tilted 37° toward the participant's anterior direction, and assuming that the radiofrequency vector follows the antenna axis, the fractions are approximately 0.60 over an exactly vertical background (0.61 for the measured vertical-only down vector) and 0.80 over mN. The tilted axis lies approximately 11° (in three dimensions) from the measured Con vector (perpendicular fraction, 0.18), and this combination was not tested (Fig. 4f). Only the vertical antenna over the vertical-only background left the oscillating field nearly parallel to the static field (perpendicular fraction, 0.01), the alignment under which robins retained migratory orientation.[6,7] Avian tests compared parallel alignment with alignments of 24° and 48° to an inclined geomagnetic field,[6–8] and we are not aware of tests over single-component backgrounds. The present point means, a reduction over Con but little change over mN where the perpendicular fraction is maximal, run counter to a simple monotonic perpendicular-component account without establishing a departure from it. In the earlier orientation study, responses to 1.260 MHz RF differed among antenna angles that also changed the amplitude at the eyes.[9]

**2. Avian amplitude dependence.** Reported avian effects include compass disruption by Larmor-frequency fields of approximately 15 nT at 24° to the static field but not at 5 nT,[8] and only slightly increased scatter under a parallel 485 nT field, which disoriented the birds when applied at 24°.[7] In a trained four-arm assay, approximately 10 nT exposure disrupted orientation in zebra finches trained without it but not in birds trained with it, and stronger narrow-band exposures altered the response pattern.[10] Another study found no significant disruption under 48 and 400 nT narrow-band exposures, whereas weak broadband noise disrupted orientation.[11] Because species, assay, training, field geometry, and spectral composition differ, these results do not define a single dose–response curve. The nominal 100 nT (root-mean-square; Methods) used here lies within the avian range, but this comparison defines neither a human response threshold nor the effective dose at a human receptor. Smaller modified-background contrasts therefore cannot be attributed to, or used to exclude, insufficient effective exposure on this basis, and amplitude and angular dependence should be tested separately under controlled conditions.

### Supplementary Note 5. Electromagnetic induction and reference-frame considerations.

This note details the induction and reference-frame considerations summarized in the Discussion. Because the posture model on which several statements rest is observation-based (Methods), these considerations identify targeted tests rather than providing confirmatory evidence.

**1. Posture-model outputs.** The following outputs were derived from the observation-based posture and induction models (Methods; Supplementary Methods 6). At the north seat, the transverse fields yielded the largest eye-fixed interaural component in every decision window (0.69–0.93, versus at most 0.72 for any other configuration) and over the whole trial (0.97, versus 0.54 for mNW and at most 0.16 for the others), and the head-fixed axis showed the same ordering. In contrast, induction before choices under the transverse fields lay within the range of the other configurations and above that under the vertical-only fields. These relations held for shorter hair-cell latencies (0.01–0.1 s), for head contributions of 60% and 100%, and when each canal normal was rotated randomly with a 6° standard deviation (20 draws) or replaced by the mean normals of Bradshaw et al.,[12] but depended on the observed mixture of cue-detection modes. Had every participant detected the same, smaller fraction of cues peripherally, other configurations would have matched or exceeded the transverse fields in some decision windows, although not over the whole trial. This occurred for the eye-fixed axis in every simulation at fractions of 0.25 or less and in some at 0.3–0.4, and for the head-fixed axis in every simulation at 0.1 or less and in some at 0.15–0.2 (five simulations of 200 participants). Had every cue been detected peripherally, induction under the transverse fields would have come within approximately 10% of that under the vertical-only fields and fallen slightly below it in some windows. At the south seat, neither measure separated the transverse pair from the other configurations.

**2. Electromagnetic induction.** In an idealized canal loop, rotation about an axis parallel to the field produces no change in flux. The observation-based posture model (Supplementary Methods 6) places most step-2 choices during head turns and tilts toward the stone containers. The fastest head movements accompanying choices (Methods) were turns from the experimenter toward the stone container after a cue detected by turning the head. At 45 µT, the induced electric field $E$ would reach the 5 nV cm$^{-1}$ detection limit of elasmobranch electroreceptors[13] only for a canal radius $r$ above approximately 0.56 cm even at the fastest of these movements, approximately 1.6–2 times the mean canal radii of 2.7–3.4 mm in humans,[12] which are themselves larger than those of pigeons (2.1–3.2 mm).[14] For most movements, the required radius was three or more times that of humans (approximately 1.1 cm at 120° s$^{-1}$). Turns toward the experimenter, which preceded the cue, could reach approximately 250° s$^{-1}$ if completed within 0.4 s and would require $r$ above approximately 5 mm.

**3. Head-fixed and retinal reference frames.** As described in section 1, the ordering that placed the transverse fields closest to the interaural axis held just before choices under the observed mixture of cue-detection modes but could fail in some decision windows if peripheral cue detection were uniformly rare. For a retinal substrate, the anatomical orientation of photoreceptor outer segments does not by itself specify the magnetic axes of the candidate molecules or the weighting of signals from different retinal regions. A reproducible left–right versus up–down difference under matched conditions would indicate a departure from symmetry around the visual axis, potentially through regional retinal weighting, molecular alignment, or downstream processing. In human vision, threshold changes occurred when the magnetic field was aligned with the line of sight.[15,16] By contrast, in the present study, fields nominally along the facing axis (mN and mS), which were also oblique to the downward gaze toward the board, did not lower the rate, possibly because different tasks weight differently oriented receptors or retinal regions.

**4. Relation to free-rotation orientation assays.** In the earlier orientation assays, participants rotated freely on a chair, over 360° or within 240°, while searching for magnetic north.[9,17] In contrast, in the stone selection task, each participant faced a nominally fixed direction relative to the field throughout a set, although head posture varied within trials (Supplementary Methods 6).

**Supplementary Methods 1. Glucose records and glucose-adjusted sensitivity analyses.**

This section details the glucose records summarized in Methods ('Pre-experiment preparation and physiological state'). A participant's initial value was compared with the mean for men after the approximately 20 h fast in the earlier study (5.1 mmol $l^{-1}$).[1] A deviation of more than 15% prompted assessment of relevant circumstances and reinforcement of the fasting requirement. At the second measurement, after full fasting had been confirmed, the mean of the first two values became that participant's reference mean. Thereafter, testing was postponed if either participant's value differed by more than 15% from his reference mean, unless a pre-test questionnaire and an overall review supported proceeding. Values for both participants were available for 394 of the 454 ten-configuration sets, all after the approximately 20 h fast. The remaining 60 sets lacked a usable value for at least one participant. Glucose was similar across the ten configurations (descriptive Kruskal–Wallis $P = 0.13$, based on the value of each set's north-seat participant because seats were exchanged within each game; repeated values of a participant were treated as independent). The glucose complete-case, glucose-adjusted, and exclusion estimates are given in Supplementary Table 3.

**Supplementary Methods 2. Radiofrequency resonance calculations and amplitude profile.**

This section details the radiofrequency calculations and measurements summarized in Methods ('Radiofrequency and light-spectrum manipulations'). Ion cyclotron resonance has been proposed as a mechanism by which weak low-frequency magnetic fields could act on cells.[18] At the 45 µT static field, however, the cyclotron frequencies of physiological ions, $f = zeB/(2\pi m)$, where $z$ is the charge number, $e$ the elementary charge, and $m$ the ion mass, are 17.7 Hz for $K^+$, 30.1 Hz for $Na^+$, and 34.5 Hz for $Ca^{2+}$, and the $^1H$ nuclear Larmor frequency is 1.92 kHz. These values are approximately three to five orders of magnitude below the applied radiofrequencies (calculated with CODATA 2022 constants[19] and standard atomic weights[20]). The spectral width at the base of each original RF peak was approximately 0.02 MHz.[1,9] For an electron with $g \approx 2.0023$, the Larmor frequency at 45 µT is approximately 1.261 MHz,[8] and 1.890 MHz corresponds to approximately 67 µT.

Along the representative path used to measure the posture dependence of the radiofrequency amplitude (Supplementary Fig. 3b), the stone containers lay on the west side, opposite the experimenter. With the forward-facing amplitude normalized to 100, the value decreased to 43.8 at the deepest container-directed posture, and the equally weighted mean across sampled postures was 71.8. The sum of sampled amplitudes on the untilted limb was approximately 60% of the total sum over the path, which is a descriptive allocation rather than a biologically calibrated exposure dose. Between the two antenna positions, 12 of the 13 sampled postures differed by 0.4 percentage points or less (values read from the plot in the measurement report to approximately ±0.1). The exception was the return 0° posture (vertical, 54.3; tilted, 57.3), which the measurement report attributed to variability in posture reproduction between runs. Amplitudes were sampled at 10° yaw intervals under the ambient background with the apparatus configuration used throughout the series. Similar normalized amplitudes at the glabella do not establish identical RF vectors or spatial amplitude distributions over the head.

**Supplementary Methods 3. Record matching and label reconciliation.**

This section details the record matching and label reconciliation summarized in Methods ('Endpoint, estimand, and primary hierarchical model'). Thirty-one of the 454 ten-configuration sets were read from original record sheets recovered outside the indexed sheet lists of the transcribed archive (recovered-record sets). After schedule changes, 21 exported records required condition-label reconciliation against the original sheets and the confirmed condition map (label-reconciled sets). The 600 analyzed sets were linked to original trial sheets: 599 matched on test date, slot, and north-seat participant, and one matched on date and participant despite a differing slot label. The 21 label-reconciled records resolved to vertical-only up (10), mW (4), mE (3), mN (2), and mS (2), all within the ten-configuration family. The identical rates under mE, mW, and near-zero (467 of 1,000 trials each) came from non-overlapping sets, and no 20-trial sequence was shared between any two of the 740 analyzed sets. The confirmed condition map had been compiled from the schedule and original sheets and checked against the previously published set-level rates.[1,2] Accordingly, the analyzed Con, near-zero, and RF1 sets reproduce the published set-level rates. The published trial-level files (Fig. 3A of refs. 1 and 2; Fig. 5A of ref. 1), however, differ from the analyzed data in six Con, three near-zero, and seven RF1 sequences. Eleven of the published sequences come from sets recorded under the other condition of mixed-condition games (Methods), including RF sessions that alternated the two frequencies, and the remaining five published sequences match no surviving original sheet. In the analyzed data, each sequence is that of the linked original trial sheet. With the analyzed sequences, the odd-case step-1 differences from Con (rate defined in the legend of Supplementary Fig. 3a) were −8.3 percentage points for near-zero and −9.7 percentage points for RF1 (bootstrap 95% CIs, −14.3 to −2.3 and −15.7 to −3.3), compared with −9.0 percentage points for both in the published files. Duplicate cohort-and-game archive labels for 24 sets were consolidated by underlying set identity. Two trials outside the ten-configuration family recorded concealed counts below the instructed minimum of 10 (0 and 1). These trials were retained as recorded, because only parity enters the analyses. Each distinct set contributed at most once to a given analysis, whereas multiple distinct sets within a pair-session were retained.

For the four added RF1 conditions, the 150 supplied trial sheets and the accompanying schedule and antenna-position records allowed reproduction of the reported sample sizes and mean outcomes, and the recovery summary also reported agreement with an independent per-set archive. In the source schedule, these four conditions, all delivered at 1.260 MHz, were coded IRF2 and HRF2, which are distinct from RF2 at 1.890 MHz.

In the light-spectrum experiment, several games with different goggle conditions shared one scheduled time slot. The goggle condition of each light-spectrum set was therefore reconstructed by elimination against the schedule and the per-condition rate distributions reported in the preprint[2] and, for >500 nm, also against its component values. Some sets were candidates for more than one goggle condition, so the reconstruction is one of several consistent with these constraints. All of these give the light-spectrum means reported here and the same per-condition rate distributions, and therefore the same s.e.m. values (Supplementary Fig. 4a). The bootstrap intervals for the component contrasts (Supplementary Fig. 4b; Supplementary Table 6) are based on the adopted assignment and may differ slightly under the other solutions.

**Supplementary Methods 4. Statistical analysis details.**

**1. Primary model and software.** This section details the specifications summarized in Methods ('Endpoint, estimand, and primary hierarchical model' to 'Physical descriptors and condition ordering'). Set-weighted means and the seven non-transverse configurations other than mN provided sensitivity versions of the primary estimand. Sampling used four chains of the no-U-turn sampler (NUTS) in NumPyro 0.22.0, with 500 warm-up and 500 retained draws per chain, target_accept=0.95, and seed 20260720. Longer chains (2,000 warm-up and 5,000 retained draws) and three further seeds gave contrasts of −4.10 to −4.08 percentage points. Differences of up to 0.05 percentage points between runs of the same specification reflect Monte Carlo error. The cohort-13-omitted refit, whose interval limit lay near zero, is reported from the longer chains. Standardized predictions retained the random effects, whereas a typical-participant version set them to zero. An independent implementation in R 4.3.3 (R Foundation for Statistical Computing, Vienna, Austria) with lme4 1.1-35.1 (Laplace approximation; parametric bootstrap with 1,000 refits) shared no code with the primary analysis, and repeated runs of the analysis code with fixed random seeds reproduced the reported estimates. Linear and generalized linear models were fitted with statsmodels 0.15.0 in Python, which also supplied the cluster-robust variances of the generalized linear models. The other cluster-robust and dyadic-overlap variances, wild-cluster bootstraps, permutation tests, and the Pólya–Gamma Gibbs sampler (with Pólya–Gamma draws from the polyagamma 2.0.2 package) used custom code (Code availability).

**2. Structural robustness.** Robustness analyses were grouped by type of computation. In addition to likelihood, prior, Gibbs, wild-cluster, nuisance, and influence checks from an earlier stage of the analysis, the present analysis included four further sets of analyses. First, it estimated the within-north-participant configuration contrast with north-participant cluster-robust (CR1), bias-reduced cluster-robust (CR2)–Satterthwaite, and dyadic-overlap CR1 inference (section 3). Second, it performed leave-one-participant, leave-one-pair, leave-one-pair-session, cohort-omission, experimenter-omission, and calendar analyses. Third, it compared mE/mW with the vertical-only down condition, whose south-seat vectors closely matched theirs, in the within-north-participant model with dyadic-overlap inference. Fourth, it evaluated restricted physical-profile models by leave-one-out prediction with pair bootstrap. The likelihood and Gibbs checks from the earlier stage were reimplemented in the ten-configuration frame. A Bradley–Terry hierarchy gave each participant a single propensity that entered positively in the north role and negatively in the south role, together with pair-session intercepts. A beta-binomial check reweighted the posterior of the primary fit under uniform priors on the intra-set correlation (upper bounds 1, 0.2, and 0.1) and also gave Bayes factors for the beta-binomial relative to the binomial likelihood ($BF_{\beta/\mathrm{bin}}$). A Pólya–Gamma Gibbs sampler was run under the primary or inverse-gamma(1, 1) variance priors. Heterogeneity of the transverse contrast across cohorts, test periods, and experimenters was assessed by wild-cluster bootstrap, likelihood-ratio tests in the mixed model, and partial pooling. Transverse and non-transverse sets were also compared within the 29 pair-sessions that contained both (mean within-pair-session difference; pair-session percentile bootstrap with 10,000 resamples and two-sided sign-flip test with 200,000 draws). The pair-session wild-cluster bootstrap used 200,000 Rademacher-weighted replicates, with a restricted-null, non-studentized two-sided *P* value and an unrestricted percentile interval.

**3. Participant-aggregated analyses.** In the available-case average, each north-seat participant's transverse and non-transverse set rates were pooled over sets before differencing, and the differences were averaged with equal participant weight (percentile bootstrap over participants with 10,000 resamples; two-sided *P*, twice the one-sided proportion of 20,000 within-participant reallocations of the transverse labels). The strict complete-case average differenced each participant's mean of mE and mW from the mean of his non-transverse configuration means (sign-flip test, 200,000 draws). The within-north-participant configuration analysis instead retained all primary sets and residualized treatment against person indicators, cohort, and seat phase. CR2 used Satterthwaite degrees of freedom. CR1 used the small-sample factor $G/(G - 1) \times (N - 1)/(N - K)$, in which G is the number of clusters, N the number of observations, and K the number of parameters, including the person indicators, so its interval can be wider than the CR2 interval. Dyadic-overlap inference formed actor cluster scores for every participant appearing in either seat and subtracted duplicate same-dyad terms once. Because these methods address different aspects of repeated participation, they are reported together, without selecting the smallest *P* value.

**4. Primitive-process models.** Each logistic model of the four role-specific endpoints included indicators for the ten configurations, seat phase, set cohort, and actor fixed effects. The equally weighted transverse − non-transverse contrast was evaluated under six covariance structures (set, actor, pair-session, dyad, two-way actor and partner, and dyadic overlap) and was Holm-adjusted across the four endpoints within each structure. For the two-way structure, negative eigenvalues of the combined variance matrix were set to zero. Each endpoint was also fitted as a binomial mixed model with configuration, seat phase, and cohort as fixed effects and crossed random intercepts for actor, partner, and set (R lme4), and the same contrast was assessed by a Wald test (Table 3). Average marginal contrasts were computed by assigning each configuration to all observed rows. Incidental-parameter bias and multiplicity across role-specific tests limit the interpretation of these models. First-12 joint-outcome and parity-interaction summaries were reconstructed from the recovered record sheets.

**5. Cross-lag and outcome-history models.** Cross-lag sensitivity analyses used person-level actor and set random-intercept logistic HMC models on trials 2–20 of each set in the ten-configuration family. Current P2 one-stone choice was predicted from preceding P1 parity, and current P1 parity from preceding P2 choice, with configuration indicators, lag, the transverse × lag interaction, actor seat, seat phase, and set cohort as predictors. Four NUTS chains used 400 warm-up and 400 retained draws. Outcome-history models used the same trials. Because the previous outcome was determined by the two preceding actions, each model of current P1 odd generation or current P2 one-stone choice included three history terms: the actor's preceding action, the partner's preceding action, and whether the actor had obtained the black stone on the preceding trial. The models also included the interactions of all three terms with the transverse indicator, together with configuration, actor seat, seat phase, and cohort, and actor and set random intercepts (four NUTS chains, 500 warm-up and 500 retained draws). With both preceding actions in the model, the previous-outcome term represents their non-additive combination that defines the outcome, not a separable effect of reward. The two transverse × previous-outcome terms were Holm-adjusted using set-clustered logistic *P* values. As a sensitivity analysis, the running score (the actor's black stones minus the partner's before the current trial, standardized) and its interaction with the transverse indicator were added. Whether the transverse difference grew within a set was tested with a trial-level model of the north-seat outcome over trials 1–20 that included a transverse × trial-position term (trial position standardized) and north-participant, south-participant, and set random intercepts.

**6. Source-level contrasts.** Welch tests and percentile bootstraps with 10,000 resamples reproduced source-level contrasts. Component, condition × parity, geometric-class, and single-reference (mE or mW versus mN) contrasts used percentile bootstraps with 3 × $10^6$ resamples drawn within configuration, to reduce Monte Carlo uncertainty in tail probabilities (Fig. 2d; Fig. 3c; Supplementary Fig. 2a; Supplementary Table 3). Their two-sided $P$ values were twice the smaller proportion of bootstrap differences on either side of zero. Trial-position standard errors in Fig. 3a were binomial within configuration, and block-wise differences in Fig. 3e used percentile bootstraps with $10^5$ resamples drawn within configuration. Near-zero was compared with the transverse pair and the non-transverse family in the same way, with $10^6$ resamples, for the total rate, the first-12 components, and the set-level proportion of one-stone choices by the south-seat participant as P2 (Supplementary Fig. 3a). The direct RF1-versus-RF2 comparison used the same set-level rates and the same two-sided Welch test and percentile bootstrap. The four added RF1 conditions were compared with their same-background RF-off references in the same way, with $10^6$ resamples drawn within condition (Supplementary Table 5). For the RF-by-background contrast, the RF1 effect over each modified background pooled its two antenna positions, and the two backgrounds were weighted equally. The difference from the RF1 effect over the ambient field was tested with a Welch-type $t$ statistic (Satterthwaite degrees of freedom) and a percentile bootstrap ($10^6$ resamples drawn within condition). A quasi-binomial model of the set counts tested this difference on the log-odds scale. After all 342 sets were linked to pseudonymized participants, the contrast was refitted in the crossed-role binomial model of the primary analysis (north-seat, south-seat, and pair-session random intercepts; seat phase and cohort), with cluster-robust binomial and within-north-participant fixed-effect models as sensitivity analyses. These source-level procedures resample sets and do not reproduce the crossed-role dependence adjustment of the primary hierarchy. Equivalence of each non-transverse configuration with the pooled remaining seven was tested with two one-sided pooled-variance Student $t$-tests against the ±5-percentage-point bound. Jeffreys–Zellner–Siow Bayes factors used a Cauchy prior scale of $r = \sqrt{2}/2$.

**7. Homogeneity and selection-aware permutation.** Homogeneity of the eight non-transverse configurations was tested with a cluster-robust Wald test (7 degrees of freedom; pair-session clusters) in a binomial generalized linear model with seat phase and cohort, and with a likelihood-ratio test in the crossed-role mixed model. The equally weighted contrast of every possible pair of configurations with the remaining eight was recomputed under 100,000 condition-label permutations. The observed transverse contrast was then referred to the permutation distribution of the most extreme pair contrast, with each contrast centered on its permutation-null mean. The five symmetric pairs were Con and vertical-component inversion, mN and mS, mE and mW, mNE and mNW, and vertical-only down and up. Differences among the three non-transverse geometric classes were tested with a cluster-robust Wald test (2 degrees of freedom; linear model of set rates with configuration, cohort, and seat phase; pair-session clusters) and with condition-label permutation within the eight non-transverse configurations. The posterior probability that mE and mW occupied the two lowest ranks was computed from the standardized predictions of the primary model.

**8. Descriptor definitions.** For each measured north-seat vector, hOrthConflict = $1 - |\cos\theta|$, where $\theta$ is the angle between the condition and ambient horizontal vectors, and $\theta$ is undefined when the horizontal component $|B_h| = 0$. Departure from the geographic north–vertical plane was quantified as $1 - |\cos\varphi| = 1 - \sqrt{(B_x^2 + B_z^2)}/|B|$. The directional extractability index was DEI = $[1 - \exp(-|B|/B_0)] \times [1 - (|B_h|/|B|)(1 - |\cos\theta|)]$, with $B_0$ = 15 µT and near-zero assigned its zero-field limit. Nominal vertical-only conditions were excluded from horizontal-axis summaries despite their small measured residual horizontal components, whereas the full measured vectors were retained for DEI. Slopes and $R^2$ of the descriptor fits (Fig. 4d; Supplementary Note 3) were computed from the condition means. The north-transverse descriptor is $(\hat{b} \cdot \hat{t}_{mN})^2$, where $\hat{b} = B/|B|$ and $\hat{t}_{mN}$ is the horizontal unit vector perpendicular to the measured mN horizontal direction. Equivalently, it is $[|B_h|^2 - (B \cdot \hat{h}_{mN})^2]/|B|^2$, where $\hat{h}_{mN}$ is the mN horizontal unit vector embedded in three dimensions. Near-zero is assigned zero by convention. This descriptor is distinct from hOrthConflict and from $1 - |\cos\varphi|$.

**Supplementary Methods 5. Coil model parameters and validation.**

This section details the coil-model parameters and checks summarized in Methods (‘Directional-gradient model’). For condition c, $\mathrm{DGI}_c = \square\|\square\ \hat{b}_c(r)\|_F\rangle_{\mathrm{ROI}}$, where $\hat{b} = B/|B|$ and $\|\cdot\|_F$ is the Frobenius norm. Angled brackets denote the mean over the interior points of a head-scale cubic grid (region of interest, ROI). The primary ROI extended ±50 mm in x, y, and z at 5 mm spacing, and ±30 mm and ±80 mm regions and a 2.5 mm grid were sensitivity checks. DGI was set to zero for Con under the assumption of a spatially uniform ambient baseline and was undefined for near-zero. Central finite differences were evaluated on these grids.[21] DGI is determined by the configuration-specific coil model and is not independent of condition identity. It was therefore retained for artifact assessment rather than receptor inference.

The modeled frames were 1.89 × 1.89, 1.89 × 1.80, and 1.98 × 1.98 m for the north–south, east–west, and vertical pairs, respectively. They were separated by 1.029, 0.980, and 1.078 m, and each carried two windings of 87.5 turns. The separations are manufacturer design values. Direct measurement of the frames agreed with these values to within approximately 10 mm, with the conductor centers lying approximately 10 mm inside each nominal frame side. Across the 27 combinations of −10, 0, and +10 mm in the three pair separations, mE DGI ranged from 21% below to 33% above its nominal value, and the maximum angular deviation ranged from 0.011° to 0.019°. This represents a sensitivity range rather than a measured uncertainty bound. The maximum deviation is also below the approximately 0.13° angular scale associated with a transverse 0.1 µT reporting increment at 45 µT.

However, tabulation increments do not specify instrument accuracy or spatial uniformity. Earlier studies with the same coil apparatus measured the field distribution in the central region of the coils[9,17] but did not report values for the head-scale region analyzed here.

For the south-seat check (Methods), the measured south-seat Con vector served as the local ambient field, and the coil field of each configuration was set from the north-seat measurements. The mean difference in total intensity (|B|) between predicted and measured vectors was 1.7 μT, and the seat separation that best fitted the ten vectors was 1.32 m, compared with the nominal 1.30 m. This check covers the two seat positions rather than the head-scale field distribution.

**Supplementary Methods 6. Task posture observations and model inputs.**

This section details the behavioral observations, model structure, and timing assumptions of the posture model summarized in Methods ('Observed task posture and induction estimates'). These observations were not recorded during the sessions. They were compiled after the experiments by the principal investigator, who did not conduct the analyzed sessions, from interviews with all experimenters who conducted them. Participants detected the experimenter's hand cue with peripheral vision while facing forward in approximately 40% of trials and by turning the head toward the experimenter, who was seated to the east, in approximately 60% (approximately 20–80% across participants). While waiting, they faced straight ahead approximately 40% of the time and looked down toward the table approximately 60% of the time, and they faced forward while the opponent acted. Within 1–3 s of the step-1 cue, P1 turned the head and eyes toward the stone container, tilted the head, and took a handful, which was placed on the board within a further 1–2 s. Approximately 80% of the step-2 choices of one or two stones appeared to be made during the 1–3 s head turn and tilt toward the container, and approximately 20% with the head toward the container and the hand in it. The head contributed approximately 80% of the gaze shift toward the containers. Phase durations were obtained from the procedure video of the earlier study[1] (its Supplementary Video S1): approximately 34 s per trial, 3–5 s of waiting before the step-1 cue, approximately 5 s from the covered handful to the step-2 cue, 2–5 s for the step-2 action, and 8–15 s for the reveal.

The observed behavior and the measured table, container, and seat geometry defined a trial-cycle model of head orientation for each seat (200 simulated participants, six trials with alternating roles), with participant-level proportions drawn from the observed ranges. For each measured seat vector (Table 1), induction was taken as the summed absolute change of $\mathbf{B}\cdot\mathbf{n}$ over the six canals, using the pooled canal unit normals cited in Methods (lateral canal plane containing the interaural axis; anterior and posterior canals approximately 40° and 50° from the midsagittal plane) in Reid's coordinates, mirrored for the right labyrinth, with Reid's horizontal plane level in the upright posture. Orientation relative to a head- or eye-fixed transverse axis was taken as the squared field component along the interaural axis. Decision windows ran from the cue, or from 0.5 or 1 s before the choice, to the choice. Head turns and tilts were assumed to follow minimum-jerk profiles over the observed durations, and induction estimates assumed rigid head rotation about an axis perpendicular to the field, ignoring head translation and neural weighting.

Decision windows were shifted by sensory latencies of 0.05–0.15 s for vision and 0.15–0.35 s for vestibular signals and integrated over the preceding 0.09–0.25 s and 0.1–0.5 s, respectively, with choices made at the container assumed to occur while the head remained oriented toward it. These ranges bracket simple reaction times to light (220 ms) and to galvanic vestibular stimulation (438 ms),[22] the delay after threshold for detecting rotation (287 ms),[23] and the critical duration of the short-wavelength visual mechanism (approximately 0.1–0.25 s).[24]

**Supplementary Data 1. Condition-level descriptive statistics of the north-seat black stone rate.**

| Condition | Sets (n) | Mean (%) | SD (pp) | s.e.m. (pp) | 95% CI of the mean (%) |
|---|---|---|---|---|---|
| *Ten-configuration family (454 sets)* | | | | | |
| Con | 50 | 50.70 | 10.40 | 1.47 | 47.74 to 53.66 |
| V-inv | 46 | 49.78 | 12.29 | 1.81 | 46.13 to 53.43 |
| mN | 48 | 51.25 | 12.44 | 1.80 | 47.64 to 54.86 |
| mS | 48 | 49.38 | 12.57 | 1.81 | 45.72 to 53.03 |
| mE | 50 | 46.70 | 9.88 | 1.40 | 43.89 to 49.51 |
| mW | 50 | 46.70 | 9.56 | 1.35 | 43.98 to 49.42 |
| mNE | 40 | 52.12 | 11.15 | 1.76 | 48.56 to 55.69 |
| mNW | 34 | 50.74 | 12.56 | 2.15 | 46.35 to 55.12 |
| V-down | 44 | 51.25 | 12.39 | 1.87 | 47.48 to 55.02 |
| V-up | 44 | 50.91 | 12.26 | 1.85 | 47.18 to 54.64 |
| *Near-zero and radiofrequency reference conditions (146 sets)* | | | | | |
| Near-zero | 50 | 46.70 | 9.46 | 1.34 | 44.01 to 49.39 |
| RF1 | 50 | 46.60 | 10.37 | 1.47 | 43.65 to 49.55 |
| RF2 | 46 | 51.96 | 12.76 | 1.88 | 48.17 to 55.75 |
| *RF1 over modified backgrounds (150 added sets)* | | | | | |
| mN, vertical antenna | 46 | 50.43 | 11.63 | 1.72 | 46.98 to 53.89 |
| mN, 37° antenna | 30 | 51.67 | 10.53 | 1.92 | 47.73 to 55.60 |
| V-down, vertical antenna | 46 | 51.63 | 12.96 | 1.91 | 47.78 to 55.48 |
| V-down, 37° antenna | 28 | 52.86 | 13.01 | 2.46 | 47.81 to 57.90 |
| *Light-spectrum experiment (140 sets and the mN sets)* | | | | | |
| No goggles (mN sets) | 48 | 51.25 | 12.44 | 1.80 | 47.64 to 54.86 |
| Non-filtered goggles | 48 | 49.90 | 12.48 | 1.80 | 46.27 to 53.52 |
| >400 nm | 46 | 50.87 | 9.85 | 1.45 | 47.94 to 53.79 |
| >500 nm | 46 | 55.33 | 9.80 | 1.44 | 52.42 to 58.24 |

Rates are north-seat black stone rates of 20-trial sets (%). SD, standard deviation across sets; s.e.m., SD/$\sqrt{n}$; CI, t-based confidence interval of the set mean. These descriptive intervals treat sets as independent and therefore do not account for repeated participation or the crossed-role structure; the inferential analyses are those reported in the main text, the Tables, and the Supplementary Information. The identical means of mE, mW, and near-zero (46.70%) arose from non-overlapping sets (Supplementary Methods 3). In the radiofrequency-background experiment, RF1 was applied over the mN or V-down background with a vertical or 37°-tilted antenna; the same-background RF-off references are the mN and V-down rows of the ten-configuration family. The no-goggles condition comprises the same sets as mN. pp, percentage points.

**Supplementary Data 2. Values underlying the figures.**

Each block lists the values plotted in the corresponding figure panel. Set-level values plotted as dots in Fig. 2a are not listed; condition means and s.e.m. values are given in the Fig. 2a block, and set-level data are available on request (Data availability). Numbers are rounded for display (three decimals, or four significant digits below 1). pp, percentage points; CI, confidence interval; HMC, Hamiltonian Monte Carlo; OR, odds ratio.

**Fig. 1d. Measured two-seat field atlas (Table 1 vectors; azimuth from geographic north, east positive)**

Azimuth is not given where the horizontal component is below 1 μT (V-down, V-up, and near-zero).

| Configuration | Geometric class | North seat | | | | | South seat | | | |
|---|---|---|---|---|---|---|---|---|---|---|
| | | Bx (μT) | By (μT) | Bz (μT) | \|B\| (μT) | Azimuth (°) | Bx (μT) | By (μT) | Bz (μT) | \|B\| (μT) |
| Con | inclined | 30.7 | −6.6 | 32.4 | 45.1 | −12 | 29.3 | −4.1 | 33.2 | 44.5 |
| V-inv | inclined | 30.6 | −6.5 | −32.5 | 45.1 | −12 | 32.4 | −5.8 | 35.2 | 48.2 |
| mN | horizontal-only | 44.7 | −5.6 | 0 | 45 | −7 | 37.6 | −6.7 | 33.1 | 50.5 |
| mS | horizontal-only | −45 | 0 | 0.1 | 45 | 180 | −2 | −6.1 | 33.6 | 34.2 |
| mE | transverse | 0 | 45.1 | 0 | 45.1 | 90 | 17.7 | −8.2 | 34.5 | 39.6 |
| mW | transverse | −0.1 | −45.1 | 0.1 | 45.1 | −90 | 17 | −3.4 | 34.6 | 38.7 |
| mNE | horizontal-only | 42 | 16.2 | 0.2 | 45 | 21 | 35.4 | −6.2 | 35.2 | 50.3 |
| mNW | horizontal-only | 30.3 | −33.2 | 0.1 | 45 | −48 | 32.1 | −4 | 35.7 | 48.2 |
| V-down | vertical-only | −0.4 | 0 | 45 | 45 | — | 15.2 | −5.5 | 34.1 | 37.7 |
| V-up | vertical-only | 0.2 | 0.1 | −45 | 45 | — | 15.7 | −5.6 | 34.5 | 38.3 |
| near-zero | outside family | 0 | 0 | 0 | 0 | — | 17.8 | −5.8 | 34.4 | 39.2 |

### Fig. 1e. Analysis population of the ten-configuration family

| Item | Value |
|---|---|
| male participants | 47 |
| pair-sessions | 182 |
| north-seat sets | 454 |
| trials | 9080 |
| north-seat sets, Con | 50 |
| north-seat sets, V-inv | 46 |
| north-seat sets, mN | 48 |
| north-seat sets, mS | 48 |
| north-seat sets, mE | 50 |
| north-seat sets, mW | 50 |
| north-seat sets, mNE | 40 |
| north-seat sets, mNW | 34 |
| north-seat sets, V-down | 44 |
| north-seat sets, V-up | 44 |

### Fig. 2a. Configuration means ± s.e.m. of the north-seat black stone rate

Equally weighted mean of the eight non-transverse configurations: 50.766%

| Configuration | Sets (n) | Mean rate (%) | s.e.m. |
|---|---|---|---|
| Con | 50 | 50.7 | 1.471 |
| V-inv | 46 | 49.783 | 1.812 |
| mN | 48 | 51.25 | 1.796 |
| mS | 48 | 49.375 | 1.815 |
| mE | 50 | 46.7 | 1.397 |
| mW | 50 | 46.7 | 1.352 |
| mNE | 40 | 52.125 | 1.762 |
| mNW | 34 | 50.735 | 2.154 |
| V-down | 44 | 51.25 | 1.869 |
| V-up | 44 | 50.909 | 1.848 |

**Fig. 2b. Transverse (T) versus non-transverse (N8) contrasts (pp; 95% intervals)**

| Row | Estimate | 95% interval, lower | 95% interval, upper |
|---|---|---|---|
| Crossed-role binomial HMC, T vs N8 (primary) | −4.054 | −6.639 | −1.44 |
| mE vs N8 | −4.127 | −7.32 | −0.7793 |
| mW vs N8 | −3.981 | −7.511 | −0.5469 |
| Laplace mixed model (lme4) | −4.057 | −6.522 | −1.495 |
| Within north participant (CR2) | −4.102 | −6.723 | −1.481 |
| Pair-session wild cluster | −4.109 | −6.39 | −1.811 |
| Within pair-session (29) | −5.948 | −9.828 | −2.155 |
| Available-case person average | −2.964 | −5.438 | −0.3494 |
| Strict complete case (38) | −3.645 | −6.147 | −1.123 |
| T vs mN (same model) | −4.535 | −8.733 | −0.4199 |
| T vs N8, cohort 13 omitted (4 × (2000 + 5000)) | −3.099 | −6.164 | −0.03932 |
| T vs mN, cohort 13 omitted | −1.077 | −5.727 | 3.517 |

**Fig. 2c. Pair contrasts (pair mean minus mean of the other eight; pp), ranked**

selection-adjusted two-sided *P*: 0.0149 (five symmetric pairs)

selection-adjusted two-sided *P*: 0.0958 (all 45 pairs)

| Rank | Pair | Contrast (pp) | Symmetric pair |
|---|---|---|---|
| 1 | mE/mW | −4.066 | Yes |
| 2 | mE/mS | −2.394 | No |
| 3 | mW/mS | −2.394 | No |
| 4 | mE/V-inv | −2.139 | No |
| 5 | mW/V-inv | −2.139 | No |
| 6 | Con/mE | −1.566 | No |
| 7 | Con/mW | −1.566 | No |
| 8 | mW/mNW | −1.544 | No |
| 9 | mE/mNW | −1.544 | No |
| 10 | mE/V-up | −1.435 | No |
| 11 | mW/V-up | −1.435 | No |
| 12 | mN/mW | −1.222 | No |
| 13 | mW/V-down | −1.222 | No |
| 14 | mE/V-down | −1.222 | No |
| 15 | mN/mE | −1.222 | No |
| 16 | mE/mNE | −0.6752 | No |
| 17 | mW/mNE | −0.6752 | No |
| 18 | mS/V-inv | −0.4674 | No |
| 19 | Con/mS | 0.106 | No |
| 20 | mS/mNW | 0.1281 | No |
| 21 | mS/V-up | 0.2367 | No |
| 22 | Con/V-inv | 0.3608 | Yes |
| 23 | mNW/V-inv | 0.3828 | No |
| 24 | mN/mS | 0.4498 | Yes |
| 25 | mS/V-down | 0.4498 | No |
| 26 | V-inv/V-up | 0.4914 | No |
| 27 | V-inv/V-down | 0.7045 | No |
| 28 | mN/V-inv | 0.7045 | No |
| 29 | Con/mNW | 0.9562 | No |
| 30 | mS/mNE | 0.9966 | No |

**Fig. 2c (continued).**

| Rank | Pair | Contrast (pp) | Symmetric pair |
|---|---|---|---|
| 31 | Con/V-up | 1.065 | No |
| 32 | mNW/V-up | 1.087 | No |
| 33 | mNE/V-inv | 1.251 | No |
| 34 | Con/mN | 1.278 | No |
| 35 | Con/V-down | 1.278 | No |
| 36 | mNW/V-down | 1.3 | No |
| 37 | mN/mNW | 1.3 | No |
| 38 | V-down/V-up | 1.409 | Yes |
| 39 | mN/V-up | 1.409 | No |
| 40 | mN/V-down | 1.622 | No |
| 41 | Con/mNE | 1.825 | No |
| 42 | mNE/mNW | 1.847 | Yes |
| 43 | mNE/V-up | 1.955 | No |
| 44 | mN/mNE | 2.169 | No |
| 45 | mNE/V-down | 2.169 | No |

**Fig. 2d. Transverse pair versus each geometric class (bootstrap 95% CI)**

rate: class means (equal weight) T 46.700; horizontal-only 50.871; vertical-only 51.080; inclined 50.241; N8 50.766

Between-class difference: Wald *P* 0.794; configuration-label permutation within N8 (100,000): *P* 0.8652

| Reference | Difference (pp) | 95% interval, lower | 95% interval, upper | *P* |
|---|---|---|---|---|
| Horizontal-only | −4.171 | −6.789 | −1.533 | 0.001981 |
| Vertical-only | −4.38 | −7.543 | −1.207 | 0.006937 |
| Inclined | −3.541 | −6.491 | −0.5978 | 0.01836 |
| All eight non-transverse | −4.066 | −6.311 | −1.792 | 0.000528 |

**Fig. 3a. P(black | odd) by trial (first 12 trials); equally weighted means, binomial s.e. within configuration**

| Trial | Transverse (%) | Transverse s.e. | Transverse *n* | Non-transverse (%) | Non-transverse s.e. | Non-transverse *n* |
|---|---|---|---|---|---|---|
| 1 | 46.841 | 6.499 | 54 | 52.323 | 3.674 | 180 |
| 2 | 31.97 | 6.472 | 52 | 56.562 | 3.77 | 173 |
| 3 | 52.484 | 7.502 | 44 | 54.472 | 3.709 | 178 |
| 4 | 50 | 7.242 | 48 | 48.92 | 3.756 | 178 |
| 5 | 43.81 | 7.328 | 46 | 52.303 | 3.669 | 171 |
| 6 | 42.292 | 7.353 | 45 | 51.126 | 3.716 | 169 |
| 7 | 50.434 | 6.342 | 57 | 46.665 | 3.877 | 166 |
| 8 | 55.736 | 7.561 | 43 | 52.435 | 3.798 | 170 |
| 9 | 35.474 | 7.109 | 45 | 49.711 | 3.883 | 169 |
| 10 | 50 | 6.954 | 52 | 48.645 | 3.66 | 183 |
| 11 | 42.308 | 6.64 | 54 | 53.372 | 3.795 | 180 |
| 12 | 47.94 | 6.758 | 54 | 49.684 | 3.718 | 179 |

**Fig. 3b. Fixed-denominator components (first 12 trials; %), mean ± s.e.m.**

Equally weighted non-transverse means: odd 27.27%, even 25.05%

| Configuration | Sets (n) | Black-in-odd mean | Black-in-odd s.e.m. | Black-in-even mean | Black-in-even s.e.m. |
|---|---|---|---|---|---|
| Con | 50 | 30.667 | 2.149 | 22 | 2.256 |
| V-inv | 46 | 23.551 | 2.568 | 29.71 | 2.586 |
| mN | 48 | 31.597 | 2.683 | 20.486 | 1.937 |
| mS | 48 | 27.083 | 2.92 | 25.694 | 2.479 |
| mE | 50 | 22 | 2.402 | 26.667 | 2.234 |
| mW | 50 | 24.667 | 2.142 | 26.333 | 2.568 |
| mNE | 40 | 28.333 | 2.682 | 27.083 | 3.306 |
| mNW | 34 | 23.529 | 2.549 | 28.431 | 2.861 |
| V-down | 44 | 25.379 | 2.892 | 24.242 | 2.734 |
| V-up | 44 | 28.03 | 2.697 | 22.727 | 2.768 |

### Fig. 3c. Source-level contrasts (percentile bootstrap 95% CI; $3 \times 10^6$ resamples)

| Row | Difference (pp) | 95% interval, lower | 95% interval, upper | *P* |
|---|---|---|---|---|
| Total rate vs N8 | −4.066 | −6.311 | −1.792 | 0.000528 |
| Odd component vs N8 | −3.938 | −7.533 | −0.3107 | 0.03353 |
| Odd component vs horizontal-only | −4.302 | −8.365 | −0.2098 | 0.03949 |
| Odd component vs vertical-only | −3.371 | −8.326 | 1.553 | 0.1807 |
| Odd component vs inclined | −3.775 | −8.275 | 0.7319 | 0.1005 |
| Even component vs N8 | 1.453 | −2.289 | 5.243 | 0.4513 |

### Fig. 3d. Outcome history (HMC odds ratios, posterior medians and 95% intervals)

Holm-adjusted *P* of the two transverse × previous-outcome terms (GLM, set clusters): M1 0.4831; M2 0.4831

| Action | Term | Odds ratio | 95% interval, lower | 95% interval, upper |
|---|---|---|---|---|
| P1 odd generation | Previous black stone | 1.054 | 0.9566 | 1.161 |
| P1 odd generation | Transverse × previous black stone | 0.9115 | 0.7395 | 1.114 |
| P2 one-stone choice | Previous black stone | 1.015 | 0.9174 | 1.122 |
| P2 one-stone choice | Transverse × previous black stone | 1.122 | 0.9084 | 1.39 |

### Fig. 3e. Transverse minus non-transverse north-seat rate by five-trial block (percentile bootstrap 95% CI; $10^5$ resamples)

Transverse × trial position OR per SD: 1.036 (0.938 to 1.144)

| Trials | Difference (pp) | 95% interval, lower | 95% interval, upper |
|---|---|---|---|
| 1-5 | −5.423 | −10.386 | −0.411 |
| 6-10 | −4.259 | −8.99 | 0.5658 |
| 11-15 | −3.859 | −8.623 | 0.9499 |
| 16-20 | −2.723 | −7.142 | 1.643 |

**Fig. 4a. Configuration mean rate against north-seat |B|**

| Configuration | Sets (n) | North \|B\| (μT) | Mean rate (%) |
|---|---|---|---|
| Con | 50 | 45.12 | 50.7 |
| mN | 48 | 45.049 | 51.25 |
| V-inv | 46 | 45.109 | 49.783 |
| mS | 48 | 45 | 49.375 |
| mNE | 40 | 45.016 | 52.125 |
| mNW | 34 | 44.948 | 50.735 |
| mE | 50 | 45.1 | 46.7 |
| mW | 50 | 45.1 | 46.7 |
| V-down | 44 | 45.002 | 51.25 |
| V-up | 44 | 45.001 | 50.909 |
| near-zero | 50 | 0 | 46.7 |

**Fig. 4b. Means of the six horizontal profiles against hOrthConflict**

slope, $R^2$: −5.53, 0.76

| Configuration | hOrthConflict | Mean rate (%) |
|---|---|---|
| mN | 0.003793 | 51.25 |
| mS | 0.02234 | 49.375 |
| mNE | 0.1635 | 52.125 |
| mNW | 0.1857 | 50.735 |
| mE | 0.7898 | 46.7 |
| mW | 0.792 | 46.7 |

**Fig. 4c. Rates of the 11 conditions against DEI**

*r*, $R^2$: 0.89, 0.79

| Configuration | DEI | Mean rate (%) |
|---|---|---|
| Con | 0.9506 | 50.7 |
| mN | 0.9468 | 51.25 |
| V-inv | 0.9506 | 49.783 |
| mS | 0.929 | 49.375 |
| mNE | 0.7949 | 52.125 |
| mNW | 0.7736 | 50.735 |
| mE | 0.1998 | 46.7 |
| mW | 0.1977 | 46.7 |
| V-down | 0.95 | 51.25 |
| V-up | 0.9492 | 50.909 |
| near-zero | 0 | 46.7 |

**Fig. 4d. Rates of the ten configurations against 1 − |cos φ|**

slope, $R^2$: −4.00, 0.77

| Configuration | Midplane 1 − \|cos φ\| | Mean rate (%) |
|---|---|---|
| Con | 0.01076 | 50.7 |
| mN | 0.007756 | 51.25 |
| V-inv | 0.01044 | 49.783 |
| mS | 0 | 49.375 |
| mNE | 0.067 | 52.125 |
| mNW | 0.3259 | 50.735 |
| mE | 1 | 46.7 |
| mW | 0.9969 | 46.7 |
| V-down | 0 | 51.25 |
| V-up | $2.469 \times 10^{-6}$ | 50.909 |

**Fig. 4e. Radiofrequency contrasts (percentile bootstrap 95% CI; $10^6$ resamples; Welch *P*)**

| Item | Estimate | 95% interval, lower | 95% interval, upper | *P* | Note |
|---|---|---|---|---|---|
| RF1 − Con | −4.1 | −8.2 | −0.1 | 0.05123 | n 50 vs 50 |
| RF2 − Con | 1.257 | −3.43 | 5.839 | 0.6001 | n 46 vs 50 |
| RF1 − RF2 | −5.357 | −9.957 | −0.6957 | 0.02727 | n 50 vs 46 |
| RF1 over mN, vertical − mN | −0.8152 | −5.643 | 3.995 | 0.7435 | n 46 vs 48 |
| RF1 over mN, 37° − mN | 0.4167 | −4.667 | 5.479 | 0.8746 | n 30 vs 48 |
| RF1 over V-down, vertical − V-down | 0.3804 | −4.827 | 5.548 | 0.8871 | n 46 vs 44 |
| RF1 over V-down, 37° − V-down | 1.607 | −4.399 | 7.5 | 0.6049 | n 28 vs 44 |
| RF1 over modified backgrounds averaged − same-background RF-off | 0.2578 | −2.905 | 3.412 | 0.8744 | Welch-type t, Satterthwaite df 184.4; shared references counted once per bootstrap replicate |
| RF1 over Con − RF1 averaged over modified backgrounds | −4.358 | −9.483 | 0.7501 | 0.1002 | Welch-type t, Satterthwaite df 212.7; shared references counted once per bootstrap replicate |

**Fig. 4f. RF antenna axes relative to the measured static field (north seat): perpendicular fractions and the related field angles**

| Item | Value |
|---|---|
| Con inclination (°); horizontal azimuth (°, east positive) | 45.9; −12.1 |
| fraction of a vertical RF axis perpendicular to the static field | Con 0.70, mN 1.00, V-down 0.01 |
| fraction of the tilted axis perpendicular to an exactly vertical field | 0.60 |
| fraction of the 37° tilted RF axis (53° from horizontal in the north-vertical plane) perpendicular to the static field | Con 0.18, mN 0.80, V-down 0.61 |
| angle between the tilted axis and the Con vector (°) | 10.6 |

## Supplementary Fig. 1. Primary-model robustness and sensitivity (pp; 95% intervals)

| Row | Estimate | 95% interval, lower | 95% interval, upper |
|---|---|---|---|
| Crossed-role binomial HMC (primary) | −4.054 | −6.639 | −1.44 |
| Prior: tight | −4.005 | −6.546 | −1.427 |
| Prior: primary N | −4.077 | −6.653 | −1.52 |
| Prior: diffuse | −4.083 | −6.681 | −1.499 |
| Prior: robust | −4.076 | −6.655 | −1.509 |
| Bradley-Terry reimplementation | −4.104 | −6.719 | −1.551 |
| 13-condition hierarchy (600 sets) | −4.009 | −6.566 | −1.466 |
| Beta-binomial reweighting; rho ~ Uniform(0, 1) | −4.055 | −6.644 | −1.341 |
| Beta-binomial reweighting; rho ~ Uniform(0, 0.2) | −4.034 | −6.715 | −1.316 |
| Beta-binomial reweighting; rho ~ Uniform(0, 0.1) | −4.032 | −6.669 | −1.316 |
| Polya-Gamma Gibbs, HalfNormal(0.5) SD priors (as primary) | −4.07 | −6.618 | −1.52 |
| Polya-Gamma Gibbs, Inverse-Gamma(1, 1) variance priors | −4.298 | −7.397 | −1.168 |
| calendar: + linear study day | −4.074 | −6.718 | −1.452 |
| calendar: + natural cubic spline of study day (2 identifiable columns) | −4.144 | −6.763 | −1.627 |
| exclude recovered-record sets (31) | −4.481 | −7.279 | −1.732 |
| exclude label-reconciled sets (21) | −4.216 | −6.804 | −1.542 |
| exclude both | −4.752 | −7.554 | −1.99 |

## Supplementary Fig. 2a. Odd − even difference of the first-12 components (mean ± s.e.m.)

Transverse − non-transverse: −5.39 (−11.13 to 0.28; bootstrap *P* 0.0625)

Transverse − mN: −14.28 (−22.94 to −5.81; bootstrap *P* 0.0008)

| Configuration | Sets (n) | Odd − even mean | s.e.m. |
|---|---|---|---|
| Con | 50 | 8.667 | 3.571 |
| V-inv | 46 | −6.159 | 4.091 |
| mN | 48 | 11.111 | 3.624 |
| mS | 48 | 1.389 | 4.264 |
| mE | 50 | −4.667 | 3.468 |
| mW | 50 | −1.667 | 3.696 |
| mNE | 40 | 1.25 | 5.147 |
| mNW | 34 | −4.902 | 4.194 |
| V-down | 44 | 1.136 | 4.348 |
| V-up | 44 | 5.303 | 4.361 |

**Supplementary Fig. 2b. Transverse × lag odds ratios (HMC medians and 95% intervals)**

| Specification | Estimate | 95% interval, lower | 95% interval, upper | Note |
|---|---|---|---|---|
| preceding P1 parity → current P2 choice: T × lag OR (median, 95% PI); Pr(OR > 1) | 1.069 | 0.8738 | 1.311 | Pr(OR > 1) 0.729; divergences 0; 4 × (400 + 400) |
| preceding P2 choice → current P1 parity: T × lag OR (median, 95% PI); Pr(OR > 1) | 0.9464 | 0.7774 | 1.15 | Pr(OR > 1) 0.303; Pr(OR < 1) 0.697; divergences 0; 4 × (400 + 400) |

**Supplementary Fig. 3a. Near-zero and the transverse pair relative to the non-transverse family (bootstrap 95% CI; $10^6$ resamples)**

| Metric | Contrast | Estimate | 95% interval, lower | 95% interval, upper | *P* |
|---|---|---|---|---|---|
| Total rate | Near-zero − N8 | −4.066 | −6.936 | −1.176 | 0.006048 |
| Total rate | T − N8 | −4.066 | −6.32 | −1.795 | 0.000496 |
| Total rate | Near-zero − T | 0 | −3.2 | 3.2 | 1 |
| Black-in-odd | Near-zero − N8 | −4.938 | −9.531 | −0.4077 | 0.03273 |
| Black-in-odd | T − N8 | −3.938 | −7.531 | −0.3058 | 0.03374 |
| Black-in-odd | Near-zero − T | −1 | −6.333 | 4.167 | 0.7167 |
| Black-in-even | Near-zero − N8 | −0.04689 | −4.849 | 4.758 | 0.9845 |
| Black-in-even | T − N8 | 1.453 | −2.285 | 5.248 | 0.4518 |
| Black-in-even | Near-zero − T | −1.5 | −7 | 4 | 0.601 |
| Odd − even | Near-zero − N8 | −4.891 | −12.614 | 2.749 | 0.2115 |
| Odd − even | T − N8 | −5.391 | −11.15 | 0.2736 | 0.06209 |
| Odd − even | Near-zero − T | 0.5 | −8.167 | 9.167 | 0.9092 |
| South P2 one-stone | Near-zero − N8 | 0.2973 | −4.78 | 5.335 | 0.9036 |
| South P2 one-stone | T − N8 | 3.997 | 0.4169 | 7.582 | 0.02851 |
| South P2 one-stone | Near-zero − T | −3.7 | −9.4 | 1.9 | 0.206 |

**Supplementary Fig. 3b. RF amplitude along the representative posture path (forward-facing = 100)**

| Step | Yaw (°) | Direction | Head tilted forward | Relative amplitude, vertical antenna | Relative amplitude, 37° antenna |
|---|---|---|---|---|---|
| 1 | 0 | forward | 0 | 100 | 100 |
| 2 | 10 | toward experimenter hand cue | 0 | 98.051 | 97.74 |
| 3 | 20 | toward experimenter hand cue | 0 | 95.502 | 95.2 |
| 4 | 30 | toward experimenter hand cue | 0 | 90.705 | 90.417 |
| 5 | 40 | toward experimenter hand cue | 0 | 88.606 | 88.325 |
| 6 | 50 | toward experimenter hand cue | 0 | 86.207 | 85.934 |
| 7 | 40 | toward experimenter hand cue | 1 | 63.868 | 63.666 |
| 8 | 30 | toward experimenter hand cue | 1 | 52.174 | 52.009 |
| 9 | 20 | toward experimenter hand cue | 1 | 56.522 | 56.343 |
| 10 | 10 | toward experimenter hand cue | 1 | 56.971 | 56.791 |
| 11 | 0 | toward experimenter hand cue | 1 | 54.273 | 57.317 |
| 12 | −10 | toward stone container | 1 | 46.926 | 46.778 |
| 13 | −20 | toward stone container | 1 | 43.778 | 43.639 |

**Supplementary Fig. 4a. Light-spectrum experiment: north-seat black stone rates, mean ± s.e.m. across the analyzed sets**

Goggle conditions were reconstructed by elimination against the schedule and the per-condition rate distributions reported in the preprint (Supplementary Methods 3). The adopted assignment is one of several consistent solutions; all of them give the same condition means, per-condition rate distributions, and s.e.m. values, whereas the bootstrap intervals of the component contrasts may differ slightly under the other solutions.

| Condition | Sets (n) | Mean rate (%) | s.e.m. |
|---|---|---|---|
| No goggles (mN sets) | 48 | 51.25 | 1.796 |
| Non-filtered goggles | 48 | 49.896 | 1.802 |
| >400 nm | 46 | 50.87 | 1.452 |
| >500 nm | 46 | 55.326 | 1.445 |

**Supplementary Fig. 4b. >500 nm versus mN, odd and even components (pp; percentile bootstrap 95% CI, $10^6$ resamples within condition)**

Computed from the analyzed mN and >500 nm sets (fixed denominator of six north-P1 trials among trials 1-12).

Goggle conditions were reconstructed by elimination against the schedule and the per-condition rate distributions reported in the preprint (Supplementary Methods 3). The adopted assignment is one of several consistent solutions; all of them give the same condition means, per-condition rate distributions, and s.e.m. values, whereas the bootstrap intervals of the component contrasts may differ slightly under the other solutions.

| Component | Difference (pp) | 95% interval, lower | 95% interval, upper |
|---|---|---|---|
| Black-in-odd | −0.4378 | −7.88 | 6.975 |
| Black-in-even | 7.775 | 1.615 | 14.13 |

**Supplementary Fig. 5. Modeled DGI (a) and maximum angular deviation (b); ±50 mm cube, 5 mm grid**

| Configuration | DGI × $10^3$ ($m^{-1}$) | Maximum angle × $10^3$ (°) |
|---|---|---|
| mN | 0.1812 | 0.9046 |
| V-inv | 0.2767 | 1.24 |
| mE | 3.177 | 13.539 |
| mW | 2.371 | 10.371 |
| mS | 0.5606 | 1.433 |
| mNE | 1.46 | 3.455 |
| mNW | 1.677 | 5.583 |
| V-down | 0.5994 | 2.591 |
| V-up | 0.6866 | 2.703 |

## Supplementary References